\documentclass[twocolumn]{aastex631}
\usepackage{booktabs}
\usepackage{multirow}
\usepackage{xcolor}
\usepackage{amsmath}
\usepackage{graphicx}
\usepackage{makecell}

\shorttitle{Effects of Point Source Injection on Real/Bogus Classification}
\shortauthors{Lee et al.}
\graphicspath{{./}{figures/}}

\begin{document}

\title{Investigating the Effects of Point Source Injection Strategies on KMTNet Real/Bogus Classification}

\correspondingauthor{Gregory S.H. Paek}
\email{gregorypaek94@gmail.com}

\author[0009-0003-0777-6717]{Dongjin Lee}
\affiliation{Pohang University of Science and Technology (POSTECH), 
Pohang 37673, Republic of Korea}

\author[0000-0002-6639-6533]{Gregory S.H. Paek}
\affiliation{Institute for Astronomy, University of Hawaii, 2680 Woodlawn Drive, Honolulu, HI 96822, USA}

\author[0000-0002-3118-8275]{Seo-Won Chang}
\affiliation{SNU Astronomy Research Center, Seoul National University, 1 Gwanak-ro, Gwanak-gu, Seoul 08826, Republic of Korea}
\affiliation{Astronomy Program, Department of Physics \& Astronomy, Seoul National University, 1 Gwanak-ro, Gwanak-gu, Seoul 08826, Republic of Korea}

\author[0009-0000-2042-2348]{Changwan Kim}
\affiliation{SNU Astronomy Research Center, Seoul National University, 1 Gwanak-ro, Gwanak-gu, Seoul 08826, Republic of Korea}
\affiliation{Astronomy Program, Department of Physics \& Astronomy, Seoul National University, 1 Gwanak-ro, Gwanak-gu, Seoul 08826, Republic of Korea}

\author[0009-0003-1280-0099]{Mankeun Jeong}
\affiliation{SNU Astronomy Research Center, Seoul National University, 1 Gwanak-ro, Gwanak-gu, Seoul 08826, Republic of Korea}
\affiliation{Astronomy Program, Department of Physics \& Astronomy, Seoul National University, 1 Gwanak-ro, Gwanak-gu, Seoul 08826, Republic of Korea}

\author{Hongjae Moon}
\affiliation{SNU Astronomy Research Center, Seoul National University, 1 Gwanak-ro, Gwanak-gu, Seoul 08826, Republic of Korea}
\affiliation{Astronomy Program, Department of Physics \& Astronomy, Seoul National University, 1 Gwanak-ro, Gwanak-gu, Seoul 08826, Republic of Korea}

\author[0009-0003-3902-5678]{Seong-Heon Lee}
\affiliation{Pohang University of Science and Technology (POSTECH), 
Pohang 37673, Republic of Korea}

\author[0000-0001-7923-0674]{Jae-Hun Jung}
\affiliation{Pohang University of Science and Technology (POSTECH), 
Pohang 37673, Republic of Korea}

\author[0000-0002-8537-6714]{Myungshin Im}
\affiliation{SNU Astronomy Research Center, Seoul National University, 1 Gwanak-ro, Gwanak-gu, Seoul 08826, Republic of Korea}
\affiliation{Astronomy Program, Department of Physics \& Astronomy, Seoul National University, 1 Gwanak-ro, Gwanak-gu, Seoul 08826, Republic of Korea}



\begin{abstract}
Recently, machine learning-based \texttt{real}/\texttt{bogus} (RB) classifiers have demonstrated effectiveness in filtering out artifacts and identifying genuine transients in real-time astronomical surveys. 
However, the rarity of transient events and the extensive human labeling required for a large number of samples pose significant challenges in constructing training datasets for RB classification. 
Given these challenges, point source injection techniques, which inject simulated point sources into optical images, provide a promising solution.
This paper presents the first detailed comparison of different point source injection strategies and their effects on classification performance within a simulation-to-reality framework. 
To this end, we first construct various training datasets based on Random Injection (RI), Near Galaxy Injection (NGI), and a combined approach by using the Korea Microlensing Telescope Network datasets. Subsequently, we train convolutional neural networks on simulated cutout samples and evaluate them on real, imbalanced datasets from gravitational wave follow-up observations for GW190814 and S230518h.
Extensive experimental results show that RI excels at asteroid detection and bogus filtering but underperforms on transients occurring near galaxies (e.g., supernovae).
In contrast, NGI is effective for detecting transients near galaxies but tends to misclassify variable stars as transients, resulting in a high false positive rate.
The combined approach effectively handles these trade-offs, thereby balancing between detection rate and false positive rate. 
Our results emphasize the importance of point source injection strategy in developing robust RB classifiers for transient (or multi-messenger) follow-up campaigns.
\end{abstract}

\keywords{Transient detection (1957); Astronomy image processing (2306); Convolutional neural networks (1938)}

\section{Introduction} \label{sec:intro}
The field of time-domain transient astronomy has experienced rapid growth in recent years, driven by the advent of wide-field survey telescopes and advanced data processing techniques. These surveys, such as the Zwicky Transient Facility (ZTF; \citealt{2019PASP..131a8002B,2019PASP..131g8001G}), the Panoramic Survey Telescope and Rapid Response System (Pan-STARRS; \citealt{2010SPIE.7733E..0EK,2016arXiv161205560C}), and the upcoming Vera C. Rubin Observatory's Legacy Survey of Space and Time (Rubin/LSST; \citealt{2019ApJ...873..111I}), have revolutionized our ability to understand the nature of transients by detecting them in statistically significant numbers and uncovering previously unknown types.

In addition to these surveys, the \textit{Korea Microlensing Telescope Network} \citep[KMTNet;][]{2016JKAS...49...37K} has contributed to gravitational wave (GW) multi-messenger astronomy, searching for an optical counterpart of GW source. KMTNet comprises three $1.6\;\mathrm{m}$ telescopes, each with a wide field of view (\( \sim 4.0\;\deg^2 \)), located at distinct longitudes in the Southern Hemisphere—specifically at Cerro Tololo Inter-American Observatory (Chile; KMTNet-CTIO), South African Astronomical Observatory (South Africa; KMTNet-SAAO), and Siding Spring Observatory (Australia; KMTNet-SSO). While the primary objective of KMTNet is the detection of exoplanets in the Galactic bulge via gravitational microlensing \citep{2016JKAS...49...37K}, its large aperture, wide field of view, and continuous monitoring of the southern sky also make it highly suitable for optical follow-up observations of GW events. As part of the Gravitational-wave Electromagnetic Counterpart Korean Observatory (GECKO) project—a network of telescopes aimed at rapidly identifying optical counterparts to GW events \citep{2020grbg.conf...25I,2024ApJ...960..113P}—KMTNet has played a pivotal role in conducting GW follow-up observations throughout the O2 to O4 observational campaigns, including events such as GW170817 \citep{2017Natur.551...71T}, GW190408, GW190412, and GW190503\_185404 \citep{2021ApJ...916...47K}, GW190425 \citep{2024ApJ...960..113P}, GW190814 \citep{2019GCN.25342....1K}, S230518h \citep{2025ApJ...981...38P}, and S240422ed \citep{2024GCN.36343....1J}.
 
Difference image analysis (DIA; \citealt{1998ApJ...503..325A,2000A&AS..144..363A,2008MNRAS.386L..77B,2012MNRAS.425.1341B,2013MNRAS.428.2275B}) has been employed to identify variable sources by subtracting a reference image from a science image, effectively isolating them by removing emissions from host galaxies. However, many artifacts, so-called \texttt{bogus}, persist due to imperfect subtraction processes that fail to accurately match the point spread function (PSF) and background features inherent to the camera. Traditionally, skilled personnel manually verify whether detected sources in subtracted images are transients (\texttt{real}) or artifacts (\texttt{bogus}), but this process is time-consuming and burdensome, especially before the transient fades. The vast volume of imaging data generated by transient science, including GW follow-up efforts aimed at covering the extensive uncertainty in GW sky localization, results in an overwhelming number of transient candidates to inspect.

Consequently, many transient surveys, including searches for GW optical counterparts, have increasingly adopted machine learning-based \texttt{real}/\texttt{bogus} (RB) classifiers to efficiently classify large volumes of transient candidates \citep{10.1093/mnras/stz2357,Hosenie2021MeerCRABMC,otrain,Corbett_2023,Acero-Cuellar_2023}. However, training these RB classifiers requires extensive labeling of cutouts\footnote{In this paper, an ``image” refers to an optical image observed, and a ``cutout" or ``sample" denotes a single- or multi-channel image cropped around a point source, used as neural network input.} by human experts, which is highly inefficient. Furthermore, the rarity of transient events leads to a scarcity of labeled genuine transient cutouts, resulting in a class imbalance problem. To address these challenges, recent studies have introduced simulation-based data generation techniques that represent point sources as PSFs and inject them into science images \citep{2015AJ....150...82G, 10.1093/mnras/staa3096, otrain,10.1093/mnras/stab633,Corbett_2023}.

When training RB classifiers with cutouts containing injected PSFs, it is essential to consider both the PSF itself and its surrounding environment to generalize the trained classifiers to real-world observations. This work is focused on developing a classifier efficient at detecting \textit{off-nuclear} transients.
Explosive transients, such as supernovae (SNe) and kilonovae (KNe) from neutron star mergers associated with gamma-ray bursts (GRBs; \citealt{2017Natur.551...80K,2019LRR....23....1M}), typically occur near or within galaxies \citep{2013ApJ...776...18F,2014ARA&A..52...43B} and are absent in reference images, which are taken prior to the transient and contain only the background environment such as the host galaxy.
Short GRBs and KNe often show large offsets from their host galaxies due to natal kicks \citep{2010ApJ...722.1946B}, and some appear hostless or are associated with dusty, infrared-luminous galaxies that remain undetected in optical surveys \citep{2022MNRAS.515.4890O}.
To reflect these scenarios, PSFs should be injected near galaxies with diverse offsets and background conditions.
In contrast, solar system objects such as asteroids and comets move across the sky independently of galaxy positions and can be simulated with randomly injected PSFs.
However, there has been a lack of analysis on how different injection strategies affect the performance of RB classifiers in realistic environments.
Most previous studies have included observed transients in training or fine-tuning \citep{10.1093/mnras/stab633, Corbett_2023} and evaluated classifiers on balanced test datasets \citep{otrain, Corbett_2023}, which may not represent the imbalanced distribution of \texttt{real} and \texttt{bogus} cutouts encountered in practice.

\begin{table*}[t]
\centering
\caption{Summary of observational information for science (\texttt{sci}) and reference (\texttt{ref}) images taken with the three KMTNet observatories. Each \texttt{sci}–\texttt{ref} pair corresponds to a matching field and filter combination. Source density is calculated from Gaia DR3 sources within the magnitude range $R_p = 16\text{–-}21$ mag.}
\label{tab:img_qc}
\setlength{\tabcolsep}{4.5pt}
\small
\begin{tabular}{ccccccccccc}
\toprule
Class & Site & Filter & Exposure & R.A. & Dec. & Date & Seeing & Depth & Source Density & Moon Separation \\
      &      &        & [sec]    & (J2000) & (J2000) & (UTC) & [arcsec] & [AB mag] & [deg$^{-2}$] & [deg] \\
\midrule
\texttt{sci} & SAAO & R & 360 & 03:37:22 & $-74{:}00{:}00$ & 2023-03-08 & 2.12 & 21.47 & 1842 & 101.23 \\
\texttt{sci} & SAAO & I & 240 & 04:05:06 & $-76{:}00{:}00$ & 2023-03-09 & 1.45 & 21.49 & 2018 & 98.71 \\
\texttt{sci} & CTIO & R & 360 & 07:44:31 & $-60{:}00{:}00$ & 2023-05-19 & 2.18 & 21.38 & 9952 & 86.41 \\
\texttt{sci} & CTIO & I & 240 & 04:05:06 & $-76{:}00{:}00$ & 2023-03-10 & 1.98 & 21.37 & 2018 & 98.67 \\
\texttt{sci} & SSO  & R & 360 & 04:04:32 & $-74{:}00{:}00$ & 2023-03-08 & 1.71 & 21.34 & 2150 & 99.51 \\
\texttt{sci} & SSO  & I & 240 & 04:05:06 & $-76{:}00{:}00$ & 2023-03-09 & 1.94 & 21.16 & 2018 & 98.77 \\
\midrule
\texttt{ref} & SAAO & R & 480 & 03:37:22 & $-74{:}00{:}00$ & 2020-10-19 & 1.80 & 22.38 & 1842 & 86.24 \\
\texttt{ref} & SSO  & I & 600 & 04:05:06 & $-76{:}00{:}00$ & 2021-01-16 & 1.69 & 22.05 & 2018 & 102.88 \\
\texttt{ref} & CTIO & R & 480 & 07:44:31 & $-60{:}00{:}00$ & 2021-02-19 & 1.79 & 22.77 & 9952 & 76.99 \\
\texttt{ref} & SSO  & I & 600 & 04:05:06 & $-76{:}00{:}00$ & 2021-01-16 & 1.69 & 22.05 & 2018 & 102.88 \\
\texttt{ref} & SSO  & R & 720 & 04:04:32 & $-74{:}00{:}00$ & 2021-01-16 & 2.69 & 22.24 & 2150 & 101.72 \\
\texttt{ref} & SSO  & I & 600 & 04:05:06 & $-76{:}00{:}00$ & 2021-01-16 & 1.69 & 22.05 & 2018 & 102.88 \\
\bottomrule
\end{tabular}
\end{table*}

In this paper, we present a detailed comparison of different point source injection strategies and their effects on RB classifier performance within a simulation-to-reality scenario. 
In addition, we develop an RB classifier specifically optimized for the KMTNet, which addresses its imaging characteristics. 
Specifically, we train convolutional neural networks (CNNs) using cutout samples generated from KMTNet images containing injected point sources that reflect the characteristics of actual point sources observed by KMTNet.
Subsequently, we evaluate the trained classifiers on real-world imbalanced observations without any additional fine-tuning on transient samples to clearly dissect the influence of different injection strategies.
The experimental results emphasize that the injection strategy is just as critical as PSF modeling in developing RB classifiers.

The rest of the paper is organized as follows. In Section \ref{sec:injection_and_tsp}, we introduce the KMTNet images used in this work, the point source injection techniques, and the transient search pipeline (TSP). Section \ref{sec:dataset} details the construction of training and test datasets. Section \ref{sec:model} specifies the architecture of RB classifiers and hyperparameters used in training the classifiers. The experimental results are presented in Section \ref{sec:experiment}. Finally, we provide a brief conclusion and discuss future work in Section \ref{sec:conclusion}.

\section{KMTNet Images, Point Source Injection, and Transient Search Pipeline} 
\label{sec:injection_and_tsp}

\subsection{KMTNet Image Reduction for Training Dataset}\label{sec:img}
The KMTNet images utilized in this study are produced through median stacking of two or three sequential exposures, each with an exposure time of $120$ seconds. As the KMTNet image sensor consists of four segmented chips separated by gaps, the multiple pointings are dithered to fill these gaps, resulting in a final stacked image with a field of view of approximately $2\times 2\:\text{deg}^{2}$. This median-stacking approach effectively preserves the PSF and enhances the signal-to-noise ratio (SNR) of extragalactic sources, while fast-moving objects such as nearby asteroids may be partially removed by the stacking process.

The training dataset used in this study comprises six science images obtained with the $R$- and $I$-band filters from each of the three KMTNet observatories: CTIO, SAAO, and SSO. These images represent a limited but representative subset of standard KMTNet Target-of-Opportunity (ToO) observations. 
Each science image has an associated reference image, which covers the same field but was observed at a different epoch. All images within a science$-$reference pair were processed using the KMTNet\_ToO pipeline\footnote{\href{https://github.com/jmk5040/KMTNet_ToO}{https://github.com/jmk5040/KMTNet\_ToO}} (Jeong et al., in preparation), which includes astrometric calibration, photometric zero-point homogenization, image stacking, and bad pixel masking.
Table~\ref{tab:img_qc} summarizes the key observational properties, including the observing site, filter, observation date, median seeing, 5$\sigma$ limiting magnitude, stellar source density, and Moon separation.

The 5$\sigma$ limiting depths range from $R = 21.3\text{–-}21.5$ AB mag and $I = 21.2\text{–-}21.5$ AB mag, with seeing values between $1.45$\arcsec and $2.18$\arcsec. For comparison, the average depths of KMTNet ToO images taken during GW follow-up campaigns were $R = 21.55$ and $I = 21.59$ AB mag (Jeong et al., in preparation), indicating that the selected images are consistent with, or slightly shallower than, typical ToO image quality.

\begin{figure*}
   \centering
   \gridline{\fig{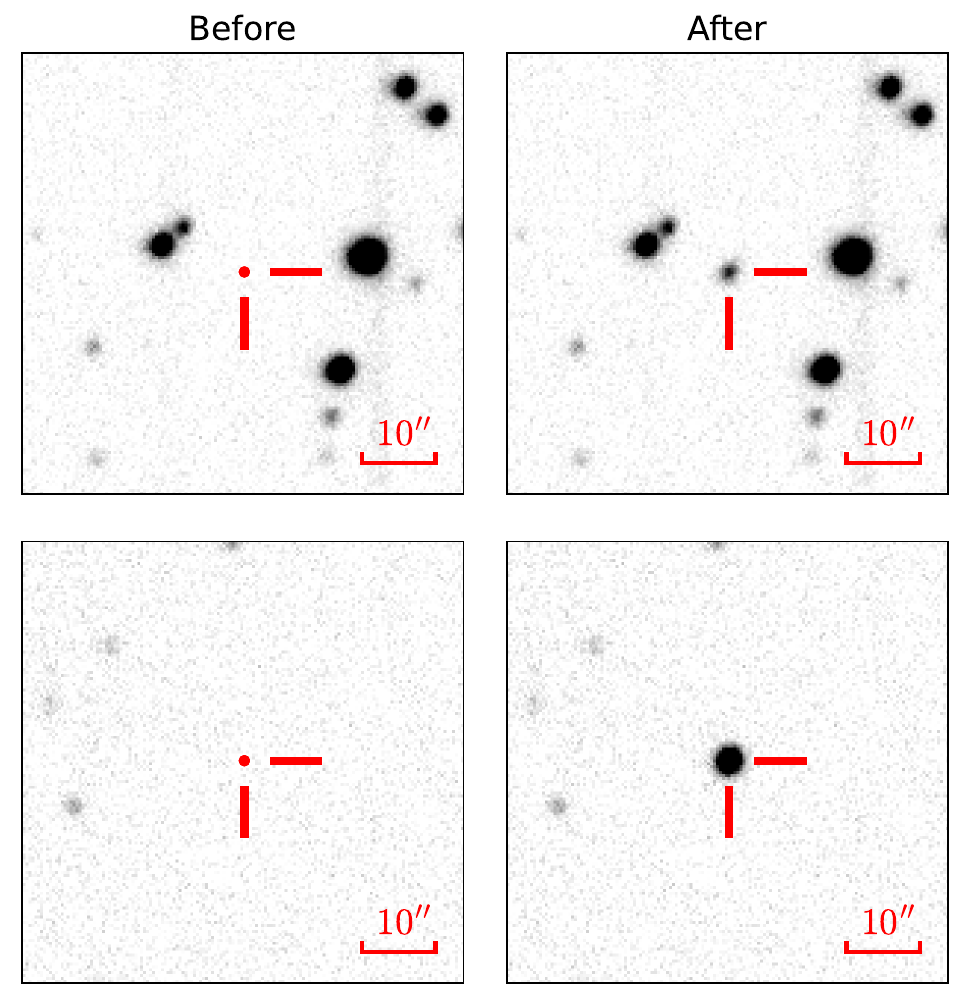}{0.49\linewidth}{(a) RI}
             \fig{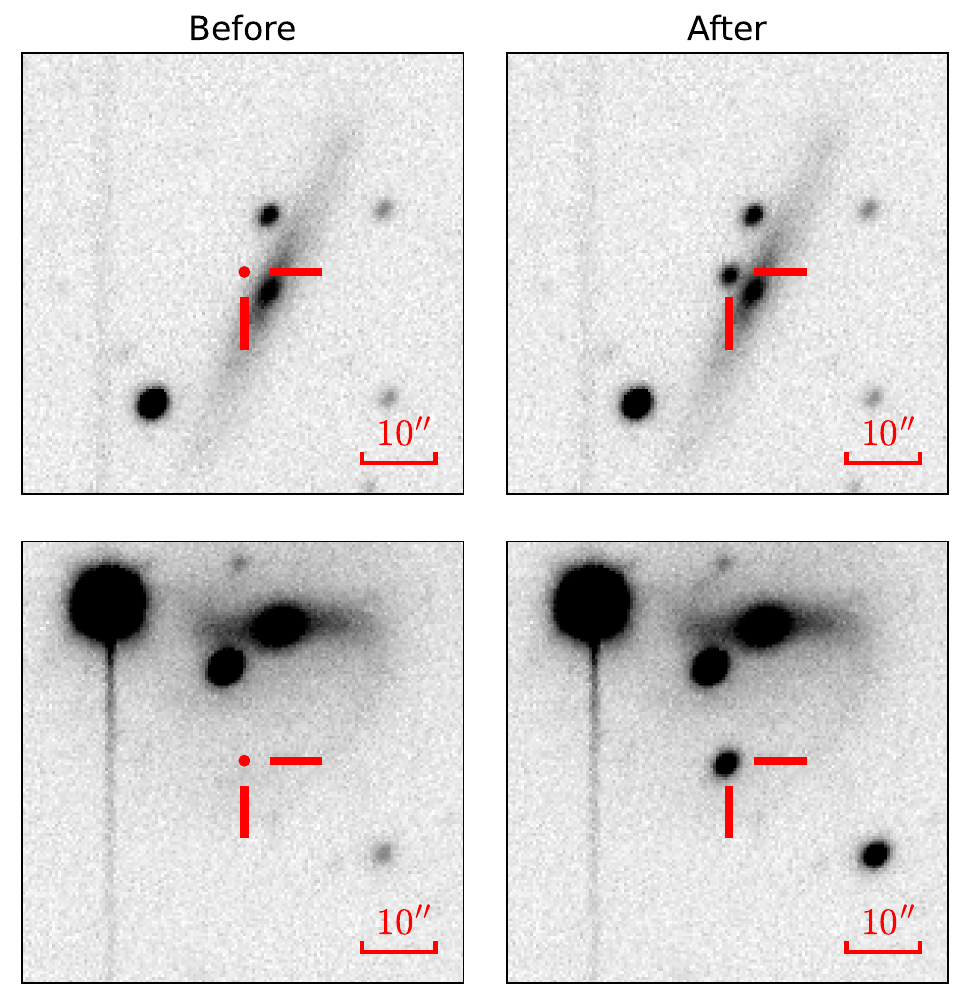}{0.49\linewidth}{(b) NGI}}
   \caption{Examples of the (a) RI and (b) NGI methods. The location of the injected source is marked with a red dot. (a) The RI method yields injected sources relatively isolated from other point sources. (b) The top example shows the injected source located approximately $4\arcsec$ away from the center of its associated galaxy, which is a relatively short distance. In contrast, the injected source in the bottom example is $19\arcsec$ away from the center of its associated galaxy, which is a significantly greater distance, as the corresponding galaxy is notably large within the image.}
   \label{fig:RI_cutout}
\end{figure*}

The stellar source density in most fields is approximately $2000 \:\text{deg}^{-2}$; however, the $R$-band image from CTIO corresponds to a relatively crowded field with a density of nearly $10,000 \:\text{deg}^{-2}$. These variations in crowding and seeing reflect the range of conditions under which KMTNet observations are typically conducted, though we acknowledge that the present images do not encompass the full diversity of observing conditions, which can affect difference imaging performance.

\subsection{Point Source Injection}\label{sec:ri}
In this section, we first explain the process of generating point sources from images using a PSF model that reflects the characteristics of transients as point sources. Then, we introduce two strategies for placing point sources: Random Injection (RI), where point sources are randomly injected across images, and Near Galaxy Injection (NGI), where point sources are injected close to galaxies (see Figure \ref{fig:RI_cutout}). 
The RI method is designed to model relatively isolated point sources, such as asteroids or hostless transients.
On the other hand, the NGI method is tailored to effectively capture transients occurring near host galaxies.

To construct the PSF model for each science image in Table \ref{tab:img_qc}, we first detect sources in the image using Source Extractor (\texttt{SExtractor}; \citealt{1996A&AS..117..393B}). We refer to the resulting catalog of detected sources as the \textit{KMTNet catalog}.
For the parameters in \texttt{SExtractor}, \texttt{PIXEL\_SCALE} is set to $0.4$ as the pixel scale of KMTNet is $0.4\arcsec\:\text{pixel}^{-1}$.
The \texttt{PHOT\_APERTURES} parameter is chosen to be $12.5$ pixels to ensure it is sufficiently larger than the average seeing of the science images, which is $2.1$ arcseconds (approximately $6$ pixels).
\texttt{MAG\_ZEROPOINT} is adjusted based on the header of each image. We use the default configuration for the remaining parameters.

From the created KMTNet catalog, we further select sources based on their \texttt{CLASS\_STAR} values and magnitudes to generate the PSF model. Here, the \texttt{CLASS\_STAR} value, ranging from 0 (extended sources) to 1 (point sources), is an estimator provided by \texttt{SExtractor} that quantifies the likelihood of a source being point-like. We select point sources with \texttt{CLASS\_STAR} values larger than $0.9$. Additionally, point sources brighter than $14$ AB mag are excluded to avoid saturation and distortion, and those fainter than the 5$\sigma$ limiting magnitude are filtered out as they are too faint to detect.

Since KMTNet images are acquired under various weather conditions, their PSF shapes vary between images. Therefore, we perform the PSF modeling individually for each image. 
Furthermore, to consider the position dependence of the PSF shapes due to the wide-field coverage of the KMTNet telescopes, each image is divided into a $4 \times 4$ grid, and a PSF model is extracted from each region of the grid.
In KMTNet images, pixel scale is $0.4\arcsec\:\text{pixel}^{-1}$ at the center and decreases to $0.385\arcsec \text{pixel}^{-1}$ at the edges.
To extract a PSF model from each region of the grid, we utilize PSF Extractor (\texttt{PSFEx}; \citealt{PSFex}). The size of the PSF model is set to $51 \times 51$ pixels ($20\arcsec \times 20\arcsec$), which is $10$ times the typical seeing, to ensure sufficient capture of the PSF wings. Here, a Moffat function is used for a better PSF modeling. 

\begin{figure}
    \centering
    \includegraphics[width=0.4\textwidth]{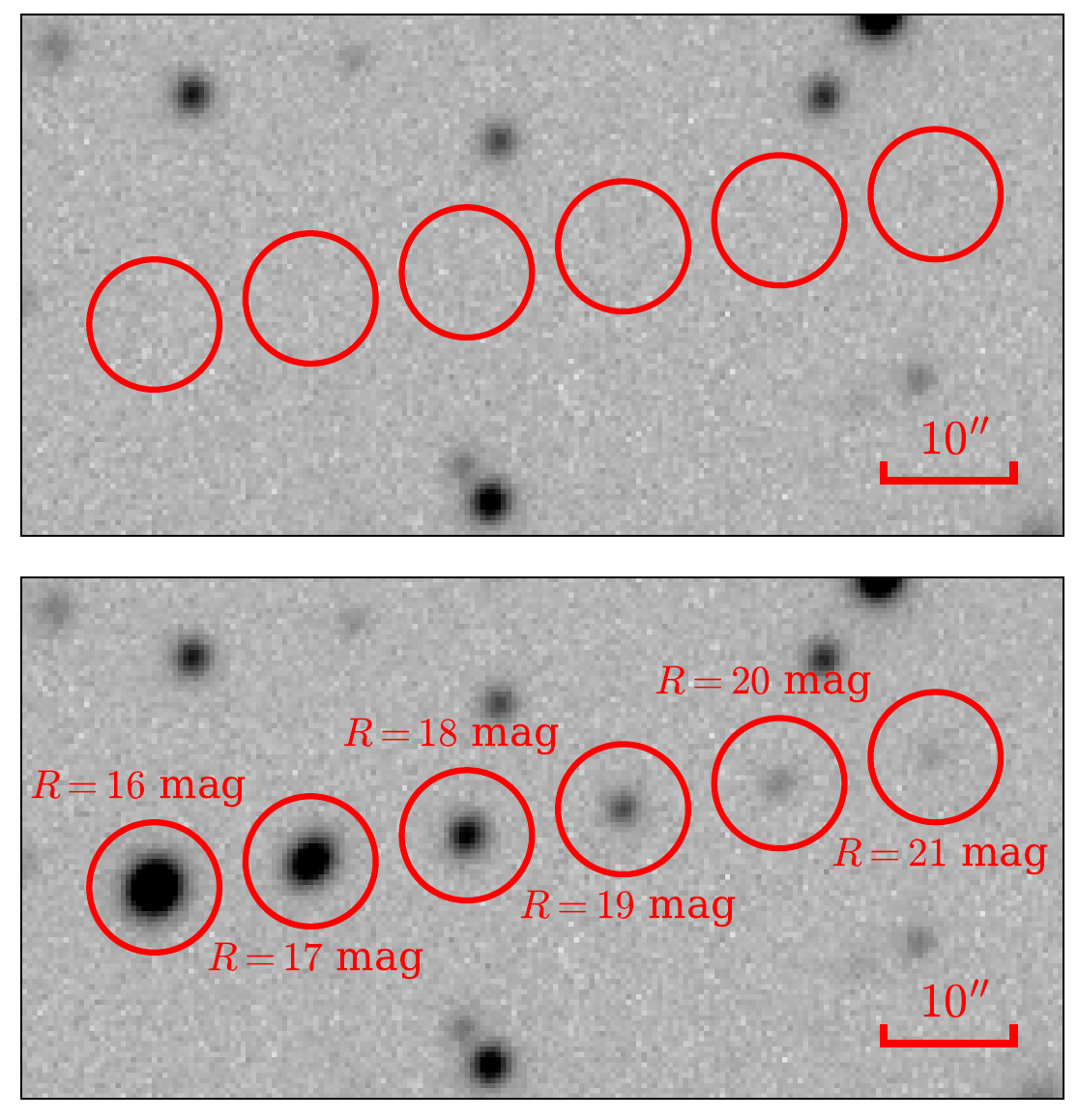}
    \caption{Examples of injected sources with $R$-band AB magnitudes from $16$ to $21$ in steps of $1$ magnitude. The image is taken from KMTNet-CTIO $R$-band image. The top figure shows the science image before injection while the bottom one illustrates injected sources with different magnitudes.}
    \label{fig:injection_ex}
\end{figure}

Subsequently, we create a pool of simulated sources for each grid region by using the extracted PSF model across magnitudes uniformly sampled from $16$ to the $5\sigma$ limiting depth.
In the previous filtering process of the KMTNet catalog, the lower limit of magnitude was set to $14$ to exclude excessively bright sources. 
For the simulated source generation, however, we adopt a limit of $16$ mag. 
This choice reflects several considerations: KNe are intrinsically faint ($M_{\rm AB}\approx-16$; \citealt{2017Natur.551...80K}), with the nearest observed event GW170817 at approximately $40$~Mpc peaking around $18$ mag (\citealt{2017ApJ...848L..13A}), making counterparts brighter than $16$ mag highly unlikely in realistic GW follow-up. 
Moreover, sources at approximately $16$ mag have sufficiently high SNR and essentially share the same PSF properties as those brighter than $14$ mag, so including them would add no substantial benefit to the training and could introduce overfitting.
Figure \ref{fig:injection_ex} shows the injected sources with different magnitudes.
Finally, the PSF model is adjusted for pixel values according to the magnitude of each simulated source, and noise uniformly sampled from $\left\{-\sqrt{\nu}, 0, \sqrt{\nu} \right\}$ is added to each pixel where $\nu$ is the pixel value.

\begin{figure}[t]
    \centering
    \includegraphics[width=0.9\linewidth]{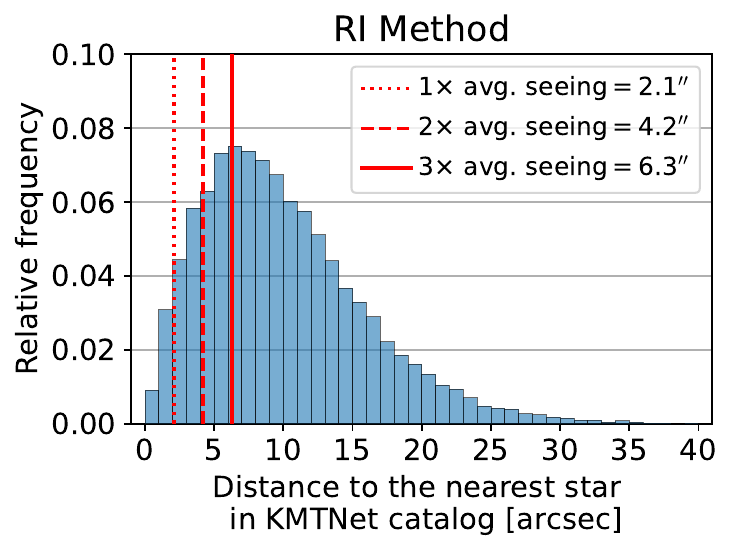}
    \caption{Histogram of angular distances between injected sources under the RI method and their nearest stars in the KMTNet catalog. Three vertical lines indicate the first, second, and third quartiles.}
    \label{fig:dist_kmtnet}
\end{figure}

For the RI method,  we inject $3000$ sources for each science image, resulting in a total of $18,000$ sources across the six images. 
To avoid overlap between two injected sources, we first uniformly sample $3000$ injection coordinates over each image. If any pair of coordinates lies closer than the size of the PSF model ($51 \times 51$ pixels), one of them is reassigned to a new random coordinate. We repeat this process until no coordinate pairs overlap to reflect the rarity of two independent transients appearing at the same location. Then, for each coordinate, we inject a point source randomly selected from the pool of simulated sources corresponding to its grid region.
Figure \ref{fig:RI_cutout}(a) illustrates examples of the RI method where injected sources are randomly distributed across the image regardless of the presence of galaxies. As shown in the bottom example, the injected point source can be located in empty space.
Importantly, even though injected sources do not overlap with each other, they may still be located near actual point sources, which enables the RI method to capture cases where transients appear close to other sources. Figure \ref{fig:dist_kmtnet} shows the histogram of distances between the injected sources under the RI method and their nearest stars in the KMTNet catalog: $4.41\%$ of injected sources are located within $2.1\arcsec$ (the average seeing) of an actual source, $15.52\%$ within twice the average seeing, and $30.02\%$ within three times the average seeing.

\begin{figure}[t]
    \centering
    \includegraphics[width=0.9\linewidth]{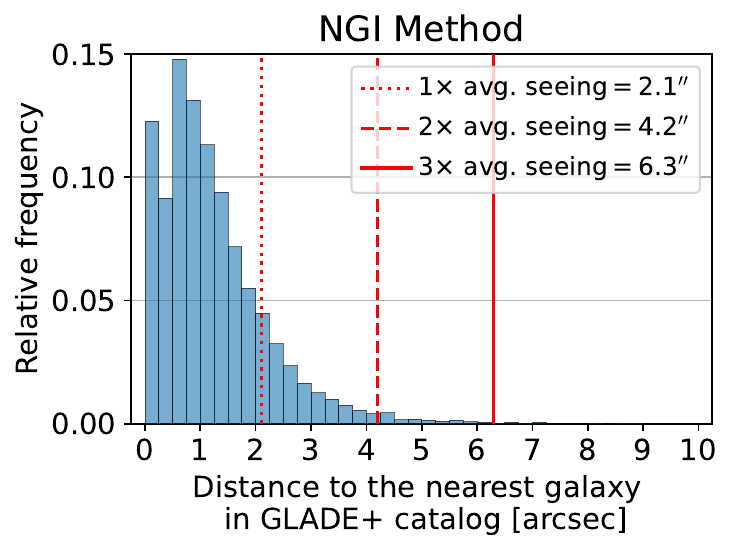}
    \caption{Histogram of angular distances between injected sources under the NGI method and their associated galaxies in the GLADE+ catalog. Three vertical lines indicate the first, second, and third quartiles.}
    \label{fig:Dist_Hist}
\end{figure}

For the NGI method, we use the GLADE+ catalog \citep{D_lya_2022}, an extended version of the GLADE galaxy catalog that integrates data from multiple astronomical surveys. It includes galaxies and quasars, which we use to identify host galaxies within the coordinate range of each science image for PSF injection.
In the NGI method, we inject one point source for each galaxy in the GLADE+ catalog.
To this end, we first categorize the GLADE+ galaxies into two groups based on whether they are included in the KMTNet catalog or not.
For a GLADE+ galaxy in the KMTNet catalog, its size in each science image is specified by the \texttt{A\_IMAGE} parameter from \texttt{SExtractor}.
Then, we determine the coordinates for injection by uniformly sampling coordinates from a square region centered on the galaxy with a side length of $2\times$\texttt{A\_IMAGE}.
For a GLADE+ galaxy not in the KMTNet catalog, a point source is injected at the coordinates of the galaxy. 
In this case, the injected source may appear isolated as the associated GLADE+ galaxy is too faint to be visible, which prompts the NGI method to marginally reflect the characteristics of asteroids and hostless transients.
Consequently, we inject $19,203$ sources using the NGI method across the six science images. Among them, $17,368$ sources are placed near GLADE+ galaxies in the KMTNet catalog. As shown in Figure \ref{fig:RI_cutout}(b), the point sources injected with the NGI method are typically located near galaxies.
Figure \ref{fig:Dist_Hist} presents the histogram of distances between injected sources under the NGI method and their associated galaxies in the GLADE+ catalog. Among them, $84.86\%$ lie within the average seeing of their associated galaxies, and $98.24\%$ within twice the average seeing.

\subsection{Transient Search Pipeline}\label{ssec:pipeline}

In this section, we describe the process of identifying transient candidates in the KMTNet images. This process utilizes sets of science and reference images. The science images are KMTNet images with injected point sources, as described in the previous section, whereas the reference images consist of those taken from earlier epochs. Before performing DIA, it is essential that the images undergo a consistent reduction pipeline. This includes steps such as astrometry, zero-point homogenization, artifact masking, and image stacking. Once the science and reference images are prepared, we can subtract one from the other and produce cutout samples of detected sources that meet specific filtering criteria.

The first step in TSP is astrometry, for which we use the \texttt{SCAMP} software \citep{2006ASPC..351..112B}. In this process, the source positions in the KMTNet images are compared with the Gaia EDR3 catalog as a reference \citep{2021A&A...649A...1G}. Based on this analysis, the appropriate coordinate system is updated in the header information. For quality assurance of the astrometry, each image is then segmented into 64 sections. We then calculate the separation of the image sources with the closest catalog sources. If fewer than three sections exhibit a root-mean-square error greater than $1$ arcsecond in separation distribution, the astrometry is deemed successful for the entire field, and the image proceeds to the stacking phase.

The next step involves photometric zero-point calibration. The wide field of view of KMTNet is read out through multiple ports, each of which exhibits slight differences in sensitivity. This causes zero-point variation across the field, requiring homogenization to ensure accurate flux scaling during stacking and subtraction (Jeong et al. in preparation). To assess the variation in zero-points among KMTNet images, we utilize the APASS DR14 for $BVR$ filters \citep{2014CoSka..43..518H} and the SkyMapper DR3 catalog for $i$ and $z$ filters \citep{2019PASA...36...33O} photometry. By comparing KMTNet photometry with these reference catalogs, we conduct a flux scaling process to standardize the zero-point across the entire field to $30$ AB mag.

In addition, various factors can create artifacts in KMTNet images, which disrupt both photometry and the process of identifying transients. To mitigate this, it is necessary to generate a defect mask that identifies the problematic regions of the pixels. The defect mask addresses several issues, including cosmic rays, bad pixels on the KMTNet image sensor, and cross-talk. Cosmic rays are detected using the \texttt{astroscrappy} module \citep{2001PASP..113.1420V}, while regions of bad pixels are identified using an image map provided by the KMTNet team. Cross-talk, which refers to the electronic interference effects that occur during the simultaneous readout of multiple channels, is managed by calculating and marking the positions affected by bright stars \citep{2016PKAS...31...35K}.

All the generated science images and defect mask images are then stacked using the \texttt{SWarp} software \citep{2010ascl.soft10068B}. We employ the median operation as the combination method, select \texttt{EQUATORIAL} for the \texttt{CELESTIAL\_TYPE}, and \texttt{TAN} for the \texttt{PROJECTION\_TYPE} to establish the coordinate projection method for the final stacked image. This process ensures that the science images have the same image projection and astrometric alignment as the reference image, preparing them for DIA.

In the processed KMTNet science images, we additionally inject point sources as described in Sections \ref{sec:ri}. During this process, noise consistent with a Poisson distribution is added. The sources are placed in regions not flagged by the defect mask to avoid problematic areas.

The next step involves the subtraction process, using the point source injected KMTNet images and the reference images. For this, we employ the \texttt{HOTPANTS} software \citep{2015ascl.soft04004B}, which facilitates image subtraction through PSF convolution and flux normalization. To optimize the subtraction process, we implement several key settings. Firstly, we specify the positions of stars that appear in both the science and reference images in a stamp file. Additionally, by comparing the full width at half maximums (FWHMs) of the science and reference images, we determine which image has a sharper PSF and perform convolution to match it to the less sharp PSF of the other image. The previously defined defect mask is also utilized during this process to exclude defective regions from affecting the convolution. Finally, considering the extensive field covered by a KMTNet image, the entire image is divided into 16 sections, with solutions computed separately for each. These steps ensure the production of a properly subtracted image.

The difference image obtained after the subtraction process is used for photometry to identify transient candidates. We perform photometry using \texttt{SExtractor}, and the outcome catalog contains all sources with flux exceeding 1.5$\sigma$ above the local background fluctuations. This catalog includes both the injected PSF sources and bogus artifacts introduced during the subtraction process.

Accurate background estimation is essential for reliable source detection. To account for spatial variations across the wide field of view, we set \texttt{BACK\_TYPE = AUTO} and \texttt{BACK\_SIZE = 64}, enabling automatic modeling of the background using a mesh grid of that size. Additionally, we use \texttt{BACKPHOTO\_TYPE = LOCAL}, which instructs \texttt{SExtractor} to estimate the background locally around each detected source, rather than relying on a global background value. These settings provide robust background modeling and improve source detection performance in difference images.

In addition to detecting transient candidates, effective artifact rejection is essential for building a clean and representative training sample. To aid this process, we perform source detection on pixel-inverted versions of both the subtracted and reference images. This approach is motivated by the fact that many subtraction artifacts exhibit over-subtracted residuals in their surroundings, which become more easily detectable in the inverted subtracted image. Likewise, performing photometry on the inverted reference image can reveal regions with artificially depressed background levels that may otherwise lead to spurious positive detections in the difference image.


All sources detected in the subtracted image are evaluated against the criteria below to determine whether they are transient candidates. A source that meets one or more of these criteria is considered unlikely to be a transient candidate and is excluded from the cutout image generation process.

\begin{enumerate}

\item The source matches a \texttt{SkyBoT} asteroid’s ephemeris position within $5$ arcseconds \citep{2006ASPC..351..367B}.
\item The source matches an inverted subtracted image detection within the FWHM, and the magnitude difference is within $0.5$ mag.
\item The source matches an inverted reference image detection within the FWHM, and the magnitude difference is within $0.5$ mag.
\item The source’s elongation is more than four times the median elongation of point sources in the science image.
\item The source’s FWHM is less than $0.7$ times or greater than $2.4$ times the median FWHM of point sources in the science image.
\item The source is saturated or overlaps with the defect mask.
\item The average background value around the source deviates by more than $2\sigma$ from the overall background value.
\item The source’s SNR is less than $5$.
\item There is a pixel within the FWHM with a value exactly equal to 1E-30.\
\end{enumerate}

The first criterion is to exclude known solar system objects. We query the \texttt{SkyBoT} service for all known asteroids at a given sky position, time, and observation location. The second and third criteria exclude sources that closely match detections in the inverted images described earlier. However, to avoid excluding valid sources that are marginally affected by subtraction residuals, this criterion is applied only when the inverted detection is bright enough to significantly influence the source flux. Therefore, we impose an additional condition that the magnitude difference must be within $0.5$ mag.

The fourth and fifth criteria exclude detections that deviate significantly from the expected PSF shape in the science image, including spurious artifacts as well as extended or moving objects such as galaxies and asteroids. The sixth criterion corresponds to cases where \texttt{SExtractor FLAGS} $\geq 4$ or \texttt{IMAFLAGS\_ISO} $> 0$. The former indicates that the source contains values above the image saturation level, while the latter means that the source overlaps with pixels in the defect mask. This criterion helps exclude artifacts near saturated sources as well as cosmic rays.

The seventh criterion removes sources in regions where over or under subtraction has occurred, and the eighth criterion excludes cases where there is no clear detection. Lastly, the nineth criterion addresses the issue that arises when \texttt{HOTPANTS} fails to achieve adequate subtraction, and thus such sources are not considered valid candidates. The overall criteria share a similar process to \texttt{gpPy}-a pipeline for SN monitoring project—for transient candidate filtering \citep{2023zndo...8321870P}.

\begin{table}[t!]
\centering
\caption{The number of \texttt{real} and \texttt{bogus} cutouts in each training dataset.}
\label{tab:training_dataset}
\centering
\begin{tabular}{cccc}
\toprule
Dataset & \texttt{real} & \texttt{bogus} & Total  \\
\midrule
RI               & $14,972$ & \multirow{4}{*}{$59,312$} & $74,284$ \\
NGI              & $12,693$ &                           & $72,005$ \\
RI+NGI           & $13,832$ &                           & $73,144$ \\
RI+NGI$^\dagger$ & $12,987$ &                           & $72,299$ \\
\bottomrule
\end{tabular}
\end{table}

In this study, the cutout samples of injected point sources that meet the criteria are employed as \texttt{real} samples in subsequent steps, while those that meet the criteria but are not from injected sources served as \texttt{bogus} samples.


\section{Dataset} \label{sec:dataset}

\subsection{Training Dataset} \label{sec:training_dataset}

To construct the training dataset for RB classification, a total of $18,000$ point sources are injected into the six science images based on the RI method, as described in Section \ref{sec:ri}. Subsequently, our TSP is applied to the science-reference image pairs and generates $65,669$ cutouts. 
Among these cutouts, $14,972$ cutouts correspond to the injected sources and are labeled as \texttt{real}.
Meanwhile, $50,697$ cutouts from sources that are not generated from injected point sources but still pass the flagging criteria are assigned as \texttt{bogus}. 
We call this dataset the \textit{RI dataset}. 

Similar to the RI dataset, we construct the \textit{NGI dataset} using the same six science-reference image pairs. Here, we inject a total of $19,203$ point sources near galaxies and our TSP produces $21,308$ cutouts where $12,693$ correspond to injected point sources, and the remaining cutouts are regarded as \texttt{bogus}. 
To minimize the potential effect of the distribution differences in \texttt{bogus} cutouts between two datasets, we take the union of \texttt{bogus} samples from both datasets when training RB classifiers. Hence, both RI and NGI datasets have the same $59,312$ \texttt{bogus} samples. Although this labeling approach, which assumes only injected sources to be \texttt{real}, may introduce label contamination by assigning genuine astronomical objects to \texttt{bogus}, it significantly reduces the human effort required for manual labeling and allows investigation of the direct effect of PSF injection on the behavior of a trained RB classifier. In practice, the degree of label contamination is relatively small, because the majority of the generated cutouts stem from artifacts. For example, in our S230518h follow-up test dataset--where all cutouts were visually inspected--only $1942$ out of $44,555$ cutouts (approximately $4.4\%$) correspond to transients, asteroids, or variable sources.

To further investigate the impact of dataset composition, we construct the \textit{RI+NGI dataset} by combining \texttt{real} cutouts from the RI and NGI datasets. To ensure a fair comparison, we randomly sample half of the \texttt{real} cutouts from each dataset, thereby balancing the number of \texttt{real} samples.
As discussed in the later Section \ref{ssec:classification_results}, an RB classifier trained on the NGI or RI+NGI dataset tends to misclassify variable stars as \texttt{real}. This is mainly due to point sources injected extremely close to galaxies, which can resemble variable stars. 
Thus, we consider another combined dataset, called the \textit{RI+NGI}$^\dagger$ \textit{dataset}, where half of the \texttt{real} cutouts are randomly sampled from the RI dataset, while the other half are selected from the NGI dataset by excluding the \texttt{real} cutouts that are close to their host galaxies (see Section \ref{sec:analysis_false_positive}). The summary of four training datasets is presented in Table \ref{tab:training_dataset}.

Each sample consists of three channels (reference, science, and subtraction cutouts), each with a size of $150 \times 150$ ($1\arcmin \times 1\arcmin$ field of view) pixels. Before feeding samples into the RB classifiers, we apply the following data augmentation techniques: Each sample is horizontally flipped with a probability of $0.5$. Subsequently, the sample is randomly resized into a size between $120 \times 120$ and $180 \times 180$ pixels using linear interpolation, then cropped into $51 \times 51$ pixels around the center. This resize-crop strategy simulates observations with different pixel scales that range from $0.83\times$ to $1.25\times$ its original pixel scale, thereby enhancing the robustness of RB classifiers to variations in point source size. Finally, Min-Max normalization is applied to each channel of the sample.

\subsection{Test Dataset} \label{sec:test_dataset}

We evaluate the performance of the RB classifier on a set of cutouts obtained from the GECKO follow-up observations for two GW events induced by the neutron star-black hole (NSBH) merger, GW190814 \citep{gw190814} and S230518h \citep{2023GCN.33813....1L,2025ApJ...981...38P}. Similar to Section \ref{sec:img}, the KMTNet images used to construct the test dataset are obtained by the median stacking of sequential exposures.

The GW190814 is one of the nearest NSBH merger systems, as indicated by its luminosity distance (241$_{-45}^{+41}$ Mpc) with high asymmetric binary mass (23.2 and 2.59 solar mass). The 2.59 solar mass object could be the heaviest NS or lightest BH but GW190814 is identified as an NSBH with high confidence. It has one of the most well-constrained sky localization, approximately $18.5\:\deg^2$ at 90\% probability comparable to that of GW170817 and KMTNet covers the whole localization area with 51 fields tiling observations over two days on the 15th and 16th. With these data, we expect to find KN, electromagnetic counterpart of GW from classifications among various transients. 
Specifically, we generate $64,513$ cutouts and compare these cutouts with GCN circulars and existing literature, resulting in $10$ cutouts from four previously identified transient events \citep{2019GCN.25336....1S,2019GCN.25355....1G,2019GCN.25373....1H,2019GCN.25379....1T,2019GCN.25362....1A,2019GCN.25423....1R,2019GCN.25669....1G,2020A&A...643A.113A}, which are used as \texttt{real} samples. In addition, we identify $161$ asteroid cutouts through catalog matching with \texttt{SkyBoT} \citep{2006ASPC..351..367B} and categorize them into two groups based on their SNR by human visual inspection, which results in $64$ cutouts visually distinct and $97$ visually ambiguous ones. Consequently, a total of $171$ \texttt{real} samples are generated from the GW190814 follow-up observations.

\begin{table}[t!]
\centering
\caption{Summary of the test dataset. The `$\#$ Cutouts' column indicates the number of cutouts that correspond to the given `Type',  collected from the follow-up observations of the associated `Event'.}
\label{tab:test_dataset}
\begin{tabular}{cccr}
\toprule
Event & Type & Label & $\#$ Cutouts  \\
\midrule
\multirow{3}{*}{GW190814} & Transient            & 1 &    10 \\
                          & Asteroid (Distinct)  & 1 &    64 \\
                          & Asteroid (Ambiguous) & 1 &    97 \\
\hline
\multirow{2}{*}{S230518h} & \texttt{bogus}       & 0 & 44427 \\
                          & \texttt{real}        & 1 &   128 \\
\hline
\multirow{2}{*}{Total}    & \texttt{bogus}       & 0 & 44427 \\
                          & \texttt{real}        & 1 &   299 \\
\bottomrule
\end{tabular}
\end{table}

\begin{table*}[htb!]
    \caption{Architecture of O$'$TRAIN \citep{otrain}. $d_{\text{in}}$ and $d_{\text{out}}$ indicate the dimension of the input and output of each layer, respectively. $k$ is the kernel size and $p$ is the dropout rate. The architecture has a total of $1,704,353$ parameters.}
    \label{table:otrain-architecture}
    \centering
    \begin{tabular}{ccccc}
        \toprule
        Layer & Operator              & Nonlinearity & $d_{\text{in}}$           & $d_{\text{out}}$ \\
        \midrule
        1    & Conv2D $(k=3)$        & ReLU         & $3 \times 51 \times 51$   & $16 \times 51 \times 51$ \\
        2    & Conv2D $(k=3)$        & ReLU         & $16 \times 51 \times 51$  & $32 \times 51 \times 51$ \\
        3    & AvgPooling2D $(k=2)$  & -            & $32 \times 51 \times 51$  & $32 \times 25 \times 25$ \\
        4    & Conv2D $(k=3)$        & ReLU         & $32 \times 25 \times 25$  & $64 \times 25 \times 25$ \\
        5    & MaxPooling2D $(k=2)$  & -            & $64 \times 25 \times 25$  & $64 \times 12 \times 12$ \\
        6    & Dropout $(p=0.3)$     & -            & $64 \times 12 \times 12$  & $64 \times 12 \times 12$ \\
        7    & Conv2D $(k=3)$        & ReLU         & $64 \times 12 \times 12$  & $128 \times 12 \times 12$ \\
        8    & MaxPooling2D $(k=2)$  & -            & $128 \times 12 \times 12$ & $128 \times 6 \times 6$ \\
        9    & Dropout $(p=0.3)$     & -            & $128 \times 6 \times 6$   & $128 \times 6 \times 6$ \\
        10   & Conv2D $(k=3)$        & ReLU         & $128 \times 6 \times 6$   & $256 \times 6 \times 6$ \\
        11   & MaxPooling2D $(k=2)$  & -            & $256 \times 6 \times 6$   & $256 \times 3 \times 3$ \\
        12   & Flatten               &              & $256 \times 3 \times 3$   & $2304$ \\
        13   & Fully-connected layer & ReLU         & $2304$                    & $512$ \\
        14   & Dropout $(p=0.3)$     & -            & $512$                     & $512$ \\
        15   & Fully-connected layer & ReLU         & $512$                     & $256$ \\
        16   & Fully-connected layer & Sigmoid      & $256$                     & $1$ \\
        \bottomrule
    \end{tabular}
\end{table*}

For the S230518h, it was the first GW event including NS in the pre-O4 run, and it has an elongated sky localization area. 
The transient candidates are selected with the following procedures from the 13 million detections in the subtracted images:
\begin{enumerate}
    \item Filtering clean sources using the source parameters from \texttt{SExtractor} with the criteria described in Section \ref{ssec:pipeline}
    \item  Excluding sources with a predicted score from the prototype RB classifier below $0.5$
    \item Manually labeling the remaining sources through visual inspection by human expert
\end{enumerate}
The prototype RB classifier employed in the second step is a prebuilt CNN-based RB classifier trained on the \texttt{autoscan} dataset \citep{2015AJ....150...82G}, which is a publicly available dataset for RB classification consisting of about $450$K simulated \texttt{real} and $450$K \texttt{bogus} samples. This prototype classifier was developed to serve as a metric for filtering out \texttt{bogus} samples when a dataset based on KMTNet observations was still under construction.
As a result of the procedures, $128$ \texttt{real} cutouts (potentially containing point source-like asteroids) and $44,427$ \texttt{bogus} cutouts are generated from the S230518h follow-up observations. In our analysis, variable stars are treated as \texttt{bogus} because our goal is to detect astronomical objects that are absent in the reference cutouts but appear in the science cutouts. Although variable stars are astronomical objects that exhibit variability between the reference and science cutouts, they are present in both, unlike the transients of interest. Consequently, the final test dataset consists of $299$ \texttt{real} cutouts and $44,427$ \texttt{bogus} cutouts, as summarized in the Total column of Table \ref{tab:test_dataset}. 


\section{Model and Training} \label{sec:model}

\begin{figure}[t!]
    \centering
    \includegraphics[width=0.8\linewidth]{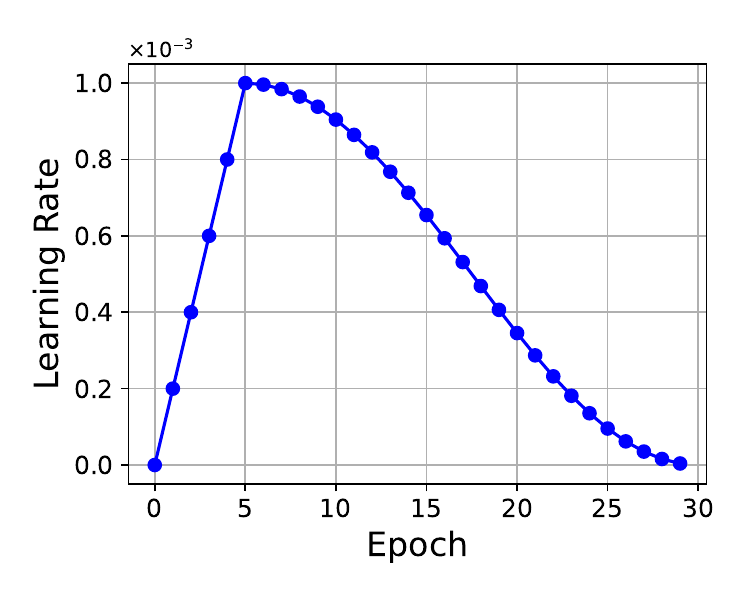}
    \caption{Learning rate schedule used in the training.}
    \label{fig:learning_rate}
\end{figure}

For RB classifiers, we employ convolutional neural networks (CNNs), which are widely used for image classification tasks \citep{7486599, 7780459, pmlr-v97-tan19a}. In specific, we adopt the architecture of O$'$TRAIN \citep{otrain} with two minor changes: (1) Our classifier takes a $51 \times 51$ three-channel cutout as an input instead of a $38 \times 38$ single-channel cutout, and (2) it predicts a probability of the input cutout belonging to the \texttt{real} class using the sigmoid function rather than using the softmax function over two output neurons. We employ the rectified linear unit (ReLU) \citep{relu} activation function for all convolutional and fully-connected layers except for the output layer, which is followed by a sigmoid function. The resulting architecture has a total of $1.7$M learnable parameters and is presented in Table \ref{table:otrain-architecture}.

\begin{figure*}[ht!]
    \centering
    \gridline{
        \fig{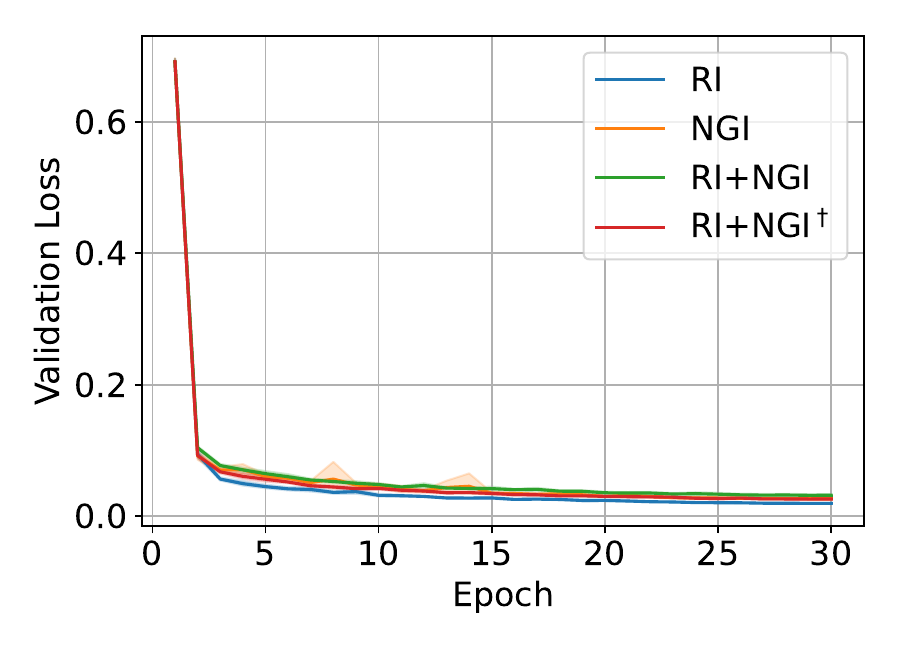}{0.38\linewidth}{(a) Validation loss}
        \fig{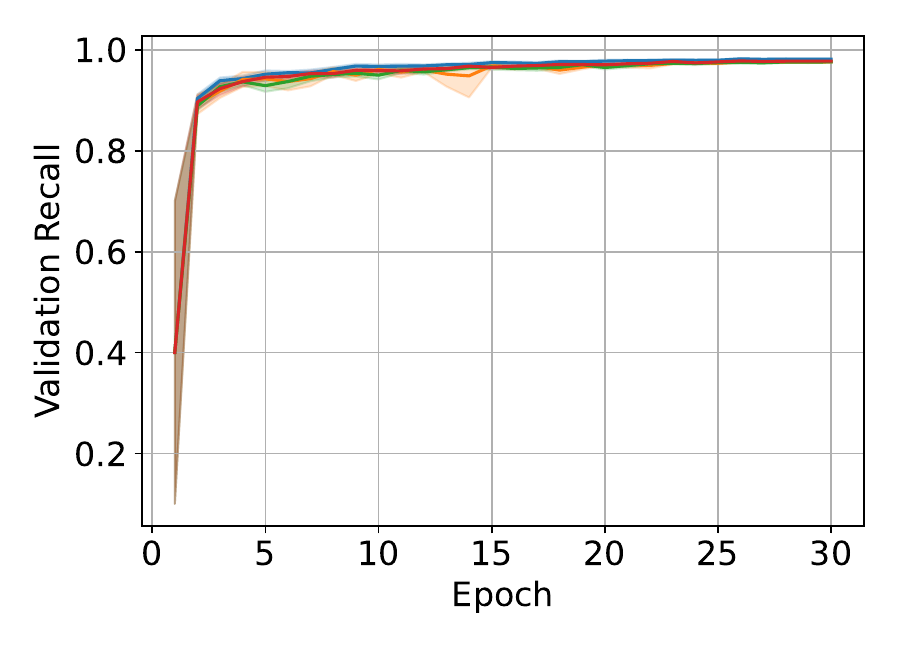}{0.38\linewidth}{(b) Validation recall}
    }
    \gridline{
        \fig{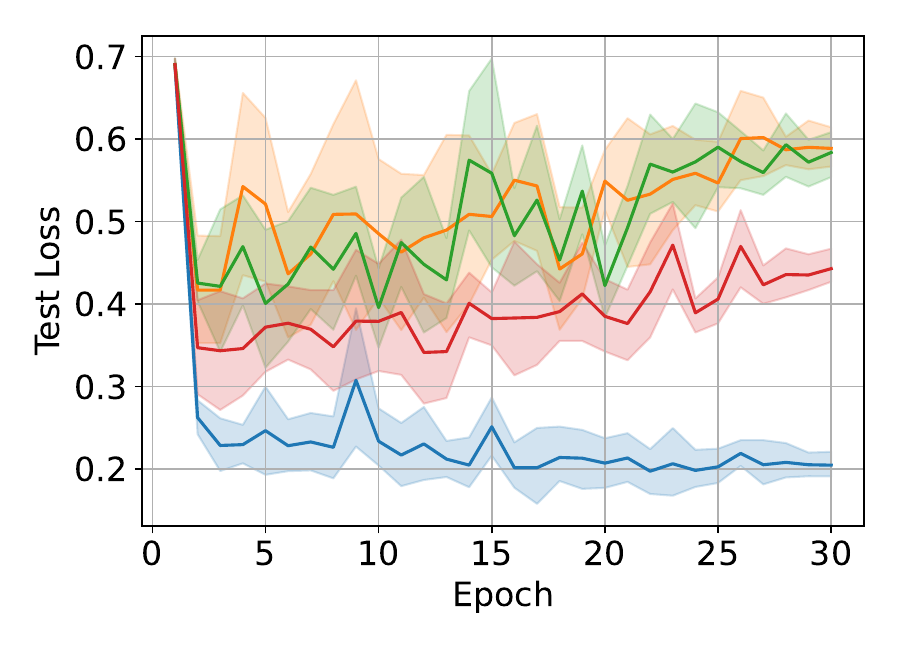}{0.38\linewidth}{(c) Test loss}
        \fig{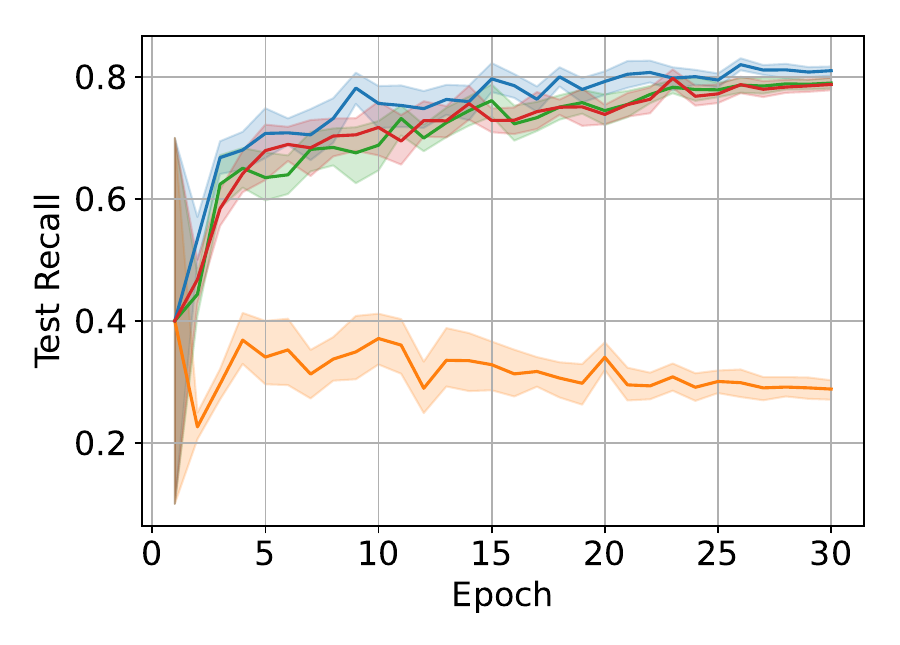}{0.38\linewidth}{(d) Test recall}
        }    
    \caption{Learning curves. We measure the values of the binary cross-entropy loss function and recall scores for the validation dataset (a)-(b) and test dataset (c)-(d) over training epochs. The solid lines indicate the average metrics and the shaded region $95\%$ confidence interval over $10$ runs.}
    \label{fig:learning_curves}
\end{figure*}

We train RB classifiers for $30$ epochs using the adaptive moment estimation (Adam) optimizer \citep{adam} where the learning rate is scheduled by cosine annealing with a linear warm-up. Specifically, the learning rate linearly increases from zero to $0.001$ for the first five epochs and then decreases to zero along a cosine curve over the remaining epochs, as shown in Figure \ref{fig:learning_rate}. This learning rate schedule enhances the stability in the early stages of training and improves convergence at the end \citep{loshchilov2017sgdr}. The size of a mini-batch is set to $256$. 


\section{Experiments}\label{sec:experiment}
Previous research has typically evaluated RB classifiers on test datasets with balanced proportions of real and bogus samples \citep{otrain,Hosenie2021MeerCRABMC,Corbett_2023}. 
However, real-world datasets produced by the TSP are highly imbalanced: the vast majority of cutouts are \texttt{bogus}, while only a few are \texttt{real}. Hence, even though an RB classifier achieves a high precision score on a balanced test dataset, it may yield a significant number of false positives during real-world deployment. Therefore, we argue that evaluating performance on an imbalanced dataset must be included. 
In Section \ref{sec:test_entire}, we present the experimental results of our RB classifiers on the entire test dataset described in Section \ref{sec:test_dataset}. We then evaluate the performance on balanced datasets by following the convention of previous works in Section \ref{sec:test_balanced}.

\subsection{Experimental setup}

We train RB classifiers on each dataset described in Section \ref{sec:training_dataset}. For reliability, all experiments are replicated $10$ times with different random configurations. A random configuration indicates a set of factors in an experiment affected by randomness such as initial network parameters, data splitting/shuffling, and data augmentation. This results in $10$ RB classifiers per training dataset and we report the performance aggregated over these $10$ RB classifiers. When training an RB classifier, we measure the values of the binary cross-entropy loss function and recall scores for the validation and test datasets for each epoch. For the validation dataset, we use a holdout scheme: we assign $20\%$ of the training dataset to a validation dataset before training and use the remaining 80\% to train the classifier. The test loss and test recall score are computed on the test dataset described in Section \ref{sec:test_dataset}. As shown in Figures \ref{fig:learning_curves}(a) and (b), all RB classifiers sufficiently converge and achieve high performance on the validation dataset. 
However, all classifiers suffer from the performance drop on the test dataset, which implies that the classifiers trained with injected point sources have a limited ability to generalize to real-world observations. 

\begin{table*}[ht!]
\centering
\caption{Evaluation of trained classifiers on the whole test dataset. The `Total' column presents the number of samples corresponding to each row. The remaining values indicate the average number of correctly classified samples over $10$ runs and the corresponding average correct rate (\%) in parentheses. A threshold value of $0.5$ is used.}
\label{tab:main_result}
\begin{tabular}{cccrrrrr}
\toprule
Event & Type & Label & Total & \multicolumn{1}{c}{RI} & \multicolumn{1}{c}{NGI} & \multicolumn{1}{c}{RI+NGI} & \multicolumn{1}{c}{RI+NGI$^\dagger$}  \\
\midrule
\multirow{3}{*}{GW190814} & Transient            & 1 &    10 &     5.0 (50.0\%) &     8.5 (85.0\%) &     9.9 (99.0\%) &     8.4 (84.0\%) \\
                          & Asteroid (Distinct)  & 1 &    64 &    62.3 (97.3\%) &    28.7 (44.8\%) &    62.6 (97.8\%) &    62.9 (98.3\%) \\
                          & Asteroid (Ambiguous) & 1 &    97 &    61.6 (63.5\%) &     0.6  (0.62\%) &    52.1 (53.7\%) &    52.6 (54.2\%) \\
\hline
\multirow{2}{*}{S230518h} & \texttt{bogus}       & 0 & 44427 & 41287.3 (92.9\%) & 33581.4 (75.6\%) & 33596.5 (75.6\%) & 36601.8 (82.4\%) \\
                          & \texttt{real}        & 1 &   128 &   116.7 (91.2\%) &    48.0 (37.5\%) &   113.6 (88.7\%) &   113.2 (88.4\%) \\
\hline
\multirow{2}{*}{Total}    & \texttt{bogus}       & 0 & 44427 & 41287.3 (92.9\%) & 33581.4 (75.6\%) & 33596.5 (75.6\%) & 36601.8 (82.4\%) \\
                          & \texttt{real}        & 1 &   299 &   245.6 (82.1\%) &    85.8 (28.7\%) &   238.2 (79.7\%) &   237.1 (79.3\%) \\
\bottomrule
\end{tabular}
\end{table*}

\subsection{Evaluation on the entire test dataset}\label{sec:test_entire}
Table \ref{tab:main_result} shows the classification results of RB classifiers trained on each training dataset. The values from the fifth to the last columns indicate the average number of corrected samples over $10$ different RB classifiers and the corresponding percentage in parentheses. The cutouts with predicted probabilities greater than a threshold of $0.5$ are regarded as \texttt{real} samples, while those with lower probabilities are categorized as \texttt{bogus}. 

\begin{figure*}[ht!]
    \centering
    \includegraphics[width=0.75\linewidth]{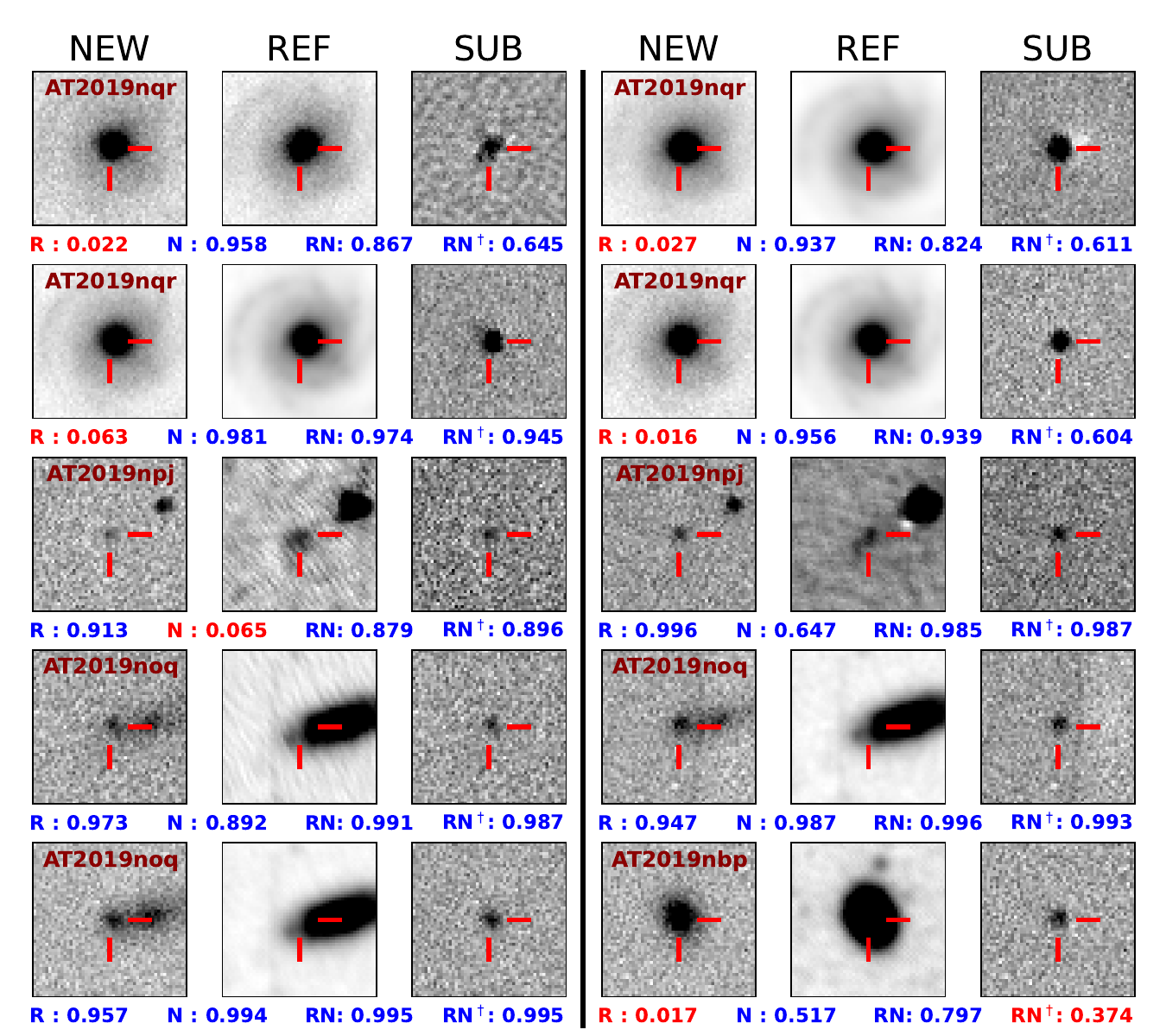}
    \caption{10 cutouts from four transient events identified by the follow-up survey for GW190814. For each sample, we visualize science (NEW), reference (REF), and subtraction (SUB) cutouts with a size of $51 \times 51$. The IRAF zscale transform is applied for visualization.
    The object name is denoted in brown at the top of each science image. The average predicted probabilities over $10$ classifiers are marked in red if the values are less than $0.5$ and in blue otherwise. \textsc{R}, \textsc{N}, \textsc{RN} and \textsc{RN}$^\dagger$ indicates RI, NGI, RI+NGI and RI+NGI$^\dagger$ datasets, respectively.}
    \label{fig:gw190814}
\end{figure*}

\begin{figure*}[ht!]
    \centering
    \includegraphics[width=0.8\linewidth]{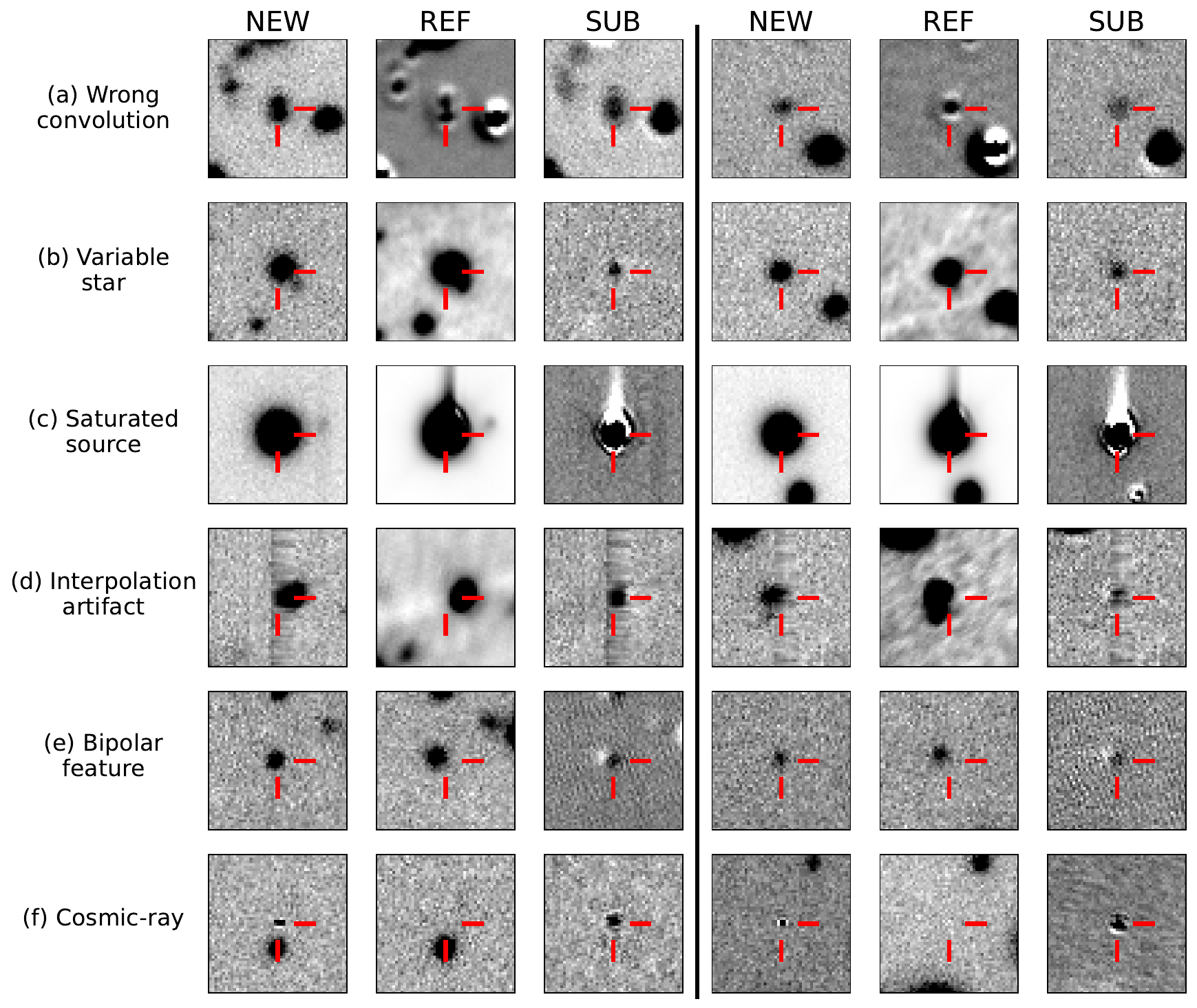}
    \caption{Six most prevalent categories of \texttt{bogus} samples in S230518h follow-up dataset.}
    \label{fig:example_bogus}
\end{figure*}

\begin{figure*}[ht!]
    \centering
    \begin{minipage}{0.32\textwidth}
        \fig{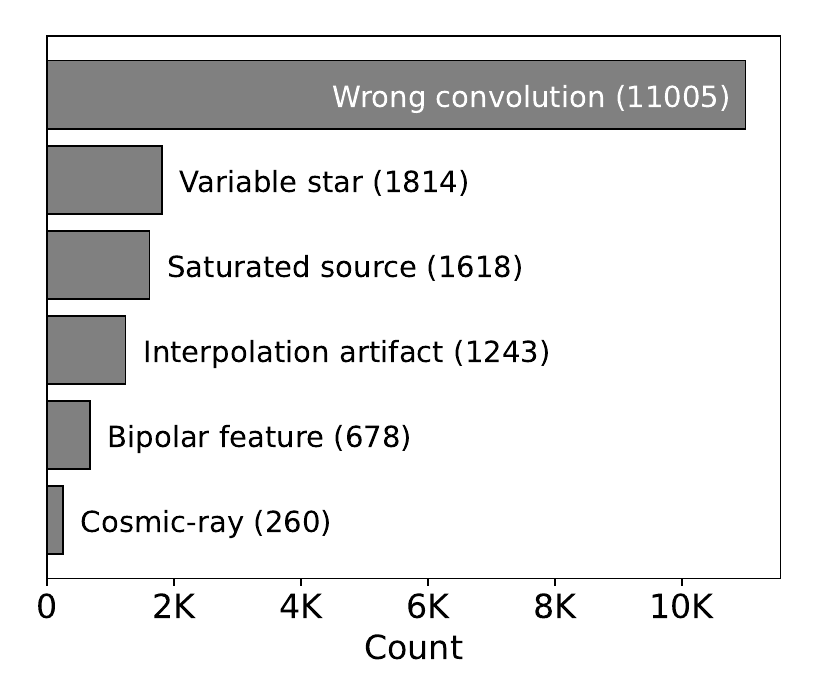}{0.99\linewidth}{(a) \texttt{bogus} distribution}
    \end{minipage}
    \begin{minipage}{0.55\textwidth}
        \gridline{\fig{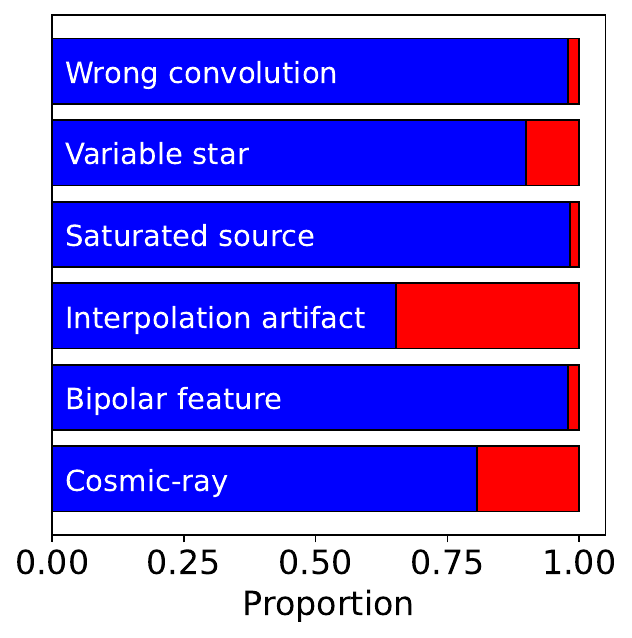}{0.48\linewidth}{(b) RI}\fig{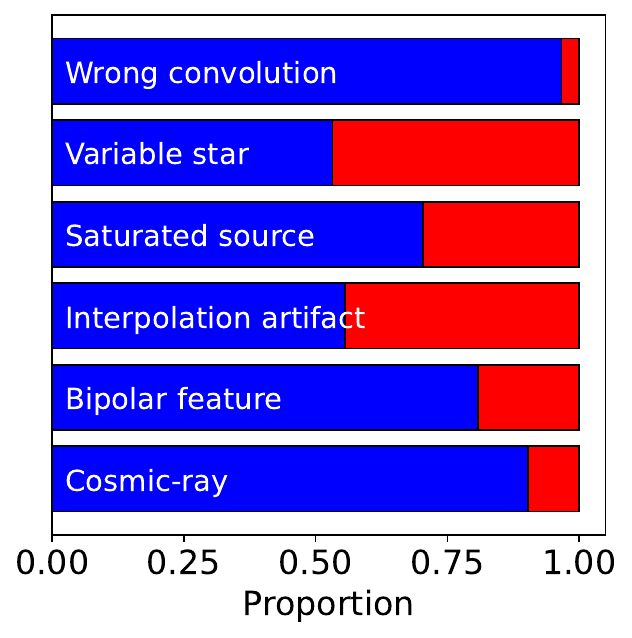}{0.48\linewidth}{(c) NGI}}
        \gridline{\fig{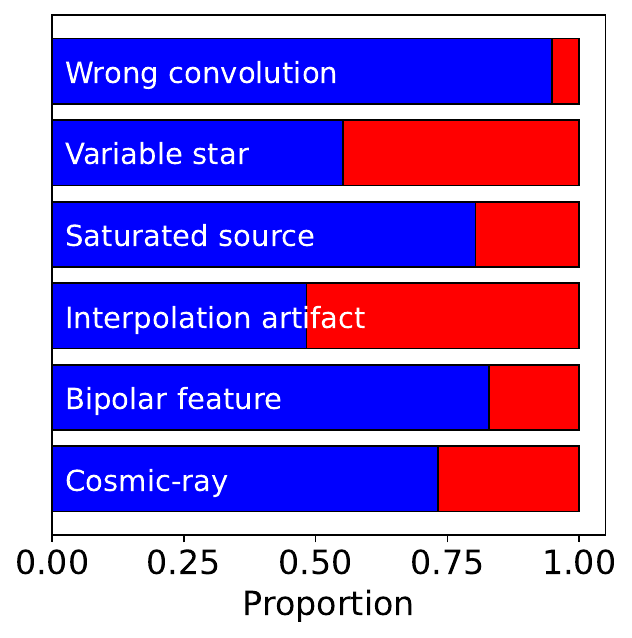}{0.48\linewidth}{(d) RI+NGI}
        \fig{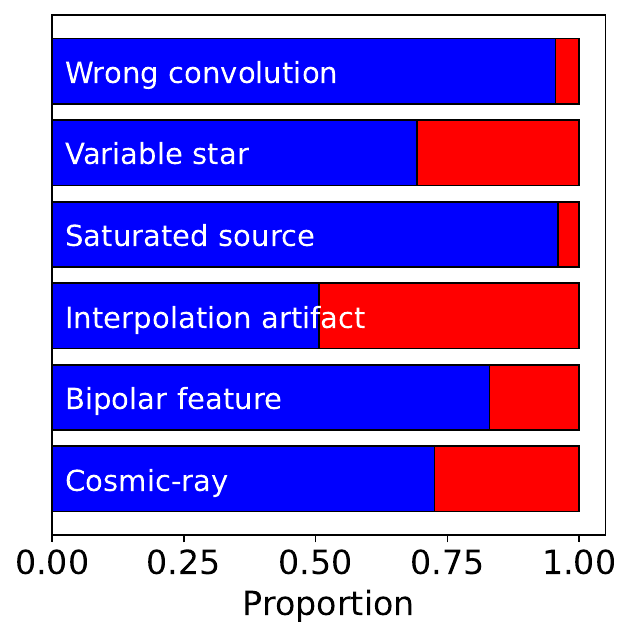}{0.48\linewidth}{(e) RI+NGI$^\dagger$}}
    \end{minipage}
    \caption{Histogram of \texttt{bogus} types in the S230518h follow-up dataset and the correct (blue) and incorrect (red) classification rates of RB classifiers. For (b)-(e), the threshold value of $0.5$ is used. For each training dataset, the rates are averaged over $10$ classifiers.}
    \label{fig:bogus_analysis}
\end{figure*}

\subsubsection{Classification results}
\label{ssec:classification_results}
There are clear distinctions in the model behaviors depending on training datasets. 
The RB classifiers trained on the RI dataset achieve strong performance in classifying asteroids and \texttt{bogus} samples while struggling to detect a significant portion ($50\%$) of transient cutouts from the follow-up observation for GW190814. 
In contrast, the classifiers trained on the NGI dataset demonstrate superior performance compared to those trained on the RI dataset in detecting transient samples from GW190814 ($85\%$). 
However, these models misclassify asteroids as \texttt{bogus} and exhibit a higher rate of false positives of $24.4 \%$ where \texttt{bogus} samples are classified as \texttt{real}. This poor performance for asteroid samples is mainly attributed to the injection strategy.
The NGI dataset contains \texttt{real} cutouts that include both a point source and its host galaxy. Therefore, the presence of a host galaxy in a cutout serves as a critical signal for the classifiers trained on the NGI dataset to classify the cutout as \texttt{real}. However, since asteroids typically appear in irregular locations independent of galaxies, the classifiers are less likely to predict asteroid samples as \texttt{real}.
Remarkably, training RB classifiers on the combined dataset enhances the capability to detect transient samples while keeping a false positive rate comparable to models trained on the NGI dataset alone.

Figure \ref{fig:gw190814} shows the transient cutouts from GW190814 follow-up observation with the predicted probabilities averaged over $10$ RB classifiers trained on each dataset. These probabilities are marked in red if the values are less than a threshold of $0.5$ and in blue otherwise. The RB classifiers trained on the RI dataset fail to detect the transient events AT2019nqr and AT2019nbp, which appeared near their host galaxies. In contrast, the RB classifiers trained on the NGI and RI+NGI datasets successfully detect such transient cutouts.

\subsubsection{Analysis of false positives}\label{sec:analysis_false_positive}

\begin{figure}[t!]
    \centering
    \gridline{
        \fig{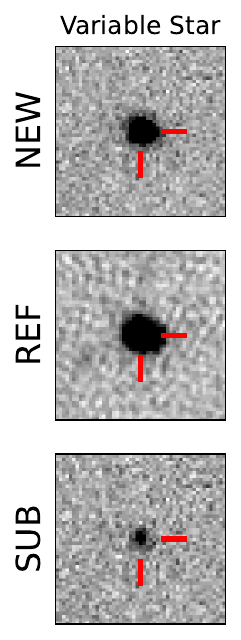}{0.0865\textwidth}{(a)}
        \fig{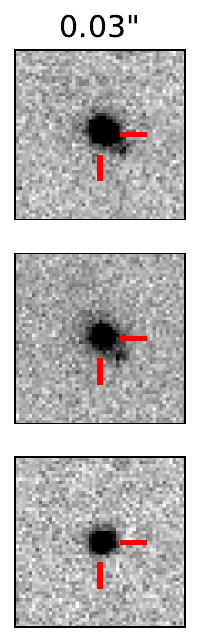}{0.072\textwidth}{(b)}
        \fig{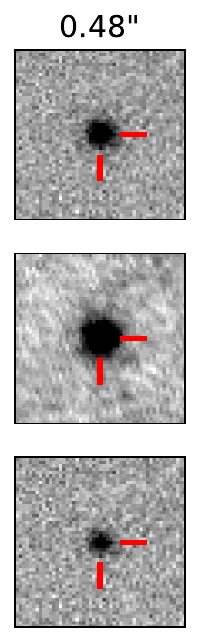}{0.072\textwidth}{(c)}
        \fig{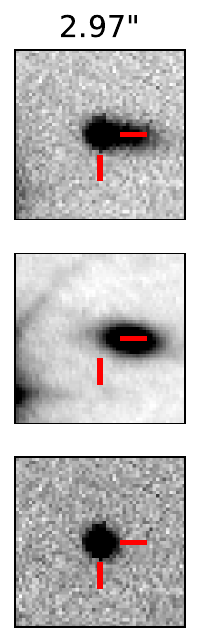}{0.072\textwidth}{(d)}
        \fig{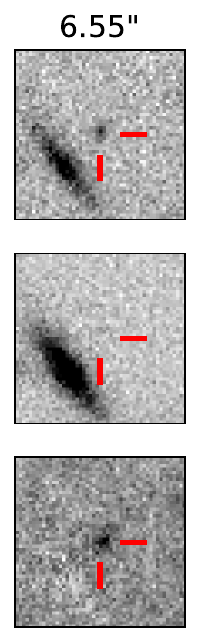}{0.072\textwidth}{(e)}
        \fig{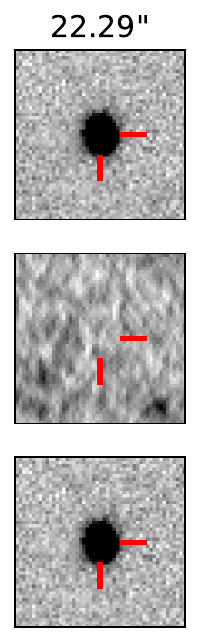}{0.072\textwidth}{(f)}
        }
    \caption{Examples of variable stars (a) in the S230518h follow-up and injected sources with different distances from galaxies (b)-(f) in the NGI dataset. The values at the top of (b)-(f) indicate the angular distances in arcseconds from the galaxies associated with injected point sources.}
    \label{fig:example_distance}
\end{figure}

Despite detecting more transient cutouts, the RB classifiers trained on either the NGI and RI+NGI dataset generate more false positives compared to those trained on the RI dataset. Thus, we further investigate the rate of false positives for each \texttt{bogus} type. For each \texttt{bogus} cutout from the S230518h follow-up, we manually assign its type out of the following six representative \text{bogus} types produced by our TSP:
\begin{itemize}
    \item \textit{Wrong convolution}: This error occurs during the convolution step in image subtraction. When the reference image is locally of poorer quality than the science image, sharpening occurs instead of smoothing, resulting in inaccurate residuals and numerous bogus detections.
    \item \textit{Variable star}: These traces appear in the subtracted image due to the actual brightening of variable stars. Such stars are visible in both the science and reference images.
    \item \textit{Saturated source}: Saturated sources exhibit truncated peak profiles and diffraction spikes, which prevent complete subtraction. This leads to the generation of various shapes of artifacts around the sources.
    \item \textit{Interpolation artifact}: Interpolation is applied during the KMTNet image processing to correct bleeding patterns or bad pixel areas by replacing their values with nearby pixel values. Sources affected by the interpolations may leave inaccurate remnants in the subtraction process.
    \item \textit{Bipolar feature}: Imprecise astrometric alignment between the science and reference images can lead to under-subtracted and over-subtracted regions during source subtraction, resulting in bogus with bipolar features.
    \item \textit{Cosmic-ray}: Cosmic rays are traces of high-energy particles from space commonly captured during image acquisition. They appear at random locations with varying intensities and are characterized by sharp, irregular shapes that differ from typical PSFs.
\end{itemize}
Illustrative examples of each \texttt{bogus} type are presented in Figure \ref{fig:example_bogus} and the histogram of \texttt{bogus} types is shown in Figure \ref{fig:bogus_analysis}(a). 
These six categories represent the most prevalent artifacts produced by our TSP on KMTNet images, which account for approximately $37.5\%$ of all \texttt{bogus} samples. The remaining \texttt{bogus} samples are grouped into a single remainder group since they are either relatively rare or difficult to assign to a specific category, and are excluded from the false positive analysis.

Figures \ref{fig:bogus_analysis}(b)-(d) show the correct and incorrect classification rates of the trained RB classifiers at a threshold values of $0.5$. It is worth mentioning that the interpolation artifact is a chronic artifact in KMTNet’s optical images, which leads to a high false positive rate across all models. 
Except for the interpolation artifact, the RB classifiers trained on the RI dataset exhibit moderate performance across most \texttt{bogus} types. 
However, the models trained on the NGI and RI+NGI datasets frequently misclassify the variable stars as \texttt{real}. This behavior can be attributed to the similarity between cutouts of a variable star and a simulated point source that is injected very close to galaxies, as shown in Figures \ref{fig:example_distance}(a)-(c).

Based on these observations, we introduce a variant of the RI+NGI dataset, denoted as RI+NGI$^\dagger$. Unlike the original RI+NGI dataset, which randomly subsamples \texttt{real} cutouts from the NGI dataset, the RI+NGI$^\dagger$ dataset selectively includes only those \texttt{real} samples where the injected sources are located at least $1.15\arcsec$ distant from their nearest galaxy. This threshold distance value is empirically determined to exclude approximately half of the \texttt{real} samples from the NGI dataset. 
As shown in the rightmost column of Table \ref{tab:main_result}, RB classifiers trained on the RI+NGI$^\dagger$ dataset can effectively reduce about 3K false positives compared to those models trained on the RI+NGI dataset while keeping comparable detection performance for \texttt{real} sources. 
Figure \ref{fig:bogus_analysis}(e) reveals that the primary causes of the reduced false positive rate are attributed to saturated sources and variable stars. Thus, the distance of injected point sources from galaxies significantly influences the behavior of the resulting RB classifiers.

\subsubsection{Model behaviors over threshold values}

\begin{figure}[t!]
    \centering
    \includegraphics[width=0.7\linewidth]{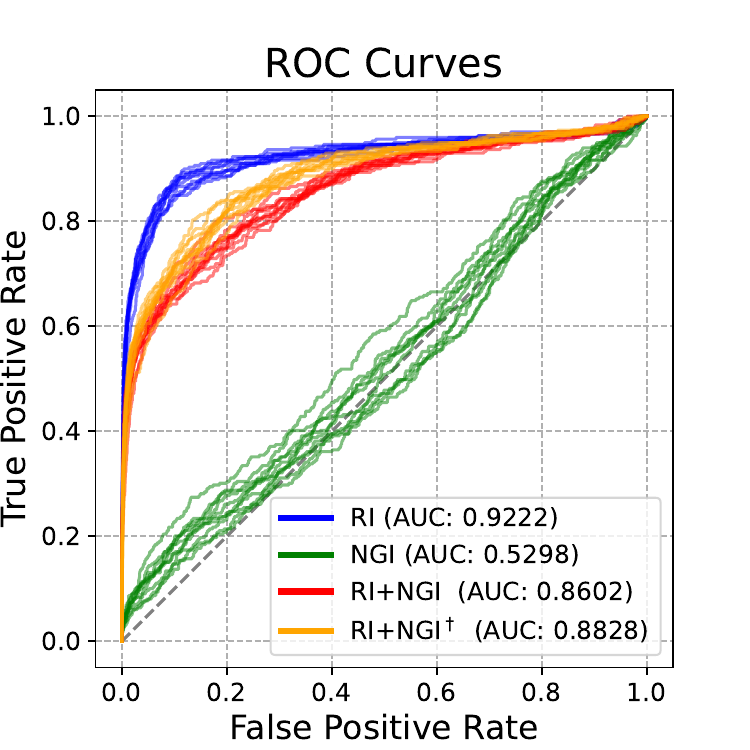}
    \caption{ROC curves of trained classifiers. For each dataset, ROC curves of $10$ classifiers trained with different random configurations are visualized. The average AUC score is shown in the legend.}
    \label{fig:roc_curves}
\end{figure}

Previous analyses have mainly focused on the classification results at a fixed threshold value of $0.5$. In this section, we provide an analysis of model behavior over threshold values. Figure \ref{fig:roc_curves} shows the receiver operating characteristic (ROC) curves for the RB classifiers and the corresponding area-under-curve (AUC) scores (averaged over $10$ curves for each dataset) in the legend. The RB classifiers trained on RI, RI+NGI, and RI+NGI$^\dagger$ datasets achieve moderate ROC AUC scores ($>0.86$), which implies that the user can adaptively balance the trade-off between the number of false positive and detected \texttt{real} sources by adjusting a threshold value. In contrast, the ROC curves of the NGI dataset approximately lie on the $y=x$ line, with a corresponding AUC score of $0.5298$, indicating the trade-off cannot be handled through threshold adjustments. This poor performance of the NGI dataset is mainly due to misclassified asteroid samples.

\begin{figure*}[t!]
    \centering
    \gridline{\fig{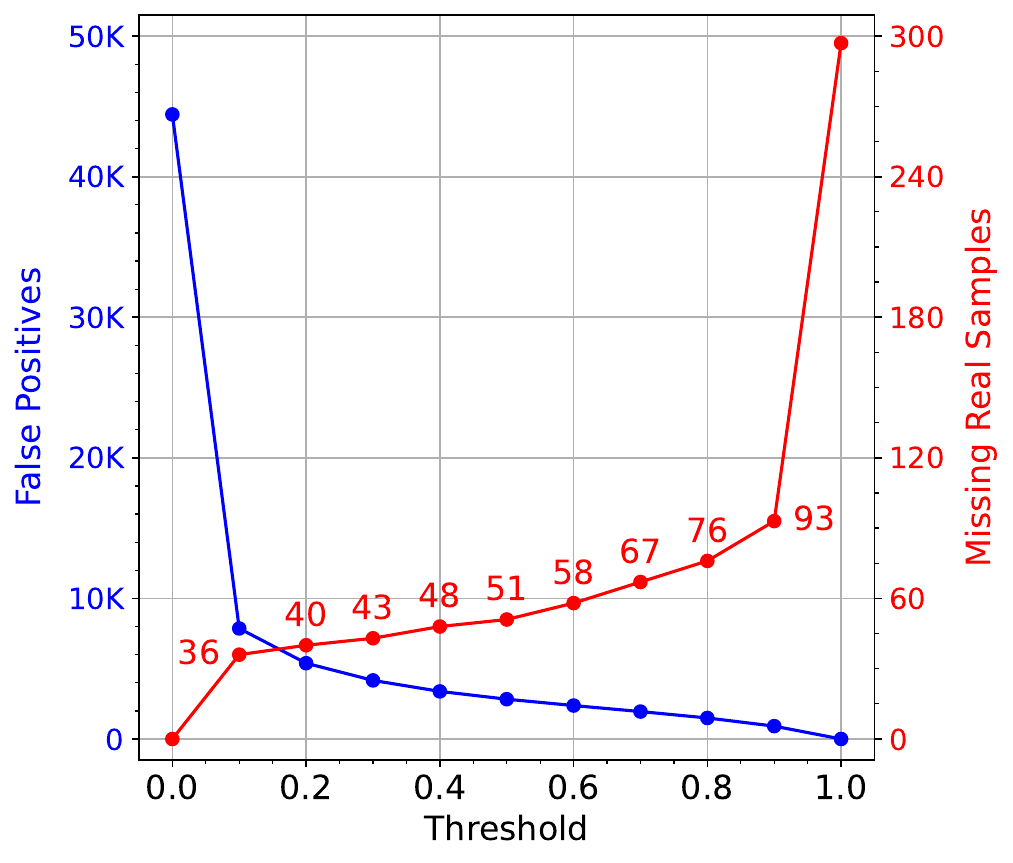}{0.4\textwidth}{(a) RI}\fig{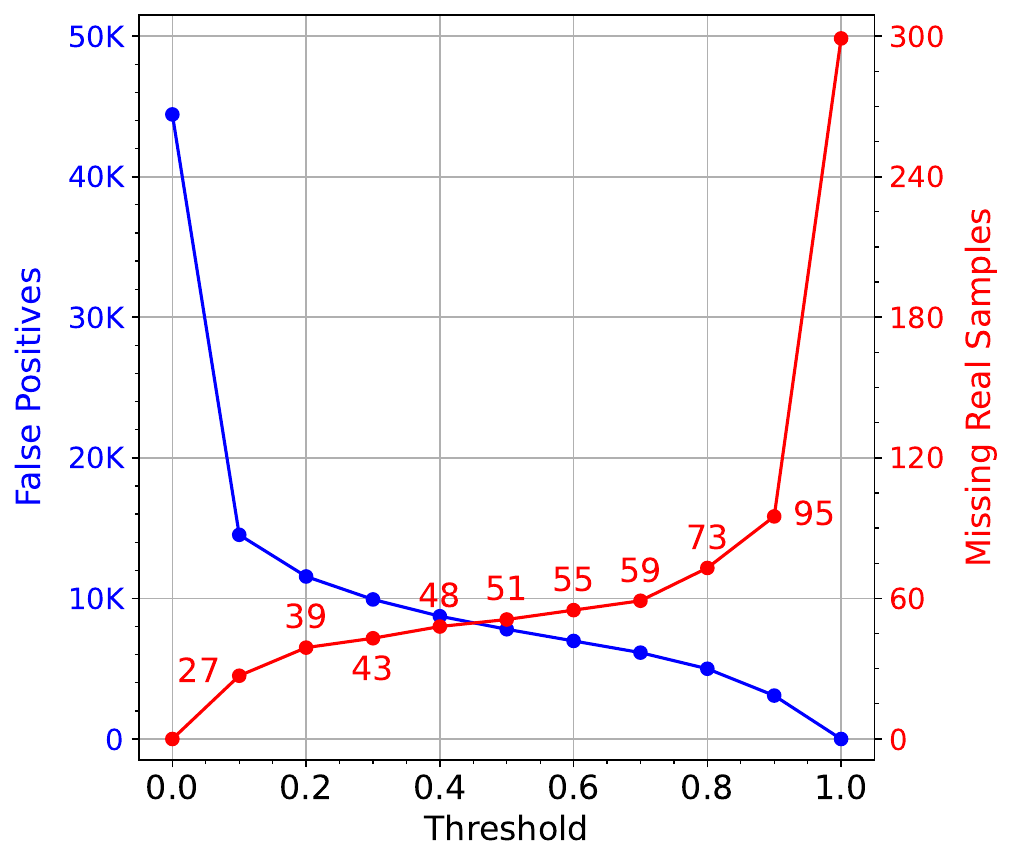}{0.4\textwidth}{(b) RI+NGI$^\dagger$}}
    \caption{The number of misclassified \texttt{bogus} (left axes) and \texttt{real} samples (right axes) of a single RB classifier trained on each of the RI (left figure) and RI+NGI$^\dagger$ (right figure) datasets over various threshold values.}
    \label{fig:threshold}
\end{figure*}

Figure \ref{fig:threshold} shows the number of false positives (blue, left axes) and misclassified \texttt{real} samples (red, right axes) of a single RB classifier trained on each of RI (left figure) and RI+NGI$^{\dagger}$ (right figure) datasets over threshold values from $0.0$ to $1.0$ with an interval of $0.1$. Using a threshold value of $0.1$ seems reasonable for both models:
the classifier trained on the RI dataset achieves $\frac{299-36}{299}\approx 0.88$ recall score with less than $10$K false positives. Similarly,  the classifier trained on the RI+NGI$^{\dagger}$ dataset obtains $\frac{299-27}{299}\approx 0.91$ recall score while filtering out about $67\%$ of \texttt{bogus} samples.

\subsection{Evaluation on the balanced test dataset} \label{sec:test_balanced}

Prior works have typically evaluated RB classifiers on balanced test datasets as performance metrics can be skewed on imbalanced datasets. This sections provides the performance of the RB classifiers on the balanced test datasets to facilitate comparison with existing studies. We construct $30$ balanced test datasets, each of which consists of $299$ entire \texttt{real} samples and $299$ \texttt{bogus} cutouts randomly sampled from the original \texttt{bogus} pools. 

Table \ref{tab:balanced_metrics} shows the average performance metrics of RB classifiers trained on each training dataset. These metrics are calculated across 30 datasets and 10 classifiers. 
In the table, accuracy is defined as the number of correctly classified samples divided by the total number of samples. Recall represents the fraction of actual positive samples (i.e., those labeled as \texttt{real}) that are correctly identified by the classifier. Precision refers to the proportion of correctly predicted positive samples among all samples predicted as positive. Intuitively speaking, recall indicates how well the model can detect actual \texttt{real} samples without missing them, while precision reflects how accurate those detections are. Therefore, recall and precision often exhibit a trade-off. Considering the balance between recall and precision, F1 score is defined as the harmonic mean of precision and recall.
The RB classifiers trained on the RI dataset consistently outperform those trained on other datasets across all metrics. These results are attributed to their ability to classify asteroids as \texttt{real} and filter out \texttt{bogus} samples. 
On the other hand, the RI+NGI$^\dagger$ dataset, which selectively excludes \texttt{bogus} samples closely located near galaxies, achieves better accuracy, precision and F1 scores compared to RI+NGI dataset. This implies the strategy for combining the RI and NGI datasets is crucial.

\begin{table}[t!]
\centering
\caption{Evaluation of trained models on balanced test datasets. The values indicate the average performance metrics across $30$ datasets and $10$ RB classifiers for each training dataset. A threshold value of $0.5$ is used.}
\label{tab:balanced_metrics}
\begin{tabular}{ccccc}
\toprule
 & \multicolumn{1}{c}{Accuracy} & \multicolumn{1}{c}{Recall} & \multicolumn{1}{c}{Precision} & \multicolumn{1}{c}{F1}  \\
\midrule
              RI & 0.8746 & 0.8214 & 0.9196 & 0.8676 \\
             NGI & 0.5187 & 0.2870 & 0.5352 & 0.3731 \\
          RI+NGI & 0.7738 & 0.7967 & 0.7624 & 0.7789 \\
RI+NGI$^\dagger$ & 0.8066 & 0.7930 & 0.8155 & 0.8040 \\
\bottomrule
\end{tabular}
\end{table}

It is important to note that these metrics alone do not fully capture the behavior of the classifier on the different astronomical objects. For instance, the recall scores in Table \ref{tab:balanced_metrics} are computed over all \texttt{real} samples, which consist mainly of asteroids and a few transients near galaxies. The high recall scores of the RI dataset are mainly attributed to asteroid samples, which are less relevant for GW follow-up observations, rather than transients near galaxies. Therefore, we argue that a detailed sample-level analysis, as conducted in Section \ref{sec:test_entire}, is essential rather than relying on a single performance metric.

\section{Conclusion and Discussion} \label{sec:conclusion}

In this paper, we explore the impact of various point source injection strategies on the behaviors of RB classifiers when applied to real-world KMTNet observations. 
Our experimental results demonstrate that RB classifiers trained on the RI dataset effectively classify asteroids as \texttt{real} while maintaining a lower false positive rate. 
In contrast, the classifiers trained on the NGI dataset exhibit superior performance in detecting transient sources of interest but tend to have a higher false positive rate. 
This trade-off between detection performance and false positive rate can be addressed by training RB classifiers using the combined datasets:
Using the RI+NGI dataset as a training dataset increases performance on both asteroids and transient samples, but it still fails to reduce false positives originating from variable stars. Consequently, training with the RI+NGI$^\dagger$ dataset, which selectively includes \texttt{real} samples from the NGI dataset based on their distances from galaxies, improves performance in bogus filtering. These results demonstrate that the injection position is just as crucial as the PSF modeling in determining the behavior of RB classifiers.

This study emphasizes the importance of tailored injection strategies in RB classification. While RI method is well-suited for randomly distributed objects like asteroids, NGI more accurately represents transients near galaxies, such as KNe as GW optical counterparts, and SNe. The combined dataset approach, incorporating the strengths of both strategies, is particularly valuable in GW follow-up campaigns, where rapid and reliable detection of potential counterparts with minimal false positives is critical.
Moreover, our evaluation framework highlights that classification performance on balanced test datasets may not generalize to real-world imbalanced data. For instance, even with high precision on a balanced test dataset, an RB classifier may exhibit a large number of false positives due to the overwhelming number of \texttt{bogus} cutouts in real-world dataset. Consequently, we argue that it is essential to include evaluations on real-world imbalanced datasets and conduct sample-level post-analysis beyond focusing solely on performance scores. 


On the other hand, our results reveal inherent limitations in each injection strategy. Specifically, NGI’s higher recall for transients with host galaxies is offset by an increased false positive rate, particularly among variable sources, while RI shows limitations in detecting galaxy-associated transients. Future work should explore adaptive injection methods that dynamically consider different astrophysical environments, such as variable stars or active galactic nuclei (AGN), allowing classifiers to generalize across a broader range of transient types and distances. Moreover, in this experiment, variable sources present in science, reference, and subtraction images are treated as \texttt{bogus} during training, potentially introducing confusion in the model, especially for injections close to galactic nuclei. This results in lower RB scores for such cases. 
One possible approach involving minimal human effort is to additionally treat known variable stars as \texttt{real} samples during training. However, this approach may provide only limited improvement since many unidentified variable sources may still be present in the \texttt{bogus} class.
A more adaptive injection approach that includes variables as distinct classes could improve the model’s ability to identify significant astrophysical events, such as tidal disruption events (TDEs) and AGN flares, which occur near galactic centers and are potential GW optical counterparts. 

In addition, the diversity of science and reference images used in this experiment presents its own limitations. The science image dataset consists of six images obtained from actual KMTNet ToO observation campaigns, covering three observatories and two filters. While these images are representative of typical KMTNet ToO conditions, they do not encompass the full diversity of observational scenarios that may arise in practice.

For example, the current experiment does not cover cases where the seeing is exceptionally poor, or where sky background gradients are present due to Moon proximity or imperfect flat-fielding. Our dataset also lacks coverage of fields with varying source densities, and we did not investigate in detail how crowding affects the quality of subtracted images. If such diverse observational conditions can be more finely characterized and incorporated into the training process, the performance of model could be further improved for broader and more realistic applications.

The reference images were constructed from images obtained at different time epoch and generally have deeper depth than the science images. These reference images were not always taken at the same observatory as their corresponding science images, and such mismatches can influence the appearance of subtraction artifacts. The impact of different science–reference observatory pairings on the distribution of bogus detections was not systematically explored in this study, which we acknowledge as a limitation of the current analysis.

To conclude, this paper investigates the PSF injection technique for KMTNet images as part of an effort to develop an RB classifier specialized for KMTNet. Nonetheless, we believe our findings can generalize to other surveys. Additionally, in this study, we use only injected point sources as \texttt{real} samples for training classifiers to dissect the direct effect of PSF injection on performance for real-world samples. While this approach provides valuable insights, further performance improvements could be achieved by fine-tuning the classifiers with a small amount of real-world data.


\begin{acknowledgments}
\section*{Acknowledgments}
This work was supported by the National Research Foundation of Korea (NRF) grants, No. 2020R1A2C3011091, No. 2021R1A2C3009648, and No. 2021M3F7A1084525, funded by the Ministry of Science and ICT (MSIT). 

G.S.H.P. acknowledges support from the Pan-STARRS project, which is a project of the Institute for Astronomy of the University of Hawaii, and is supported by the NASA SSO Near Earth Observation Program under grants 80NSSC18K0971, NNX14AM74G, NNX12AR65G, NNX13AQ47G, NNX08AR22G, 80NSSC21K1572, and by the State of Hawaii.

SWC acknowledges the support from the Basic Science Research Program through the NRF funded by the Ministry of Education (RS-2023-00245013).

This research has made use of the KMTNet system operated by the Korea Astronomy and Space Science Institute (KASI) at three host sites of CTIO in Chile, SAAO in South Africa, and SSO in Australia.
Data transfer from the host site to KASI was supported by the Korea Research Environment Open NETwork (KREONET).

This work has made use of data from the European Space Agency (ESA) mission
{\it Gaia} (\url{https://www.cosmos.esa.int/gaia}), processed by the {\it Gaia}
Data Processing and Analysis Consortium (DPAC,
\url{https://www.cosmos.esa.int/web/gaia/dpac/consortium}). Funding for the DPAC
has been provided by national institutions, in particular the institutions
participating in the {\it Gaia} Multilateral Agreement.

\end{acknowledgments}

\facilities{KMTNet}

\software{
        astropy \citep{2013A&A...558A..33A,2018AJ....156..123A,2022ApJ...935..167A},
        HOTPANTS \citep{2015ascl.soft04004B}, 
        SExtractor \citep{1996A&AS..117..393B},
        SCAMP \citep{2010ascl.soft10063B},
        SWarp \citep{2010ascl.soft10068B}}

\bibliography{references}{}

\begin{thebibliography}{}
\expandafter\ifx\csname natexlab\endcsname\relax\def\natexlab#1{#1}\fi
\providecommand{\url}[1]{\href{#1}{#1}}
\providecommand{\dodoi}[1]{doi:~\href{http://doi.org/#1}{\nolinkurl{#1}}}
\providecommand{\doeprint}[1]{\href{http://ascl.net/#1}{\nolinkurl{http://ascl.net/#1}}}
\providecommand{\doarXiv}[1]{\href{https://arxiv.org/abs/#1}{\nolinkurl{https://arxiv.org/abs/#1}}}

\bibitem[{{Abbott} {et~al.}(2017){Abbott}, {Abbott}, {Abbott}, {Acernese}, {Ackley}, {Adams}, {Adams}, {Addesso}, {Adhikari}, {Adya}, {Affeldt}, {Afrough}, {Agarwal}, {Agathos}, {Agatsuma}, {Aggarwal}, {Aguiar}, {Aiello}, {Ain}, {Ajith}, {Allen}, {Allen}, {Allocca}, {Aloy}, {Altin}, {Amato}, {Ananyeva}, {Anderson}, {Anderson}, {Angelova}, {Antier}, {Appert}, {Arai}, {Araya}, {Areeda}, {Arnaud}, {Arun}, {Ascenzi}, {Ashton}, {Ast}, {Aston}, {Astone}, {Atallah}, {Aufmuth}, {Aulbert}, {AultONeal}, {Austin}, {Avila-Alvarez}, {Babak}, {Bacon}, {Bader}, {Bae}, {Baker}, {Baldaccini}, {Ballardin}, {Ballmer}, {Banagiri}, {Barayoga}, {Barclay}, {Barish}, {Barker}, {Barkett}, {Barone}, {Barr}, {Barsotti}, {Barsuglia}, {Barta}, {Bartlett}, {Bartos}, {Bassiri}, {Basti}, {Batch}, {Bawaj}, {Bayley}, {Bazzan}, {B{\'e}csy}, {Beer}, {Bejger}, {Belahcene}, {Bell}, {Berger}, {Bergmann}, {Bero}, {Berry}, {Bersanetti}, {Bertolini}, {Betzwieser}, {Bhagwat}, {Bhandare}, {Bilenko}, {Billingsley}, {Billman}, {Birch}, {Birney},
  {Birnholtz}, {Biscans}, {Biscoveanu}, {Bisht}, {Bitossi}, {Biwer}, {Bizouard}, {Blackburn}, {Blackman}, {Blair}, {Blair}, {Blair}, {Bloemen}, {Bock}, {Bode}, {Boer}, {Bogaert}, {Bohe}, {Bondu}, {Bonilla}, {Bonnand}, {Boom}, {Bork}, {Boschi}, {Bose}, {Bossie}, {Bouffanais}, {Bozzi}, {Bradaschia}, {Brady}, {Branchesi}, {Brau}, {Briant}, {Brillet}, {Brinkmann}, {Brisson}, {Brockill}, {Broida}, {Brooks}, {Brown}, {Brown}, {Brunett}, {Buchanan}, {Buikema}, {Bulik}, {Bulten}, {Buonanno}, {Buskulic}, {Buy}, {Byer}, {Cabero}, {Cadonati}, {Cagnoli}, {Cahillane}, {Calder{\'o}n Bustillo}, {Callister}, {Calloni}, {Camp}, {Canepa}, {Canizares}, {Cannon}, {Cao}, {Cao}, {Capano}, {Capocasa}, {Carbognani}, {Caride}, {Carney}, {Casanueva Diaz}, {Casentini}, {Caudill}, {Cavagli{\`a}}, {Cavalier}, {Cavalieri}, {Cella}, {Cepeda}, {Cerd{\'a}-Dur{\'a}n}, {Cerretani}, {Cesarini}, {Chamberlin}, {Chan}, {Chao}, {Charlton}, {Chase}, {Chassande-Mottin}, {Chatterjee}, {Chatziioannou}, {Cheeseboro}, {Chen}, {Chen}, {Chen}, {Cheng},
  {Chia}, {Chincarini}, {Chiummo}, {Chmiel}, {Cho}, {Cho}, {Chow}, {Christensen}, {Chu}, {Chua}, {Chua}, {Chung}, {Chung}, {Ciani}, {Ciolfi}, {Cirelli}, {Cirone}, {Clara}, {Clark}, {Clearwater}, {Cleva}, {Cocchieri}, {Coccia}, {Cohadon}, {Cohen}, {Colla}, {Collette}, {Cominsky}, {Constancio}, {Conti}, {Cooper}, {Corban}, {Corbitt}, {Cordero-Carri{\'o}n}, {Corley}, {Cornish}, {Corsi}, {Cortese}, {Costa}, {Coughlin}, {Coughlin}, {Coulon}, {Countryman}, {Couvares}, {Covas}, {Cowan}, {Coward}, {Cowart}, {Coyne}, {Coyne}, {Creighton}, {Creighton}, {Cripe}, {Crowder}, {Cullen}, {Cumming}, {Cunningham}, {Cuoco}, {Dal Canton}, {D{\'a}lya}, {Danilishin}, {D'Antonio}, {Danzmann}, {Dasgupta}, {Da Silva Costa}, {Dattilo}, {Dave}, {Davier}, {Davis}, {Daw}, {Day}, {De}, {DeBra}, {Degallaix}, {De Laurentis}, {Del{\'e}glise}, {Del Pozzo}, {Demos}, {Denker}, {Dent}, {De Pietri}, {Dergachev}, {De Rosa}, {DeRosa}, {De Rossi}, {DeSalvo}, {de Varona}, {Devenson}, {Dhurandhar}, {D{\'\i}az}, {Di Fiore}, {Di Giovanni}, {Di
  Girolamo}, {Di Lieto}, {Di Pace}, {Di Palma}, {Di Renzo}, {Doctor}, {Dolique}, {Donovan}, {Dooley}, {Doravari}, {Dorrington}, {Douglas}, {Dovale {\'A}lvarez}, {Downes}, {Drago}, {Dreissigacker}, {Driggers}, {Du}, {Ducrot}, {Dupej}, {Dwyer}, {Edo}, {Edwards}, {Effler}, {Eggenstein}, {Ehrens}, {Eichholz}, {Eikenberry}, {Eisenstein}, {Essick}, {Estevez}, {Etienne}, {Etzel}, {Evans}, {Evans}, {Factourovich}, {Fafone}, {Fair}, {Fairhurst}, {Fan}, {Farinon}, {Farr}, {Farr}, {Fauchon-Jones}, {Favata}, {Fays}, {Fee}, {Fehrmann}, {Feicht}, {Fejer}, {Fernandez-Galiana}, {Ferrante}, {Ferreira}, {Ferrini}, {Fidecaro}, {Finstad}, {Fiori}, {Fiorucci}, {Fishbach}, {Fisher}, {Fitz-Axen}, {Flaminio}, {Fletcher}, {Fong}, {Font}, {Forsyth}, {Forsyth}, {Fournier}, {Frasca}, {Frasconi}, {Frei}, {Freise}, {Frey}, {Frey}, {Fries}, {Fritschel}, {Frolov}, {Fulda}, {Fyffe}, {Gabbard}, {Gadre}, {Gaebel}, {Gair}, {Gammaitoni}, {Ganija}, {Gaonkar}, {Garcia-Quiros}, {Garufi}, {Gateley}, {Gaudio}, {Gaur}, {Gayathri}, {Gehrels}, {Gemme},
  {Genin}, {Gennai}, {George}, {George}, {Gergely}, {Germain}, {Ghonge}, {Ghosh}, {Ghosh}, {Ghosh}, {Giaime}, {Giardina}, {Giazotto}, {Gill}, {Glover}, {Goetz}, {Goetz}, {Gomes}, {Goncharov}, {Gonz{\'a}lez}, {Gonzalez Castro}, {Gopakumar}, {Gorodetsky}, {Gossan}, {Gosselin}, {Gouaty}, {Grado}, {Graef}, {Granata}, {Grant}, {Gras}, {Gray}, {Greco}, {Green}, {Gretarsson}, {Groot}, {Grote}, {Grunewald}, {Gruning}, {Guidi}, {Guo}, {Gupta}, {Gupta}, {Gushwa}, {Gustafson}, {Gustafson}, {Halim}, {Hall}, {Hall}, {Hamilton}, {Hammond}, {Haney}, {Hanke}, {Hanks}, {Hanna}, {Hannam}, {Hannuksela}, {Hanson}, {Hardwick}, {Harms}, {Harry}, {Harry}, {Hart}, {Haster}, {Haughian}, {Healy}, {Heidmann}, {Heintze}, {Heitmann}, {Hello}, {Hemming}, {Hendry}, {Heng}, {Hennig}, {Heptonstall}, {Heurs}, {Hild}, {Hinderer}, {Hoak}, {Hofman}, {Holt}, {Holz}, {Hopkins}, {Horst}, {Hough}, {Houston}, {Howell}, {Hreibi}, {Hu}, {Huerta}, {Huet}, {Hughey}, {Husa}, {Huttner}, {Huynh-Dinh}, {Indik}, {Inta}, {Intini}, {Isa}, {Isac}, {Isi}, {Iyer},
  {Izumi}, {Jacqmin}, {Jani}, {Jaranowski}, {Jawahar}, {Jim{\'e}nez-Forteza}, {Johnson}, {Johnson-McDaniel}, {Jones}, {Jones}, {Jonker}, {Ju}, {Junker}, {Kalaghatgi}, {Kalogera}, {Kamai}, {Kandhasamy}, {Kang}, {Kanner}, {Kapadia}, {Karki}, {Karvinen}, {Kasprzack}, {Kastaun}, {Katolik}, {Katsavounidis}, {Katzman}, {Kaufer}, {Kawabe}, {K{\'e}f{\'e}lian}, {Keitel}, {Kemball}, {Kennedy}, {Kent}, {Key}, {Khalili}, {Khan}, {Khan}, {Khan}, {Khazanov}, {Kijbunchoo}, {Kim}, {Kim}, {Kim}, {Kim}, {Kim}, {Kim}, {Kimbrell}, {King}, {King}, {Kinley-Hanlon}, {Kirchhoff}, {Kissel}, {Kleybolte}, {Klimenko}, {Knowles}, {Koch}, {Koehlenbeck}, {Koley}, {Kondrashov}, {Kontos}, {Korobko}, {Korth}, {Kowalska}, {Kozak}, {Kr{\"a}mer}, {Kringel}, {Krishnan}, {Kr{\'o}lak}, {Kuehn}, {Kumar}, {Kumar}, {Kumar}, {Kuo}, {Kutynia}, {Kwang}, {Lackey}, {Lai}, {Landry}, {Lang}, {Lange}, {Lantz}, {Lanza}, {Lartaux-Vollard}, {Lasky}, {Laxen}, {Lazzarini}, {Lazzaro}, {Leaci}, {Leavey}, {Lee}, {Lee}, {Lee}, {Lee}, {Lee}, {Lehmann}, {Lenon},
  {Leonardi}, {Leroy}, {Letendre}, {Levin}, {Li}, {Linker}, {Littenberg}, {Liu}, {Lo}, {Lockerbie}, {London}, {Lord}, {Lorenzini}, {Loriette}, {Lormand}, {Losurdo}, {Lough}, {Lousto}, {Lovelace}, {L{\"u}ck}, {Lumaca}, {Lundgren}, {Lynch}, {Ma}, {Macas}, {Macfoy}, {Machenschalk}, {MacInnis}, {Macleod}, {Maga{\~n}a Hernandez}, {Maga{\~n}a-Sandoval}, {Maga{\~n}a Zertuche}, {Magee}, {Majorana}, {Maksimovic}, {Man}, {Mandic}, {Mangano}, {Mansell}, {Manske}, {Mantovani}, {Marchesoni}, {Marion}, {M{\'a}rka}, {M{\'a}rka}, {Markakis}, {Markosyan}, {Markowitz}, {Maros}, {Marquina}, {Martelli}, {Martellini}, {Martin}, {Martin}, {Martynov}, {Mason}, {Massera}, {Masserot}, {Massinger}, {Masso-Reid}, {Mastrogiovanni}, {Matas}, {Matichard}, {Matone}, {Mavalvala}, {Mazumder}, {McCarthy}, {McClelland}, {McCormick}, {McCuller}, {McGuire}, {McIntyre}, {McIver}, {McManus}, {McNeill}, {McRae}, {McWilliams}, {Meacher}, {Meadors}, {Mehmet}, {Meidam}, {Mejuto-Villa}, {Melatos}, {Mendell}, {Mercer}, {Merilh}, {Merzougui}, {Meshkov},
  {Messenger}, {Messick}, {Metzdorff}, {Meyers}, {Miao}, {Michel}, {Middleton}, {Mikhailov}, {Milano}, {Miller}, {Miller}, {Miller}, {Millhouse}, {Milovich-Goff}, {Minazzoli}, {Minenkov}, {Ming}, {Mishra}, {Mitra}, {Mitrofanov}, {Mitselmakher}, {Mittleman}, {Moffa}, {Moggi}, {Mogushi}, {Mohan}, {Mohapatra}, {Montani}, {Moore}, {Moraru}, {Moreno}, {Morriss}, {Mours}, {Mow-Lowry}, {Mueller}, {Muir}, {Mukherjee}, {Mukherjee}, {Mukherjee}, {Mukund}, {Mullavey}, {Munch}, {Mu{\~n}iz}, {Muratore}, {Murray}, {Napier}, {Nardecchia}, {Naticchioni}, {Nayak}, {Neilson}, {Nelemans}, {Nelson}, {Nery}, {Neunzert}, {Nevin}, {Newport}, {Newton}, {Ng}, {Nguyen}, {Nichols}, {Nielsen}, {Nissanke}, {Nitz}, {Noack}, {Nocera}, {Nolting}, {North}, {Nuttall}, {Oberling}, {O'Dea}, {Ogin}, {Oh}, {Oh}, {Ohme}, {Okada}, {Oliver}, {Oppermann}, {Oram}, {O'Reilly}, {Ormiston}, {Ortega}, {O'Shaughnessy}, {Ossokine}, {Ottaway}, {Overmier}, {Owen}, {Pace}, {Page}, {Page}, {Pai}, {Pai}, {Palamos}, {Palashov}, {Palomba}, {Pal-Singh}, {Pan},
  {Pan}, {Pang}, {Pang}, {Pankow}, {Pannarale}, {Pant}, {Paoletti}, {Paoli}, {Papa}, {Parida}, {Parker}, {Pascucci}, {Pasqualetti}, {Passaquieti}, {Passuello}, {Patil}, {Patricelli}, {Pearlstone}, {Pedraza}, {Pedurand}, {Pekowsky}, {Pele}, {Penn}, {Perez}, {Perreca}, {Perri}, {Pfeiffer}, {Phelps}, {Piccinni}, {Pichot}, {Piergiovanni}, {Pierro}, {Pillant}, {Pinard}, {Pinto}, {Pirello}, {Pitkin}, {Poe}, {Poggiani}, {Popolizio}, {Porter}, {Post}, {Powell}, {Prasad}, {Pratt}, {Pratten}, {Predoi}, {Prestegard}, {Prijatelj}, {Principe}, {Privitera}, {Prodi}, {Prokhorov}, {Puncken}, {Punturo}, {Puppo}, {P{\"u}rrer}, {Qi}, {Quetschke}, {Quintero}, {Quitzow-James}, {Raab}, {Rabeling}, {Radkins}, {Raffai}, {Raja}, {Rajan}, {Rajbhandari}, {Rakhmanov}, {Ramirez}, {Ramos-Buades}, {Rapagnani}, {Raymond}, {Razzano}, {Read}, {Regimbau}, {Rei}, {Reid}, {Reitze}, {Ren}, {Reyes}, {Ricci}, {Ricker}, {Rieger}, {Riles}, {Rizzo}, {Robertson}, {Robie}, {Robinet}, {Rocchi}, {Rolland}, {Rollins}, {Roma}, {Romano}, {Romel}, {Romie},
  {Rosi{\'n}ska}, {Ross}, {Rowan}, {R{\"u}diger}, {Ruggi}, {Rutins}, {Ryan}, {Sachdev}, {Sadecki}, {Sadeghian}, {Sakellariadou}, {Salconi}, {Saleem}, {Salemi}, {Samajdar}, {Sammut}, {Sampson}, {Sanchez}, {Sanchez}, {Sanchis-Gual}, {Sandberg}, {Sanders}, {Sassolas}, {Sathyaprakash}, {Saulson}, {Sauter}, {Savage}, {Sawadsky}, {Schale}, {Scheel}, {Scheuer}, {Schmidt}, {Schmidt}, {Schnabel}, {Schofield}, {Sch{\"o}nbeck}, {Schreiber}, {Schuette}, {Schulte}, {Schutz}, {Schwalbe}, {Scott}, {Scott}, {Seidel}, {Sellers}, {Sengupta}, {Sentenac}, {Sequino}, {Sergeev}, {Shaddock}, {Shaffer}, {Shah}, {Shahriar}, {Shaner}, {Shao}, {Shapiro}, {Shawhan}, {Sheperd}, {Shoemaker}, {Shoemaker}, {Siellez}, {Siemens}, {Sieniawska}, {Sigg}, {Silva}, {Singer}, {Singh}, {Singhal}, {Sintes}, {Slagmolen}, {Smith}, {Smith}, {Smith}, {Somala}, {Son}, {Sonnenberg}, {Sorazu}, {Sorrentino}, {Souradeep}, {Spencer}, {Srivastava}, {Staats}, {Staley}, {Steinke}, {Steinlechner}, {Steinlechner}, {Steinmeyer}, {Stevenson}, {Stone}, {Stops},
  {Strain}, {Stratta}, {Strigin}, {Strunk}, {Sturani}, {Stuver}, {Summerscales}, {Sun}, {Sunil}, {Suresh}, {Sutton}, {Swinkels}, {Szczepa{\'n}czyk}, {Tacca}, {Tait}, {Talbot}, {Talukder}, {Tanner}, {T{\'a}pai}, {Taracchini}, {Tasson}, {Taylor}, {Taylor}, {Tewari}, {Theeg}, {Thies}, {Thomas}, {Thomas}, {Thomas}, {Thorne}, {Thorne}, {Thrane}, {Tiwari}, {Tiwari}, {Tokmakov}, {Toland}, {Tonelli}, {Tornasi}, {Torres-Forn{\'e}}, {Torrie}, {T{\"o}yr{\"a}}, {Travasso}, {Traylor}, {Trinastic}, {Tringali}, {Trozzo}, {Tsang}, {Tse}, {Tso}, {Tsukada}, {Tsuna}, {Tuyenbayev}, {Ueno}, {Ugolini}, {Unnikrishnan}, {Urban}, {Usman}, {Vahlbruch}, {Vajente}, {Valdes}, {van Bakel}, {van Beuzekom}, {van den Brand}, {Van Den Broeck}, {Vander-Hyde}, {van der Schaaf}, {van Heijningen}, {van Veggel}, {Vardaro}, {Varma}, {Vass}, {Vas{\'u}th}, {Vecchio}, {Vedovato}, {Veitch}, {Veitch}, {Venkateswara}, {Venugopalan}, {Verkindt}, {Vetrano}, {Vicer{\'e}}, {Viets}, {Vinciguerra}, {Vine}, {Vinet}, {Vitale}, {Vo}, {Vocca}, {Vorvick},
  {Vyatchanin}, {Wade}, {Wade}, {Wade}, {Walet}, {Walker}, {Wallace}, {Walsh}, {Wang}, {Wang}, {Wang}, {Wang}, {Wang}, {Ward}, {Warner}, {Was}, {Watchi}, {Weaver}, {Wei}, {Weinert}, {Weinstein}, {Weiss}, {Wen}, {Wessel}, {We{\ss}els}, {Westerweck}, {Westphal}, {Wette}, {Whelan}, {Whitcomb}, {Whiting}, {Whittle}, {Wilken}, {Williams}, {Williams}, {Williamson}, {Willis}, {Willke}, {Wimmer}, {Winkler}, {Wipf}, {Wittel}, {Woan}, {Woehler}, {Wofford}, {Wong}, {Worden}, {Wright}, {Wu}, {Wysocki}, {Xiao}, {Yamamoto}, {Yancey}, {Yang}, {Yap}, {Yazback}, {Yu}, {Yu}, {Yvert}, {Zadro{\.z}ny}, {Zanolin}, {Zelenova}, {Zendri}, {Zevin}, {Zhang}, {Zhang}, {Zhang}, {Zhang}, {Zhao}, {Zhou}, {Zhou}, {Zhu}, {Zhu}, {Zimmerman}, {Zucker}, {Zweizig}, {(LIGO Scientific Collaboration}, {Virgo Collaboration}, {Burns}, {Veres}, {Kocevski}, {Racusin}, {Goldstein}, {Connaughton}, {Briggs}, {Blackburn}, {Hamburg}, {Hui}, {von Kienlin}, {McEnery}, {Preece}, {Wilson-Hodge}, {Bissaldi}, {Cleveland}, {Gibby}, {Giles}, {Kippen}, {McBreen},
  {Meegan}, {Paciesas}, {Poolakkil}, {Roberts}, {Stanbro}, {Gamma-ray Burst Monitor}, {Savchenko}, {Ferrigno}, {Kuulkers}, {Bazzano}, {Bozzo}, {Brandt}, {Chenevez}, {Courvoisier}, {Diehl}, {Domingo}, {Hanlon}, {Jourdain}, {Laurent}, {Lebrun}, {Lutovinov}, {Mereghetti}, {Natalucci}, {Rodi}, {Roques}, {Sunyaev}, {Ubertini}, \& {(INTEGRAL}}]{2017ApJ...848L..13A}
{Abbott}, B.~P., {Abbott}, R., {Abbott}, T.~D., {et~al.} 2017, \apjl, 848, L13, \dodoi{10.3847/2041-8213/aa920c}

\bibitem[{Abbott {et~al.}(2020)Abbott, Abbott, Abraham, Acernese, Ackley, Adams, Adhikari, Adya, Affeldt, Agathos, Agatsuma, Aggarwal, Aguiar, Aich, Aiello, Ain, Ajith, Akcay, Allen, Allocca, Altin, Amato, Anand, Ananyeva, Anderson, Anderson, Angelova, Ansoldi, Antier, Appert, Arai, Araya, Areeda, Arène, Arnaud, Aronson, Arun, Asali, Ascenzi, Ashton, Aston, Astone, Aubin, Aufmuth, AultONeal, Austin, Avendano, Babak, Bacon, Badaracco, Bader, Bae, Baer, Baird, Baldaccini, Ballardin, Ballmer, Bals, Balsamo, Baltus, Banagiri, Bankar, Bankar, Barayoga, Barbieri, Barish, Barker, Barkett, Barneo, Barone, Barr, Barsotti, Barsuglia, Barta, Bartlett, Bartos, Bassiri, Basti, Bawaj, Bayley, Bazzan, Bécsy, Bejger, Belahcene, Bell, Beniwal, Benjamin, Benkel, Bentley, Bergamin, Berger, Bergmann, Bernuzzi, Berry, Bersanetti, Bertolini, Betzwieser, Bhandare, Bhandari, Bidler, Biggs, Bilenko, Billingsley, Birney, Birnholtz, Biscans, Bischi, Biscoveanu, Bisht, Bissenbayeva, Bitossi, Bizouard, Blackburn, Blackman, Blair,
  Blair, Blair, Bobba, Bode, Boer, Boetzel, Bogaert, Bondu, Bonilla, Bonnand, Booker, Boom, Bork, Boschi, Bose, Bossilkov, Bosveld, Bouffanais, Bozzi, Bradaschia, Brady, Bramley, Branchesi, Brau, Breschi, Briant, Briggs, Brighenti, Brillet, Brinkmann, Brito, Brockill, Brooks, Brooks, Brown, Brunett, Bruno, Bruntz, Buikema, Bulik, Bulten, Buonanno, Buskulic, Byer, Cabero, Cadonati, Cagnoli, Cahillane, Bustillo, Callaghan, Callister, Calloni, Camp, Canepa, Cannon, Cao, Cao, Carapella, Carbognani, Caride, Carney, Carullo, Diaz, Casentini, Castañeda, Caudill, Cavaglià, Cavalier, Cavalieri, Cella, Cerdá-Durán, Cesarini, Chaibi, Chakravarti, Chan, Chan, Chao, Charlton, Chase, Chassande-Mottin, Chatterjee, Chaturvedi, Chatziioannou, Chen, Chen, Chen, Cheng, Cheong, Chia, Chiadini, Chierici, Chincarini, Chiummo, Cho, Cho, Cho, Christensen, Chu, Chua, Chung, Chung, Ciani, Ciecielag, Cieślar, Ciobanu, Ciolfi, Cipriano, Cirone, Clara, Clark, Clearwater, Clesse, Cleva, Coccia, Cohadon, Cohen, Colleoni, Collette,
  Collins, Colpi, Constancio, Conti, Cooper, Corban, Corbitt, Cordero-Carrión, Corezzi, Corley, Cornish, Corre, Corsi, Cortese, Costa, Cotesta, Coughlin, Coughlin, Coulon, Countryman, Couvares, Covas, Coward, Cowart, Coyne, Coyne, Creighton, Creighton, Cripe, Croquette, Crowder, Cudell, Cullen, Cumming, Cummings, Cunningham, Cuoco, Curylo, Canton, Dálya, Dana, Daneshgaran-Bajastani, D’Angelo, Danilishin, D’Antonio, Danzmann, Darsow-Fromm, Dasgupta, Datrier, Dattilo, Dave, Davier, Davies, Davis, Daw, DeBra, Deenadayalan, Degallaix, Laurentis, Deléglise, Delfavero, Lillo, Pozzo, DeMarchi, D’Emilio, Demos, Dent, Pietri, Rosa, Rossi, DeSalvo, de~Varona, Dhurandhar, Díaz, Diaz-Ortiz, Dietrich, Fiore, Fronzo, Giorgio, Giovanni, Giovanni, Girolamo, Lieto, Ding, Pace, Palma, Renzo, Divakarla, Dmitriev, Doctor, Donovan, Dooley, Doravari, Dorrington, Downes, Drago, Driggers, Du, Ducoin, Dupej, Durante, D’Urso, Dwyer, Easter, Eddolls, Edelman, Edo, Edy, Effler, Ehrens, Eichholz, Eikenberry, Eisenmann,
  Eisenstein, Ejlli, Errico, Essick, Estelles, Estevez, Etienne, Etzel, Evans, Evans, Ewing, Fafone, Fairhurst, Fan, Farinon, Farr, Farr, Fauchon-Jones, Favata, Fays, Fazio, Feicht, Fejer, Feng, Fenyvesi, Ferguson, Fernandez-Galiana, Ferrante, Ferreira, Ferreira, Fidecaro, Fiori, Fiorucci, Fishbach, Fisher, Fittipaldi, Fitz-Axen, Fiumara, Flaminio, Floden, Flynn, Fong, Font, Forsyth, Fournier, Frasca, Frasconi, Frei, Freise, Frey, Frey, Fritschel, Frolov, Fronzè, Fulda, Fyffe, Gabbard, Gadre, Gaebel, Gair, Galaudage, Ganapathy, Ganguly, Gaonkar, García-Quirós, Garufi, Gateley, Gaudio, Gayathri, Gemme, Genin, Gennai, George, George, Gergely, Ghonge, Ghosh, Ghosh, Ghosh, Giacomazzo, Giaime, Giardina, Gibson, Gier, Gill, Glanzer, Gniesmer, Godwin, Goetz, Goetz, Gohlke, Goncharov, González, Gopakumar, Gossan, Gosselin, Gouaty, Grace, Grado, Granata, Grant, Gras, Grassia, Gray, Gray, Greco, Green, Green, Gretarsson, Griggs, Grignani, Grimaldi, Grimm, Grote, Grunewald, Gruning, Guidi, Guimaraes, Guixé, Gulati,
  Guo, Gupta, Gupta, Gupta, Gustafson, Gustafson, Haegel, Halim, Hall, Hamilton, Hammond, Haney, Hanke, Hanks, Hanna, Hannam, Hannuksela, Hansen, Hanson, Harder, Hardwick, Haris, Harms, Harry, Harry, Hasskew, Haster, Haughian, Hayes, Healy, Heidmann, Heintze, Heinze, Heitmann, Hellman, Hello, Hemming, Hendry, Heng, Hennes, Hennig, Heurs, Hild, Hinderer, Hoback, Hochheim, Hofgard, Hofman, Holgado, Holland, Holt, Holz, Hopkins, Horst, Hough, Howell, Hoy, Huang, Hübner, Huerta, Huet, Hughey, Hui, Husa, Huttner, Huxford, Huynh-Dinh, Idzkowski, Iess, Inchauspe, Ingram, Intini, Isac, Isi, Iyer, Jacqmin, Jadhav, Jadhav, James, Jani, Janthalur, Jaranowski, Jariwala, Jaume, Jenkins, Jiang, Johns, Johnson-McDaniel, Jones, Jones, Jones, Jones, Jones, Jonker, Ju, Junker, Kalaghatgi, Kalogera, Kamai, Kandhasamy, Kang, Kanner, Kapadia, Karki, Kashyap, Kasprzack, Kastaun, Katsanevas, Katsavounidis, Katzman, Kaufer, Kawabe, Kéfélian, Keitel, Keivani, Kennedy, Key, Khadka, Khalili, Khan, Khan, Khan, Khazanov, Khetan,
  Khursheed, Kijbunchoo, Kim, Kim, Kim, Kim, Kim, Kim, Kim, Kimball, King, Kinley-Hanlon, Kirchhoff, Kissel, Kleybolte, Klimenko, Knowles, Knyazev, Koch, Koehlenbeck, Koekoek, Koley, Kondrashov, Kontos, Koper, Korobko, Korth, Kovalam, Kozak, Kringel, Krishnendu, Królak, Krupinski, Kuehn, Kumar, Kumar, Kumar, Kumar, Kumar, Kuo, Kutynia, Lackey, Laghi, Lalande, Lam, Lamberts, Landry, Landry, Lane, Lang, Lange, Lantz, Lanza, Rosa, Lartaux-Vollard, Lasky, Laxen, Lazzarini, Lazzaro, Leaci, Leavey, Lecoeuche, Lee, Lee, Lee, Lee, Lee, Lehmann, Leroy, Letendre, Levin, Li, Li, li, Li, Li, Linde, Linker, Linley, Littenberg, Liu, Liu, Llorens-Monteagudo, Lo, Lockwood, London, Longo, Lorenzini, Loriette, Lormand, Losurdo, Lough, Lousto, Lovelace, Lück, Lumaca, Lundgren, Ma, Macas, Macfoy, MacInnis, Macleod, MacMillan, Macquet, Hernandez, Magaña-Sandoval, Magee, Majorana, Maksimovic, Malik, Man, Mandic, Mangano, Mansell, Manske, Mantovani, Mapelli, Marchesoni, Marion, Márka, Márka, Markakis, Markosyan, Markowitz,
  Maros, Marquina, Marsat, Martelli, Martin, Martin, Martinez, Martynov, Masalehdan, Mason, Massera, Masserot, Massinger, Masso-Reid, Mastrogiovanni, Matas, Matichard, Mavalvala, Maynard, McCann, McCarthy, McClelland, McCormick, McCuller, McGuire, McIsaac, McIver, McManus, McRae, McWilliams, Meacher, Meadors, Mehmet, Mehta, Villa, Melatos, Mendell, Mercer, Mereni, Merfeld, Merilh, Merritt, Merzougui, Meshkov, Messenger, Messick, Metzdorff, Meyers, Meylahn, Mhaske, Miani, Miao, Michaloliakos, Michel, Middleton, Milano, Miller, Millhouse, Mills, Milotti, Milovich-Goff, Minazzoli, Minenkov, Mishkin, Mishra, Mistry, Mitra, Mitrofanov, Mitselmakher, Mittleman, Mo, Mogushi, Mohapatra, Mohite, Molina-Ruiz, Mondin, Montani, Moore, Moraru, Morawski, Moreno, Morisaki, Mours, Mow-Lowry, Mozzon, Muciaccia, Mukherjee, Mukherjee, Mukherjee, Mukherjee, Mukund, Mullavey, Munch, Muñiz, Murray, Nagar, Nardecchia, Naticchioni, Nayak, Neil, Neilson, Nelemans, Nelson, Nery, Neunzert, Ng, Ng, Nguyen, Nguyen, Nichols, Nichols,
  Nissanke, Nocera, Noh, North, Nothard, Nuttall, Oberling, O’Brien, Oganesyan, Ogin, Oh, Oh, Ohme, Ohta, Okada, Oliver, Olivetto, Oppermann, Oram, O’Reilly, Ormiston, Ortega, O’Shaughnessy, Ossokine, Osthelder, Ottaway, Overmier, Owen, Pace, Pagano, Page, Pagliaroli, Pai, Pai, Palamos, Palashov, Palomba, Pan, Panda, Pang, Pankow, Pannarale, Pant, Paoletti, Paoli, Parida, Parker, Pascucci, Pasqualetti, Passaquieti, Passuello, Patricelli, Payne, Pearlstone, Pechsiri, Pedersen, Pedraza, Pele, Penn, Perego, Perez, Périgois, Perreca, Perriès, Petermann, Pfeiffer, Phelps, Phukon, Piccinni, Pichot, Piendibene, Piergiovanni, Pierro, Pillant, Pinard, Pinto, Piotrzkowski, Pirello, Pitkin, Plastino, Poggiani, Pong, Ponrathnam, Popolizio, Porter, Powell, Prajapati, Prasai, Prasanna, Pratten, Prestegard, Principe, Prodi, Prokhorov, Punturo, Puppo, Pürrer, Qi, Quetschke, Quinonez, Raab, Raaijmakers, Radkins, Radulesco, Raffai, Rafferty, Raja, Rajan, Rajbhandari, Rakhmanov, Ramirez, Ramos-Buades, Rana, Rao,
  Rapagnani, Raymond, Razzano, Read, Regimbau, Rei, Reid, Reitze, Rettegno, Ricci, Richardson, Richardson, Ricker, Riemenschneider, Riles, Rizzo, Robertson, Robinet, Rocchi, Rodriguez-Soto, Rolland, Rollins, Roma, Romanelli, Romano, Romel, Romero-Shaw, Romie, Rose, Rose, Rose, Rosińska, Rosofsky, Ross, Rowan, Rowlinson, Roy, Roy, Roy, Ruggi, Rutins, Ryan, Sachdev, Sadecki, Sakellariadou, Salafia, Salconi, Saleem, Salemi, Samajdar, Sanchez, Sanchez, Sanchis-Gual, Sanders, Santiago, Santos, Sarin, Sassolas, Sathyaprakash, Sauter, Savage, Savant, Sawant, Sayah, Schaetzl, Schale, Scheel, Scheuer, Schmidt, Schnabel, Schofield, Schönbeck, Schreiber, Schulte, Schutz, Schwarm, Schwartz, Scott, Scott, Seidel, Sellers, Sengupta, Sennett, Sentenac, Sequino, Sergeev, Setyawati, Shaddock, Shaffer, Shahriar, Sharma, Sharma, Shawhan, Shen, Shikauchi, Shink, Shoemaker, Shoemaker, Shukla, ShyamSundar, Siellez, Sieniawska, Sigg, Singer, Singh, Singh, Singha, Singhal, Sintes, Sipala, Skliris, Slagmolen, Slaven-Blair, Smetana,
  Smith, Smith, Somala, Son, Soni, Sorazu, Sordini, Sorrentino, Souradeep, Sowell, Spencer, Spera, Srivastava, Srivastava, Staats, Stachie, Standke, Steer, Steinhoff, Steinke, Steinlechner, Steinlechner, Steinmeyer, Stevenson, Stocks, Stops, Stover, Strain, Stratta, Strunk, Sturani, Stuver, Sudhagar, Sudhir, Summerscales, Sun, Sunil, Sur, Suresh, Sutton, Swinkels, Szczepańczyk, Tacca, Tait, Talbot, Tanasijczuk, Tanner, Tao, Tápai, Tapia, Martin, Tasson, Taylor, Tenorio, Terkowski, Thirugnanasambandam, Thomas, Thomas, Thompson, Thondapu, Thorne, Thrane, Tinsman, Saravanan, Tiwari, Tiwari, Tiwari, Toland, Tonelli, Tornasi, Torres-Forné, Torrie, e~Melo, Töyrä, Trail, Travasso, Traylor, Tringali, Tripathee, Trovato, Trudeau, Tsang, Tse, Tso, Tsukada, Tsuna, Tsutsui, Turconi, Ubhi, Ueno, Ugolini, Unnikrishnan, Urban, Usman, Utina, Vahlbruch, Vajente, Valdes, Valentini, van Bakel, van Beuzekom, van~den Brand, Broeck, Vander-Hyde, van~der Schaaf, Heijningen, van Veggel, Vardaro, Varma, Vass, Vasúth, Vecchio,
  Vedovato, Veitch, Veitch, Venkateswara, Venugopalan, Verkindt, Veske, Vetrano, Viceré, Viets, Vinciguerra, Vine, Vinet, Vitale, Vivanco, Vo, Vocca, Vorvick, Vyatchanin, Wade, Wade, Wade, Walet, Walker, Wallace, Wallace, Walsh, Wang, Wang, Wang, Ward, Warden, Warner, Was, Watchi, Weaver, Wei, Weinert, Weinstein, Weiss, Wellmann, Wen, Weßels, Westhouse, Wette, Whelan, Whiting, Whittle, Wilken, Williams, Willis, Willke, Winkler, Wipf, Wittel, Woan, Woehler, Wofford, Wong, Wright, Wu, Wysocki, Xiao, Yamamoto, Yang, Yang, Yang, Yap, Yazback, Yeeles, Yu, Yu, Yuen, Zadrożny, Zadrożny, Zanolin, Zelenova, Zendri, Zevin, Zhang, Zhang, Zhang, Zhao, Zhao, Zhou, Zhou, Zhu, Zimmerman, Zucker, Zweizig, Collaboration, \& Collaboration}]{gw190814}
Abbott, R., Abbott, T.~D., Abraham, S., {et~al.} 2020, The Astrophysical Journal Letters, 896, L44, \dodoi{10.3847/2041-8213/ab960f}

\bibitem[{Acero-Cuellar {et~al.}(2023)Acero-Cuellar, Bianco, Dobler, Sako, Qu, \& Collaboration}]{Acero-Cuellar_2023}
Acero-Cuellar, T., Bianco, F., Dobler, G., {et~al.} 2023, The Astronomical Journal, 166, 115, \dodoi{10.3847/1538-3881/ace9d8}

\bibitem[{{Ackley} {et~al.}(2020){Ackley}, {Amati}, {Barbieri}, {Bauer}, {Benetti}, {Bernardini}, {Bhirombhakdi}, {Botticella}, {Branchesi}, {Brocato}, {Bruun}, {Bulla}, {Campana}, {Cappellaro}, {Castro-Tirado}, {Chambers}, {Chaty}, {Chen}, {Ciolfi}, {Coleiro}, {Copperwheat}, {Covino}, {Cutter}, {D'Ammando}, {D'Avanzo}, {De Cesare}, {D'Elia}, {Della Valle}, {Denneau}, {De Pasquale}, {Dhillon}, {Dyer}, {Elias-Rosa}, {Evans}, {Eyles-Ferris}, {Fiore}, {Fraser}, {Fruchter}, {Fynbo}, {Galbany}, {Gall}, {Galloway}, {Getman}, {Ghirlanda}, {Gillanders}, {Gomboc}, {Gompertz}, {Gonz{\'a}lez-Fern{\'a}ndez}, {Gonz{\'a}lez-Gait{\'a}n}, {Grado}, {Greco}, {Gromadzki}, {Groot}, {Guti{\'e}rrez}, {Heikkil{\"a}}, {Heintz}, {Hjorth}, {Hu}, {Huber}, {Inserra}, {Izzo}, {Japelj}, {Jerkstrand}, {Jin}, {Jonker}, {Kankare}, {Kann}, {Kennedy}, {Kim}, {Klose}, {Kool}, {Kotak}, {Kuncarayakti}, {Lamb}, {Leloudas}, {Levan}, {Longo}, {Lowe}, {Lyman}, {Magnier}, {Maguire}, {Maiorano}, {Mandel}, {Mapelli}, {Mattila}, {McBrien}, {Melandri},
  {Micha{\l}owski}, {Milvang-Jensen}, {Moran}, {Nicastro}, {Nicholl}, {Nicuesa Guelbenzu}, {Nuttal}, {Oates}, {O'Brien}, {Onori}, {Palazzi}, {Patricelli}, {Perego}, {Torres}, {Perley}, {Pian}, {Pignata}, {Piranomonte}, {Poshyachinda}, {Possenti}, {Pumo}, {Quirola-V{\'a}squez}, {Ragosta}, {Ramsay}, {Rau}, {Rest}, {Reynolds}, {Rosetti}, {Rossi}, {Rosswog}, {Sabha}, {Sagu{\'e}s Carracedo}, {Salafia}, {Salmon}, {Salvaterra}, {Savaglio}, {Sbordone}, {Schady}, {Schipani}, {Schultz}, {Schweyer}, {Smartt}, {Smith}, {Smith}, {Sollerman}, {Srivastav}, {Stanway}, {Starling}, {Steeghs}, {Stratta}, {Stubbs}, {Tanvir}, {Testa}, {Thrane}, {Tonry}, {Turatto}, {Ulaczyk}, {van der Horst}, {Vergani}, {Walton}, {Watson}, {Wiersema}, {Wiik}, {Wyrzykowski}, {Yang}, {Yi}, \& {Young}}]{2020A&A...643A.113A}
{Ackley}, K., {Amati}, L., {Barbieri}, C., {et~al.} 2020, \aap, 643, A113, \dodoi{10.1051/0004-6361/202037669}

\bibitem[{{Alard}(2000)}]{2000A&AS..144..363A}
{Alard}, C. 2000, \aaps, 144, 363, \dodoi{10.1051/aas:2000214}

\bibitem[{{Alard} \& {Lupton}(1998)}]{1998ApJ...503..325A}
{Alard}, C., \& {Lupton}, R.~H. 1998, \apj, 503, 325, \dodoi{10.1086/305984}

\bibitem[{{Andreoni} {et~al.}(2019){Andreoni}, {Goldstein}, {Ahumada}, {Anand}, {Bulla}, {Dahiwale}, {de}, {Dhawan}, {Kasliwal}, {Kong}, {Perley}, {Sharma}, {Sollerman}, {Tzanidakis}, {Zhang}, {Cenko}, {Copperwheat}, {Coughlin}, {Kaplan}, \& {Growth Collaboration}}]{2019GCN.25362....1A}
{Andreoni}, I., {Goldstein}, D.~A., {Ahumada}, T., {et~al.} 2019, GRB Coordinates Network, 25362, 1

\bibitem[{{Astropy Collaboration} {et~al.}(2013){Astropy Collaboration}, {Robitaille}, {Tollerud}, {Greenfield}, {Droettboom}, {Bray}, {Aldcroft}, {Davis}, {Ginsburg}, {Price-Whelan}, {Kerzendorf}, {Conley}, {Crighton}, {Barbary}, {Muna}, {Ferguson}, {Grollier}, {Parikh}, {Nair}, {Unther}, {Deil}, {Woillez}, {Conseil}, {Kramer}, {Turner}, {Singer}, {Fox}, {Weaver}, {Zabalza}, {Edwards}, {Azalee Bostroem}, {Burke}, {Casey}, {Crawford}, {Dencheva}, {Ely}, {Jenness}, {Labrie}, {Lim}, {Pierfederici}, {Pontzen}, {Ptak}, {Refsdal}, {Servillat}, \& {Streicher}}]{2013A&A...558A..33A}
{Astropy Collaboration}, {Robitaille}, T.~P., {Tollerud}, E.~J., {et~al.} 2013, \aap, 558, A33, \dodoi{10.1051/0004-6361/201322068}

\bibitem[{{Astropy Collaboration} {et~al.}(2018){Astropy Collaboration}, {Price-Whelan}, {Sip{\H{o}}cz}, {G{\"u}nther}, {Lim}, {Crawford}, {Conseil}, {Shupe}, {Craig}, {Dencheva}, {Ginsburg}, {VanderPlas}, {Bradley}, {P{\'e}rez-Su{\'a}rez}, {de Val-Borro}, {Aldcroft}, {Cruz}, {Robitaille}, {Tollerud}, {Ardelean}, {Babej}, {Bach}, {Bachetti}, {Bakanov}, {Bamford}, {Barentsen}, {Barmby}, {Baumbach}, {Berry}, {Biscani}, {Boquien}, {Bostroem}, {Bouma}, {Brammer}, {Bray}, {Breytenbach}, {Buddelmeijer}, {Burke}, {Calderone}, {Cano Rodr{\'\i}guez}, {Cara}, {Cardoso}, {Cheedella}, {Copin}, {Corrales}, {Crichton}, {D'Avella}, {Deil}, {Depagne}, {Dietrich}, {Donath}, {Droettboom}, {Earl}, {Erben}, {Fabbro}, {Ferreira}, {Finethy}, {Fox}, {Garrison}, {Gibbons}, {Goldstein}, {Gommers}, {Greco}, {Greenfield}, {Groener}, {Grollier}, {Hagen}, {Hirst}, {Homeier}, {Horton}, {Hosseinzadeh}, {Hu}, {Hunkeler}, {Ivezi{\'c}}, {Jain}, {Jenness}, {Kanarek}, {Kendrew}, {Kern}, {Kerzendorf}, {Khvalko}, {King}, {Kirkby}, {Kulkarni},
  {Kumar}, {Lee}, {Lenz}, {Littlefair}, {Ma}, {Macleod}, {Mastropietro}, {McCully}, {Montagnac}, {Morris}, {Mueller}, {Mumford}, {Muna}, {Murphy}, {Nelson}, {Nguyen}, {Ninan}, {N{\"o}the}, {Ogaz}, {Oh}, {Parejko}, {Parley}, {Pascual}, {Patil}, {Patil}, {Plunkett}, {Prochaska}, {Rastogi}, {Reddy Janga}, {Sabater}, {Sakurikar}, {Seifert}, {Sherbert}, {Sherwood-Taylor}, {Shih}, {Sick}, {Silbiger}, {Singanamalla}, {Singer}, {Sladen}, {Sooley}, {Sornarajah}, {Streicher}, {Teuben}, {Thomas}, {Tremblay}, {Turner}, {Terr{\'o}n}, {van Kerkwijk}, {de la Vega}, {Watkins}, {Weaver}, {Whitmore}, {Woillez}, {Zabalza}, \& {Astropy Contributors}}]{2018AJ....156..123A}
{Astropy Collaboration}, {Price-Whelan}, A.~M., {Sip{\H{o}}cz}, B.~M., {et~al.} 2018, \aj, 156, 123, \dodoi{10.3847/1538-3881/aabc4f}

\bibitem[{{Astropy Collaboration} {et~al.}(2022){Astropy Collaboration}, {Price-Whelan}, {Lim}, {Earl}, {Starkman}, {Bradley}, {Shupe}, {Patil}, {Corrales}, {Brasseur}, {N{\"o}the}, {Donath}, {Tollerud}, {Morris}, {Ginsburg}, {Vaher}, {Weaver}, {Tocknell}, {Jamieson}, {van Kerkwijk}, {Robitaille}, {Merry}, {Bachetti}, {G{\"u}nther}, {Aldcroft}, {Alvarado-Montes}, {Archibald}, {B{\'o}di}, {Bapat}, {Barentsen}, {Baz{\'a}n}, {Biswas}, {Boquien}, {Burke}, {Cara}, {Cara}, {Conroy}, {Conseil}, {Craig}, {Cross}, {Cruz}, {D'Eugenio}, {Dencheva}, {Devillepoix}, {Dietrich}, {Eigenbrot}, {Erben}, {Ferreira}, {Foreman-Mackey}, {Fox}, {Freij}, {Garg}, {Geda}, {Glattly}, {Gondhalekar}, {Gordon}, {Grant}, {Greenfield}, {Groener}, {Guest}, {Gurovich}, {Handberg}, {Hart}, {Hatfield-Dodds}, {Homeier}, {Hosseinzadeh}, {Jenness}, {Jones}, {Joseph}, {Kalmbach}, {Karamehmetoglu}, {Ka{\l}uszy{\'n}ski}, {Kelley}, {Kern}, {Kerzendorf}, {Koch}, {Kulumani}, {Lee}, {Ly}, {Ma}, {MacBride}, {Maljaars}, {Muna}, {Murphy}, {Norman},
  {O'Steen}, {Oman}, {Pacifici}, {Pascual}, {Pascual-Granado}, {Patil}, {Perren}, {Pickering}, {Rastogi}, {Roulston}, {Ryan}, {Rykoff}, {Sabater}, {Sakurikar}, {Salgado}, {Sanghi}, {Saunders}, {Savchenko}, {Schwardt}, {Seifert-Eckert}, {Shih}, {Jain}, {Shukla}, {Sick}, {Simpson}, {Singanamalla}, {Singer}, {Singhal}, {Sinha}, {Sip{\H{o}}cz}, {Spitler}, {Stansby}, {Streicher}, {{\v{S}}umak}, {Swinbank}, {Taranu}, {Tewary}, {Tremblay}, {de Val-Borro}, {Van Kooten}, {Vasovi{\'c}}, {Verma}, {de Miranda Cardoso}, {Williams}, {Wilson}, {Winkel}, {Wood-Vasey}, {Xue}, {Yoachim}, {Zhang}, {Zonca}, \& {Astropy Project Contributors}}]{2022ApJ...935..167A}
{Astropy Collaboration}, {Price-Whelan}, A.~M., {Lim}, P.~L., {et~al.} 2022, \apj, 935, 167, \dodoi{10.3847/1538-4357/ac7c74}

\bibitem[{{Becker}(2015)}]{2015ascl.soft04004B}
{Becker}, A. 2015, {HOTPANTS: High Order Transform of PSF ANd Template Subtraction}, Astrophysics Source Code Library, record ascl:1504.004

\bibitem[{{Becker} {et~al.}(2012){Becker}, {Homrighausen}, {Connolly}, {Genovese}, {Owen}, {Bickerton}, \& {Lupton}}]{2012MNRAS.425.1341B}
{Becker}, A.~C., {Homrighausen}, D., {Connolly}, A.~J., {et~al.} 2012, \mnras, 425, 1341, \dodoi{10.1111/j.1365-2966.2012.21542.x}

\bibitem[{{Bellm} {et~al.}(2019){Bellm}, {Kulkarni}, {Graham}, {Dekany}, {Smith}, {Riddle}, {Masci}, {Helou}, {Prince}, {Adams}, {Barbarino}, {Barlow}, {Bauer}, {Beck}, {Belicki}, {Biswas}, {Blagorodnova}, {Bodewits}, {Bolin}, {Brinnel}, {Brooke}, {Bue}, {Bulla}, {Burruss}, {Cenko}, {Chang}, {Connolly}, {Coughlin}, {Cromer}, {Cunningham}, {De}, {Delacroix}, {Desai}, {Duev}, {Eadie}, {Farnham}, {Feeney}, {Feindt}, {Flynn}, {Franckowiak}, {Frederick}, {Fremling}, {Gal-Yam}, {Gezari}, {Giomi}, {Goldstein}, {Golkhou}, {Goobar}, {Groom}, {Hacopians}, {Hale}, {Henning}, {Ho}, {Hover}, {Howell}, {Hung}, {Huppenkothen}, {Imel}, {Ip}, {Ivezi{\'c}}, {Jackson}, {Jones}, {Juric}, {Kasliwal}, {Kaspi}, {Kaye}, {Kelley}, {Kowalski}, {Kramer}, {Kupfer}, {Landry}, {Laher}, {Lee}, {Lin}, {Lin}, {Lunnan}, {Giomi}, {Mahabal}, {Mao}, {Miller}, {Monkewitz}, {Murphy}, {Ngeow}, {Nordin}, {Nugent}, {Ofek}, {Patterson}, {Penprase}, {Porter}, {Rauch}, {Rebbapragada}, {Reiley}, {Rigault}, {Rodriguez}, {van Roestel}, {Rusholme}, {van
  Santen}, {Schulze}, {Shupe}, {Singer}, {Soumagnac}, {Stein}, {Surace}, {Sollerman}, {Szkody}, {Taddia}, {Terek}, {Van Sistine}, {van Velzen}, {Vestrand}, {Walters}, {Ward}, {Ye}, {Yu}, {Yan}, \& {Zolkower}}]{2019PASP..131a8002B}
{Bellm}, E.~C., {Kulkarni}, S.~R., {Graham}, M.~J., {et~al.} 2019, \pasp, 131, 018002, \dodoi{10.1088/1538-3873/aaecbe}

\bibitem[{{Berger}(2010)}]{2010ApJ...722.1946B}
{Berger}, E. 2010, \apj, 722, 1946, \dodoi{10.1088/0004-637X/722/2/1946}

\bibitem[{{Berger}(2014)}]{2014ARA&A..52...43B}
---. 2014, \araa, 52, 43, \dodoi{10.1146/annurev-astro-081913-035926}

\bibitem[{{Berthier} {et~al.}(2006){Berthier}, {Vachier}, {Thuillot}, {Fernique}, {Ochsenbein}, {Genova}, {Lainey}, \& {Arlot}}]{2006ASPC..351..367B}
{Berthier}, J., {Vachier}, F., {Thuillot}, W., {et~al.} 2006, in Astronomical Society of the Pacific Conference Series, Vol. 351, Astronomical Data Analysis Software and Systems XV, ed. C.~{Gabriel}, C.~{Arviset}, D.~{Ponz}, \& S.~{Enrique}, 367

\bibitem[{{Bertin}(2006)}]{2006ASPC..351..112B}
{Bertin}, E. 2006, in Astronomical Society of the Pacific Conference Series, Vol. 351, Astronomical Data Analysis Software and Systems XV, ed. C.~{Gabriel}, C.~{Arviset}, D.~{Ponz}, \& S.~{Enrique}, 112

\bibitem[{{Bertin}(2010{\natexlab{a}})}]{2010ascl.soft10068B}
{Bertin}, E. 2010{\natexlab{a}}, {SWarp: Resampling and Co-adding FITS Images Together}, Astrophysics Source Code Library, record ascl:1010.068

\bibitem[{{Bertin}(2010{\natexlab{b}})}]{2010ascl.soft10063B}
---. 2010{\natexlab{b}}, {SCAMP: Automatic Astrometric and Photometric Calibration}, Astrophysics Source Code Library, record ascl:1010.063

\bibitem[{{Bertin}(2011)}]{PSFex}
{Bertin}, E. 2011, in Astronomical Society of the Pacific Conference Series, Vol. 442, Astronomical Data Analysis Software and Systems XX, ed. I.~N. {Evans}, A.~{Accomazzi}, D.~J. {Mink}, \& A.~H. {Rots}, 435

\bibitem[{{Bertin} \& {Arnouts}(1996)}]{1996A&AS..117..393B}
{Bertin}, E., \& {Arnouts}, S. 1996, \aaps, 117, 393, \dodoi{10.1051/aas:1996164}

\bibitem[{{Bramich}(2008)}]{2008MNRAS.386L..77B}
{Bramich}, D.~M. 2008, \mnras, 386, L77, \dodoi{10.1111/j.1745-3933.2008.00464.x}

\bibitem[{{Bramich} {et~al.}(2013){Bramich}, {Horne}, {Albrow}, {Tsapras}, {Snodgrass}, {Street}, {Hundertmark}, {Kains}, {Arellano Ferro}, {Figuera}, \& {Giridhar}}]{2013MNRAS.428.2275B}
{Bramich}, D.~M., {Horne}, K., {Albrow}, M.~D., {et~al.} 2013, \mnras, 428, 2275, \dodoi{10.1093/mnras/sts184}

\bibitem[{{Chambers} {et~al.}(2016){Chambers}, {Magnier}, {Metcalfe}, {Flewelling}, {Huber}, {Waters}, {Denneau}, {Draper}, {Farrow}, {Finkbeiner}, {Holmberg}, {Koppenhoefer}, {Price}, {Rest}, {Saglia}, {Schlafly}, {Smartt}, {Sweeney}, {Wainscoat}, {Burgett}, {Chastel}, {Grav}, {Heasley}, {Hodapp}, {Jedicke}, {Kaiser}, {Kudritzki}, {Luppino}, {Lupton}, {Monet}, {Morgan}, {Onaka}, {Shiao}, {Stubbs}, {Tonry}, {White}, {Ba{\~n}ados}, {Bell}, {Bender}, {Bernard}, {Boegner}, {Boffi}, {Botticella}, {Calamida}, {Casertano}, {Chen}, {Chen}, {Cole}, {Deacon}, {Frenk}, {Fitzsimmons}, {Gezari}, {Gibbs}, {Goessl}, {Goggia}, {Gourgue}, {Goldman}, {Grant}, {Grebel}, {Hambly}, {Hasinger}, {Heavens}, {Heckman}, {Henderson}, {Henning}, {Holman}, {Hopp}, {Ip}, {Isani}, {Jackson}, {Keyes}, {Koekemoer}, {Kotak}, {Le}, {Liska}, {Long}, {Lucey}, {Liu}, {Martin}, {Masci}, {McLean}, {Mindel}, {Misra}, {Morganson}, {Murphy}, {Obaika}, {Narayan}, {Nieto-Santisteban}, {Norberg}, {Peacock}, {Pier}, {Postman}, {Primak}, {Rae}, {Rai},
  {Riess}, {Riffeser}, {Rix}, {R{\"o}ser}, {Russel}, {Rutz}, {Schilbach}, {Schultz}, {Scolnic}, {Strolger}, {Szalay}, {Seitz}, {Small}, {Smith}, {Soderblom}, {Taylor}, {Thomson}, {Taylor}, {Thakar}, {Thiel}, {Thilker}, {Unger}, {Urata}, {Valenti}, {Wagner}, {Walder}, {Walter}, {Watters}, {Werner}, {Wood-Vasey}, \& {Wyse}}]{2016arXiv161205560C}
{Chambers}, K.~C., {Magnier}, E.~A., {Metcalfe}, N., {et~al.} 2016, arXiv e-prints, arXiv:1612.05560, \dodoi{10.48550/arXiv.1612.05560}

\bibitem[{Corbett {et~al.}(2023)Corbett, Carney, Gonzalez, Fors, Galliher, Glazier, Howard, Law, Quimby, Ratzloff, \& Soto}]{Corbett_2023}
Corbett, H., Carney, J., Gonzalez, R., {et~al.} 2023, The Astrophysical Journal Supplement Series, 265, 63, \dodoi{10.3847/1538-4365/acbd41}

\bibitem[{Duev {et~al.}(2019)Duev, Mahabal, Masci, Graham, Rusholme, Walters, Karmarkar, Frederick, Kasliwal, Rebbapragada, \& Ward}]{10.1093/mnras/stz2357}
Duev, D.~A., Mahabal, A., Masci, F.~J., {et~al.} 2019, Monthly Notices of the Royal Astronomical Society, 489, 3582, \dodoi{10.1093/mnras/stz2357}

\bibitem[{Dálya {et~al.}(2022)Dálya, Díaz, Bouchet, Frei, Jasche, Lavaux, Macas, Mukherjee, Pálfi, de Souza, Wandelt, Bilicki, \& Raffai}]{D_lya_2022}
Dálya, G., Díaz, R., Bouchet, F.~R., {et~al.} 2022, Monthly Notices of the Royal Astronomical Society, 514, 1403–1411, \dodoi{10.1093/mnras/stac1443}

\bibitem[{{Fong} \& {Berger}(2013)}]{2013ApJ...776...18F}
{Fong}, W., \& {Berger}, E. 2013, \apj, 776, 18, \dodoi{10.1088/0004-637X/776/1/18}

\bibitem[{{Gaia Collaboration} {et~al.}(2021){Gaia Collaboration}, {Brown}, {Vallenari}, {Prusti}, {de Bruijne}, {Babusiaux}, {Biermann}, {Creevey}, {Evans}, {Eyer}, {Hutton}, {Jansen}, {Jordi}, {Klioner}, {Lammers}, {Lindegren}, {Luri}, {Mignard}, {Panem}, {Pourbaix}, {Randich}, {Sartoretti}, {Soubiran}, {Walton}, {Arenou}, {Bailer-Jones}, {Bastian}, {Cropper}, {Drimmel}, {Katz}, {Lattanzi}, {van Leeuwen}, {Bakker}, {Cacciari}, {Casta{\~n}eda}, {De Angeli}, {Ducourant}, {Fabricius}, {Fouesneau}, {Fr{\'e}mat}, {Guerra}, {Guerrier}, {Guiraud}, {Jean-Antoine Piccolo}, {Masana}, {Messineo}, {Mowlavi}, {Nicolas}, {Nienartowicz}, {Pailler}, {Panuzzo}, {Riclet}, {Roux}, {Seabroke}, {Sordo}, {Tanga}, {Th{\'e}venin}, {Gracia-Abril}, {Portell}, {Teyssier}, {Altmann}, {Andrae}, {Bellas-Velidis}, {Benson}, {Berthier}, {Blomme}, {Brugaletta}, {Burgess}, {Busso}, {Carry}, {Cellino}, {Cheek}, {Clementini}, {Damerdji}, {Davidson}, {Delchambre}, {Dell'Oro}, {Fern{\'a}ndez-Hern{\'a}ndez}, {Galluccio}, {Garc{\'\i}a-Lario},
  {Garcia-Reinaldos}, {Gonz{\'a}lez-N{\'u}{\~n}ez}, {Gosset}, {Haigron}, {Halbwachs}, {Hambly}, {Harrison}, {Hatzidimitriou}, {Heiter}, {Hern{\'a}ndez}, {Hestroffer}, {Hodgkin}, {Holl}, {Jan{\ss}en}, {Jevardat de Fombelle}, {Jordan}, {Krone-Martins}, {Lanzafame}, {L{\"o}ffler}, {Lorca}, {Manteiga}, {Marchal}, {Marrese}, {Moitinho}, {Mora}, {Muinonen}, {Osborne}, {Pancino}, {Pauwels}, {Petit}, {Recio-Blanco}, {Richards}, {Riello}, {Rimoldini}, {Robin}, {Roegiers}, {Rybizki}, {Sarro}, {Siopis}, {Smith}, {Sozzetti}, {Ulla}, {Utrilla}, {van Leeuwen}, {van Reeven}, {Abbas}, {Abreu Aramburu}, {Accart}, {Aerts}, {Aguado}, {Ajaj}, {Altavilla}, {{\'A}lvarez}, {{\'A}lvarez Cid-Fuentes}, {Alves}, {Anderson}, {Anglada Varela}, {Antoja}, {Audard}, {Baines}, {Baker}, {Balaguer-N{\'u}{\~n}ez}, {Balbinot}, {Balog}, {Barache}, {Barbato}, {Barros}, {Barstow}, {Bartolom{\'e}}, {Bassilana}, {Bauchet}, {Baudesson-Stella}, {Becciani}, {Bellazzini}, {Bernet}, {Bertone}, {Bianchi}, {Blanco-Cuaresma}, {Boch}, {Bombrun}, {Bossini},
  {Bouquillon}, {Bragaglia}, {Bramante}, {Breedt}, {Bressan}, {Brouillet}, {Bucciarelli}, {Burlacu}, {Busonero}, {Butkevich}, {Buzzi}, {Caffau}, {Cancelliere}, {C{\'a}novas}, {Cantat-Gaudin}, {Carballo}, {Carlucci}, {Carnerero}, {Carrasco}, {Casamiquela}, {Castellani}, {Castro-Ginard}, {Castro Sampol}, {Chaoul}, {Charlot}, {Chemin}, {Chiavassa}, {Cioni}, {Comoretto}, {Cooper}, {Cornez}, {Cowell}, {Crifo}, {Crosta}, {Crowley}, {Dafonte}, {Dapergolas}, {David}, {David}, {de Laverny}, {De Luise}, {De March}, {De Ridder}, {de Souza}, {de Teodoro}, {de Torres}, {del Peloso}, {del Pozo}, {Delbo}, {Delgado}, {Delgado}, {Delisle}, {Di Matteo}, {Diakite}, {Diener}, {Distefano}, {Dolding}, {Eappachen}, {Edvardsson}, {Enke}, {Esquej}, {Fabre}, {Fabrizio}, {Faigler}, {Fedorets}, {Fernique}, {Fienga}, {Figueras}, {Fouron}, {Fragkoudi}, {Fraile}, {Franke}, {Gai}, {Garabato}, {Garcia-Gutierrez}, {Garc{\'\i}a-Torres}, {Garofalo}, {Gavras}, {Gerlach}, {Geyer}, {Giacobbe}, {Gilmore}, {Girona}, {Giuffrida}, {Gomel}, {Gomez},
  {Gonzalez-Santamaria}, {Gonz{\'a}lez-Vidal}, {Granvik}, {Guti{\'e}rrez-S{\'a}nchez}, {Guy}, {Hauser}, {Haywood}, {Helmi}, {Hidalgo}, {Hilger}, {H{\l}adczuk}, {Hobbs}, {Holland}, {Huckle}, {Jasniewicz}, {Jonker}, {Juaristi Campillo}, {Julbe}, {Karbevska}, {Kervella}, {Khanna}, {Kochoska}, {Kontizas}, {Kordopatis}, {Korn}, {Kostrzewa-Rutkowska}, {Kruszy{\'n}ska}, {Lambert}, {Lanza}, {Lasne}, {Le Campion}, {Le Fustec}, {Lebreton}, {Lebzelter}, {Leccia}, {Leclerc}, {Lecoeur-Taibi}, {Liao}, {Licata}, {Lindstr{\o}m}, {Lister}, {Livanou}, {Lobel}, {Madrero Pardo}, {Managau}, {Mann}, {Marchant}, {Marconi}, {Marcos Santos}, {Marinoni}, {Marocco}, {Marshall}, {Martin Polo}, {Mart{\'\i}n-Fleitas}, {Masip}, {Massari}, {Mastrobuono-Battisti}, {Mazeh}, {McMillan}, {Messina}, {Michalik}, {Millar}, {Mints}, {Molina}, {Molinaro}, {Moln{\'a}r}, {Montegriffo}, {Mor}, {Morbidelli}, {Morel}, {Morris}, {Mulone}, {Munoz}, {Muraveva}, {Murphy}, {Musella}, {Noval}, {Ord{\'e}novic}, {Orr{\`u}}, {Osinde}, {Pagani}, {Pagano},
  {Palaversa}, {Palicio}, {Panahi}, {Pawlak}, {Pe{\~n}alosa Esteller}, {Penttil{\"a}}, {Piersimoni}, {Pineau}, {Plachy}, {Plum}, {Poggio}, {Poretti}, {Poujoulet}, {Pr{\v{s}}a}, {Pulone}, {Racero}, {Ragaini}, {Rainer}, {Raiteri}, {Rambaux}, {Ramos}, {Ramos-Lerate}, {Re Fiorentin}, {Regibo}, {Reyl{\'e}}, {Ripepi}, {Riva}, {Rixon}, {Robichon}, {Robin}, {Roelens}, {Rohrbasser}, {Romero-G{\'o}mez}, {Rowell}, {Royer}, {Rybicki}, {Sadowski}, {Sagrist{\`a} Sell{\'e}s}, {Sahlmann}, {Salgado}, {Salguero}, {Samaras}, {Sanchez Gimenez}, {Sanna}, {Santove{\~n}a}, {Sarasso}, {Schultheis}, {Sciacca}, {Segol}, {Segovia}, {S{\'e}gransan}, {Semeux}, {Shahaf}, {Siddiqui}, {Siebert}, {Siltala}, {Slezak}, {Smart}, {Solano}, {Solitro}, {Souami}, {Souchay}, {Spagna}, {Spoto}, {Steele}, {Steidelm{\"u}ller}, {Stephenson}, {S{\"u}veges}, {Szabados}, {Szegedi-Elek}, {Taris}, {Tauran}, {Taylor}, {Teixeira}, {Thuillot}, {Tonello}, {Torra}, {Torra}, {Turon}, {Unger}, {Vaillant}, {van Dillen}, {Vanel}, {Vecchiato}, {Viala}, {Vicente},
  {Voutsinas}, {Weiler}, {Wevers}, {Wyrzykowski}, {Yoldas}, {Yvard}, {Zhao}, {Zorec}, {Zucker}, {Zurbach}, \& {Zwitter}}]{2021A&A...649A...1G}
{Gaia Collaboration}, {Brown}, A.~G.~A., {Vallenari}, A., {et~al.} 2021, \aap, 649, A1, \dodoi{10.1051/0004-6361/202039657}

\bibitem[{{Goldstein} {et~al.}(2019){Goldstein}, {Perley}, {Andreoni}, {Kasliwal}, \& {Growth Collaboration}}]{2019GCN.25355....1G}
{Goldstein}, D.~A., {Perley}, D., {Andreoni}, I., {Kasliwal}, M.~M., \& {Growth Collaboration}. 2019, GRB Coordinates Network, 25355, 1

\bibitem[{{Goldstein} {et~al.}(2015){Goldstein}, {D'Andrea}, {Fischer}, {Foley}, {Gupta}, {Kessler}, {Kim}, {Nichol}, {Nugent}, {Papadopoulos}, {Sako}, {Smith}, {Sullivan}, {Thomas}, {Wester}, {Wolf}, {Abdalla}, {Banerji}, {Benoit-L{\'e}vy}, {Bertin}, {Brooks}, {Carnero Rosell}, {Castander}, {da Costa}, {Covarrubias}, {DePoy}, {Desai}, {Diehl}, {Doel}, {Eifler}, {Fausti Neto}, {Finley}, {Flaugher}, {Fosalba}, {Frieman}, {Gerdes}, {Gruen}, {Gruendl}, {James}, {Kuehn}, {Kuropatkin}, {Lahav}, {Li}, {Maia}, {Makler}, {March}, {Marshall}, {Martini}, {Merritt}, {Miquel}, {Nord}, {Ogando}, {Plazas}, {Romer}, {Roodman}, {Sanchez}, {Scarpine}, {Schubnell}, {Sevilla-Noarbe}, {Smith}, {Soares-Santos}, {Sobreira}, {Suchyta}, {Swanson}, {Tarle}, {Thaler}, \& {Walker}}]{2015AJ....150...82G}
{Goldstein}, D.~A., {D'Andrea}, C.~B., {Fischer}, J.~A., {et~al.} 2015, \aj, 150, 82, \dodoi{10.1088/0004-6256/150/3/82}

\bibitem[{{Grado} {et~al.}(2019){Grado}, {Cappellaro}, {Brocato}, {Covino}, {Getman}, {Greco}, {D'Avanzo}, {Rossi}, {Palazzi}, {Yang}, \& {Grawita Collaboration}}]{2019GCN.25669....1G}
{Grado}, A., {Cappellaro}, E., {Brocato}, E., {et~al.} 2019, GRB Coordinates Network, 25669, 1

\bibitem[{{Graham} {et~al.}(2019){Graham}, {Kulkarni}, {Bellm}, {Adams}, {Barbarino}, {Blagorodnova}, {Bodewits}, {Bolin}, {Brady}, {Cenko}, {Chang}, {Coughlin}, {De}, {Eadie}, {Farnham}, {Feindt}, {Franckowiak}, {Fremling}, {Gezari}, {Ghosh}, {Goldstein}, {Golkhou}, {Goobar}, {Ho}, {Huppenkothen}, {Ivezi{\'c}}, {Jones}, {Juric}, {Kaplan}, {Kasliwal}, {Kelley}, {Kupfer}, {Lee}, {Lin}, {Lunnan}, {Mahabal}, {Miller}, {Ngeow}, {Nugent}, {Ofek}, {Prince}, {Rauch}, {van Roestel}, {Schulze}, {Singer}, {Sollerman}, {Taddia}, {Yan}, {Ye}, {Yu}, {Barlow}, {Bauer}, {Beck}, {Belicki}, {Biswas}, {Brinnel}, {Brooke}, {Bue}, {Bulla}, {Burruss}, {Connolly}, {Cromer}, {Cunningham}, {Dekany}, {Delacroix}, {Desai}, {Duev}, {Feeney}, {Flynn}, {Frederick}, {Gal-Yam}, {Giomi}, {Groom}, {Hacopians}, {Hale}, {Helou}, {Henning}, {Hover}, {Hillenbrand}, {Howell}, {Hung}, {Imel}, {Ip}, {Jackson}, {Kaspi}, {Kaye}, {Kowalski}, {Kramer}, {Kuhn}, {Landry}, {Laher}, {Mao}, {Masci}, {Monkewitz}, {Murphy}, {Nordin}, {Patterson}, {Penprase},
  {Porter}, {Rebbapragada}, {Reiley}, {Riddle}, {Rigault}, {Rodriguez}, {Rusholme}, {van Santen}, {Shupe}, {Smith}, {Soumagnac}, {Stein}, {Surace}, {Szkody}, {Terek}, {Van Sistine}, {van Velzen}, {Vestrand}, {Walters}, {Ward}, {Zhang}, \& {Zolkower}}]{2019PASP..131g8001G}
{Graham}, M.~J., {Kulkarni}, S.~R., {Bellm}, E.~C., {et~al.} 2019, \pasp, 131, 078001, \dodoi{10.1088/1538-3873/ab006c}

\bibitem[{He {et~al.}(2016)He, Zhang, Ren, \& Sun}]{7780459}
He, K., Zhang, X., Ren, S., \& Sun, J. 2016, in 2016 IEEE Conference on Computer Vision and Pattern Recognition (CVPR), 770--778, \dodoi{10.1109/CVPR.2016.90}

\bibitem[{{Henden} \& {Munari}(2014)}]{2014CoSka..43..518H}
{Henden}, A., \& {Munari}, U. 2014, Contributions of the Astronomical Observatory Skalnate Pleso, 43, 518

\bibitem[{{Herner} {et~al.}(2019){Herner}, {Palmese}, {Soares-Santos}, {Tucker}, {Allam}, {Annis}, {Garcia}, {Morgan}, {Bachmann}, {Brout}, \& {Desgw Team}}]{2019GCN.25373....1H}
{Herner}, K., {Palmese}, A., {Soares-Santos}, M., {et~al.} 2019, GRB Coordinates Network, 25373, 1

\bibitem[{Hosenie {et~al.}(2021)Hosenie, Bloemen, Groot, Lyon, Scheers, Stappers, Stoppa, Vreeswijk, de~Wet, Wolt, K{\"o}rding, McBride, le~Poole, Paterson, Pieterse, \& Woudt}]{Hosenie2021MeerCRABMC}
Hosenie, Z., Bloemen, S., Groot, P.~J., {et~al.} 2021, Experimental Astronomy, 51, 319 .
\newblock \url{https://api.semanticscholar.org/CorpusID:233443813}

\bibitem[{{Im} {et~al.}(2020){Im}, {Kim}, \& {Paek}}]{2020grbg.conf...25I}
{Im}, M., {Kim}, J., \& {Paek}, G. S.~H. 2020, in Gamma-ray Bursts in the Gravitational Wave Era 2019, ed. T.~{Sakamoto}, M.~{Serino}, \& S.~{Sugita}, 25--27

\bibitem[{{Ivezi{\'c}} {et~al.}(2019){Ivezi{\'c}}, {Kahn}, {Tyson}, {Abel}, {Acosta}, {Allsman}, {Alonso}, {AlSayyad}, {Anderson}, {Andrew}, {Angel}, {Angeli}, {Ansari}, {Antilogus}, {Araujo}, {Armstrong}, {Arndt}, {Astier}, {Aubourg}, {Auza}, {Axelrod}, {Bard}, {Barr}, {Barrau}, {Bartlett}, {Bauer}, {Bauman}, {Baumont}, {Bechtol}, {Bechtol}, {Becker}, {Becla}, {Beldica}, {Bellavia}, {Bianco}, {Biswas}, {Blanc}, {Blazek}, {Blandford}, {Bloom}, {Bogart}, {Bond}, {Booth}, {Borgland}, {Borne}, {Bosch}, {Boutigny}, {Brackett}, {Bradshaw}, {Brandt}, {Brown}, {Bullock}, {Burchat}, {Burke}, {Cagnoli}, {Calabrese}, {Callahan}, {Callen}, {Carlin}, {Carlson}, {Chandrasekharan}, {Charles-Emerson}, {Chesley}, {Cheu}, {Chiang}, {Chiang}, {Chirino}, {Chow}, {Ciardi}, {Claver}, {Cohen-Tanugi}, {Cockrum}, {Coles}, {Connolly}, {Cook}, {Cooray}, {Covey}, {Cribbs}, {Cui}, {Cutri}, {Daly}, {Daniel}, {Daruich}, {Daubard}, {Daues}, {Dawson}, {Delgado}, {Dellapenna}, {de Peyster}, {de Val-Borro}, {Digel}, {Doherty}, {Dubois},
  {Dubois-Felsmann}, {Durech}, {Economou}, {Eifler}, {Eracleous}, {Emmons}, {Fausti Neto}, {Ferguson}, {Figueroa}, {Fisher-Levine}, {Focke}, {Foss}, {Frank}, {Freemon}, {Gangler}, {Gawiser}, {Geary}, {Gee}, {Geha}, {Gessner}, {Gibson}, {Gilmore}, {Glanzman}, {Glick}, {Goldina}, {Goldstein}, {Goodenow}, {Graham}, {Gressler}, {Gris}, {Guy}, {Guyonnet}, {Haller}, {Harris}, {Hascall}, {Haupt}, {Hernandez}, {Herrmann}, {Hileman}, {Hoblitt}, {Hodgson}, {Hogan}, {Howard}, {Huang}, {Huffer}, {Ingraham}, {Innes}, {Jacoby}, {Jain}, {Jammes}, {Jee}, {Jenness}, {Jernigan}, {Jevremovi{\'c}}, {Johns}, {Johnson}, {Johnson}, {Jones}, {Juramy-Gilles}, {Juri{\'c}}, {Kalirai}, {Kallivayalil}, {Kalmbach}, {Kantor}, {Karst}, {Kasliwal}, {Kelly}, {Kessler}, {Kinnison}, {Kirkby}, {Knox}, {Kotov}, {Krabbendam}, {Krughoff}, {Kub{\'a}nek}, {Kuczewski}, {Kulkarni}, {Ku}, {Kurita}, {Lage}, {Lambert}, {Lange}, {Langton}, {Le Guillou}, {Levine}, {Liang}, {Lim}, {Lintott}, {Long}, {Lopez}, {Lotz}, {Lupton}, {Lust}, {MacArthur}, {Mahabal},
  {Mandelbaum}, {Markiewicz}, {Marsh}, {Marshall}, {Marshall}, {May}, {McKercher}, {McQueen}, {Meyers}, {Migliore}, {Miller}, {Mills}, {Miraval}, {Moeyens}, {Moolekamp}, {Monet}, {Moniez}, {Monkewitz}, {Montgomery}, {Morrison}, {Mueller}, {Muller}, {Mu{\~n}oz Arancibia}, {Neill}, {Newbry}, {Nief}, {Nomerotski}, {Nordby}, {O'Connor}, {Oliver}, {Olivier}, {Olsen}, {O'Mullane}, {Ortiz}, {Osier}, {Owen}, {Pain}, {Palecek}, {Parejko}, {Parsons}, {Pease}, {Peterson}, {Peterson}, {Petravick}, {Libby Petrick}, {Petry}, {Pierfederici}, {Pietrowicz}, {Pike}, {Pinto}, {Plante}, {Plate}, {Plutchak}, {Price}, {Prouza}, {Radeka}, {Rajagopal}, {Rasmussen}, {Regnault}, {Reil}, {Reiss}, {Reuter}, {Ridgway}, {Riot}, {Ritz}, {Robinson}, {Roby}, {Roodman}, {Rosing}, {Roucelle}, {Rumore}, {Russo}, {Saha}, {Sassolas}, {Schalk}, {Schellart}, {Schindler}, {Schmidt}, {Schneider}, {Schneider}, {Schoening}, {Schumacher}, {Schwamb}, {Sebag}, {Selvy}, {Sembroski}, {Seppala}, {Serio}, {Serrano}, {Shaw}, {Shipsey}, {Sick}, {Silvestri},
  {Slater}, {Smith}, {Smith}, {Sobhani}, {Soldahl}, {Storrie-Lombardi}, {Stover}, {Strauss}, {Street}, {Stubbs}, {Sullivan}, {Sweeney}, {Swinbank}, {Szalay}, {Takacs}, {Tether}, {Thaler}, {Thayer}, {Thomas}, {Thornton}, {Thukral}, {Tice}, {Trilling}, {Turri}, {Van Berg}, {Vanden Berk}, {Vetter}, {Virieux}, {Vucina}, {Wahl}, {Walkowicz}, {Walsh}, {Walter}, {Wang}, {Wang}, {Warner}, {Wiecha}, {Willman}, {Winters}, {Wittman}, {Wolff}, {Wood-Vasey}, {Wu}, {Xin}, {Yoachim}, \& {Zhan}}]{2019ApJ...873..111I}
{Ivezi{\'c}}, {\v{Z}}., {Kahn}, S.~M., {Tyson}, J.~A., {et~al.} 2019, \apj, 873, 111, \dodoi{10.3847/1538-4357/ab042c}

\bibitem[{{Jeong} {et~al.}(2024){Jeong}, {Im}, {Paek}, {Chang}, {Lee}, {Kim}, {Lee}, \& {Gecko Team}}]{2024GCN.36343....1J}
{Jeong}, M., {Im}, M., {Paek}, G. S.~H., {et~al.} 2024, GRB Coordinates Network, 36343, 1

\bibitem[{{Kaiser} {et~al.}(2010){Kaiser}, {Burgett}, {Chambers}, {Denneau}, {Heasley}, {Jedicke}, {Magnier}, {Morgan}, {Onaka}, \& {Tonry}}]{2010SPIE.7733E..0EK}
{Kaiser}, N., {Burgett}, W., {Chambers}, K., {et~al.} 2010, in Society of Photo-Optical Instrumentation Engineers (SPIE) Conference Series, Vol. 7733, Ground-based and Airborne Telescopes III, ed. L.~M. {Stepp}, R.~{Gilmozzi}, \& H.~J. {Hall}, 77330E, \dodoi{10.1117/12.859188}

\bibitem[{{Kasen} {et~al.}(2017){Kasen}, {Metzger}, {Barnes}, {Quataert}, \& {Ramirez-Ruiz}}]{2017Natur.551...80K}
{Kasen}, D., {Metzger}, B., {Barnes}, J., {Quataert}, E., \& {Ramirez-Ruiz}, E. 2017, \nat, 551, 80, \dodoi{10.1038/nature24453}

\bibitem[{Killestein {et~al.}(2021)Killestein, Lyman, Steeghs, Ackley, Dyer, Ulaczyk, Cutter, Mong, Galloway, Dhillon, O’Brien, Ramsay, Poshyachinda, Kotak, Breton, Nuttall, Pallé, Pollacco, Thrane, Aukkaravittayapun, Awiphan, Burhanudin, Chote, Chrimes, Daw, Duffy, Eyles-Ferris, Gompertz, Heikkilä, Irawati, Kennedy, Levan, Littlefair, Makrygianni, Mata Sánchez, Mattila, Maund, McCormac, Mkrtichian, Mullaney, Rol, Sawangwit, Stanway, Starling, Strøm, Tooke, Wiersema, \& Williams}]{10.1093/mnras/stab633}
Killestein, T.~L., Lyman, J., Steeghs, D., {et~al.} 2021, Monthly Notices of the Royal Astronomical Society, 503, 4838, \dodoi{10.1093/mnras/stab633}

\bibitem[{{Kim} {et~al.}(2019){Kim}, {Im}, {Lee}, {Kim}, {Paek}, {Lim}, {Choi}, {Hwang}, {Kim}, {Park}, {Lee}, \& {et al.}}]{2019GCN.25342....1K}
{Kim}, J., {Im}, M., {Lee}, C.-U., {et~al.} 2019, GRB Coordinates Network, 25342, 1

\bibitem[{{Kim} {et~al.}(2021){Kim}, {Im}, {Paek}, {Lee}, {Kim}, {Chang}, {Choi}, {Hwang}, {Kang}, {Kim}, {Kim}, {Lee}, {Lim}, {Seo}, \& {Sung}}]{2021ApJ...916...47K}
{Kim}, J., {Im}, M., {Paek}, G. S.~H., {et~al.} 2021, \apj, 916, 47, \dodoi{10.3847/1538-4357/ac0446}

\bibitem[{{Kim} {et~al.}(2016{\natexlab{a}}){Kim}, {Lee}, {Park}, {Kim}, {Cha}, {Lee}, {Han}, {Chun}, \& {Yuk}}]{2016JKAS...49...37K}
{Kim}, S.-L., {Lee}, C.-U., {Park}, B.-G., {et~al.} 2016{\natexlab{a}}, Journal of Korean Astronomical Society, 49, 37, \dodoi{10.5303/JKAS.2016.49.1.37}

\bibitem[{{Kim} {et~al.}(2016{\natexlab{b}}){Kim}, {Cha}, {Lee}, {Kim}, {Park}, {Lee}, {Park}, {Kyeong}, \& {Chun}}]{2016PKAS...31...35K}
{Kim}, S.-L., {Cha}, S.-M., {Lee}, C.-U., {et~al.} 2016{\natexlab{b}}, Publication of Korean Astronomical Society, 31, 35, \dodoi{10.5303/PKAS.2016.31.3.035}

\bibitem[{Kingma \& Ba(2015)}]{adam}
Kingma, D.~P., \& Ba, J. 2015, in 3rd International Conference on Learning Representations, {ICLR} 2015, San Diego, CA, USA, May 7-9, 2015, Conference Track Proceedings, ed. Y.~Bengio \& Y.~LeCun.
\newblock \url{http://arxiv.org/abs/1412.6980}

\bibitem[{{Ligo Scientific Collaboration} {et~al.}(2023){Ligo Scientific Collaboration}, {VIRGO Collaboration}, \& {Kagra Collaboration}}]{2023GCN.33813....1L}
{Ligo Scientific Collaboration}, {VIRGO Collaboration}, \& {Kagra Collaboration}. 2023, GRB Coordinates Network, 33813, 1

\bibitem[{Liu \& Deng(2015)}]{7486599}
Liu, S., \& Deng, W. 2015, in 2015 3rd IAPR Asian Conference on Pattern Recognition (ACPR), 730--734, \dodoi{10.1109/ACPR.2015.7486599}

\bibitem[{Loshchilov \& Hutter(2017)}]{loshchilov2017sgdr}
Loshchilov, I., \& Hutter, F. 2017, in International Conference on Learning Representations.
\newblock \url{https://openreview.net/forum?id=Skq89Scxx}

\bibitem[{{Makhlouf, K.} {et~al.}(2022){Makhlouf, K.}, {Turpin, D.}, {Corre, D.}, {Karpov, S.}, {Kann, D. A.}, \& {Klotz, A.}}]{otrain}
{Makhlouf, K.}, {Turpin, D.}, {Corre, D.}, {et~al.} 2022, A\&A, 664, A81, \dodoi{10.1051/0004-6361/202142952}

\bibitem[{{Metzger}(2019)}]{2019LRR....23....1M}
{Metzger}, B.~D. 2019, Living Reviews in Relativity, 23, 1, \dodoi{10.1007/s41114-019-0024-0}

\bibitem[{Mong {et~al.}(2020)Mong, Ackley, Galloway, Killestein, Lyman, Steeghs, Dhillon, O’Brien, Ramsay, Poshyachinda, Kotak, Nuttall, Pallé, Pollacco, Thrane, Dyer, Ulaczyk, Cutter, McCormac, Chote, Levan, Marsh, Stanway, Gompertz, Wiersema, Chrimes, Obradovic, Mullaney, Daw, Littlefair, Maund, Makrygianni, Burhanudin, Starling, Eyles-Ferris, Tooke, Duffy, Aukkaravittayapun, Sawangwit, Awiphan, Mkrtichian, Irawati, Mattila, Heikkilä, Breton, Kennedy, Mata~Sánchez, \& Rol}]{10.1093/mnras/staa3096}
Mong, Y.-L., Ackley, K., Galloway, D.~K., {et~al.} 2020, Monthly Notices of the Royal Astronomical Society, 499, 6009, \dodoi{10.1093/mnras/staa3096}

\bibitem[{Nair \& Hinton(2010)}]{relu}
Nair, V., \& Hinton, G.~E. 2010, in Proceedings of the 27th International Conference on International Conference on Machine Learning, ICML'10 (Madison, WI, USA: Omnipress), 807–814

\bibitem[{{O'Connor} {et~al.}(2022){O'Connor}, {Troja}, {Dichiara}, {Beniamini}, {Cenko}, {Kouveliotou}, {Gonz{\'a}lez}, {Durbak}, {Gatkine}, {Kutyrev}, {Sakamoto}, {S{\'a}nchez-Ram{\'\i}rez}, \& {Veilleux}}]{2022MNRAS.515.4890O}
{O'Connor}, B., {Troja}, E., {Dichiara}, S., {et~al.} 2022, \mnras, 515, 4890, \dodoi{10.1093/mnras/stac1982}

\bibitem[{{Onken} {et~al.}(2019){Onken}, {Wolf}, {Bessell}, {Chang}, {Da Costa}, {Luvaul}, {Mackey}, {Schmidt}, \& {Shao}}]{2019PASA...36...33O}
{Onken}, C.~A., {Wolf}, C., {Bessell}, M.~S., {et~al.} 2019, \pasa, 36, e033, \dodoi{10.1017/pasa.2019.27}

\bibitem[{{Paek}(2023)}]{2023zndo...8321870P}
{Paek}, G. S.~H. 2023, {SilverRon/GeckoDigestor: v0.1 Preliminary Release}, 0.1,  Zenodo, \dodoi{10.5281/zenodo.8321870}

\bibitem[{{Paek} {et~al.}(2024){Paek}, {Im}, {Kim}, {Lim}, {Park}, {Choi}, {Kim}, {Barbieri}, {Salafia}, {Paek}, {Shin}, {Seo}, {Lee}, {Lee}, {Kim}, \& {Sung}}]{2024ApJ...960..113P}
{Paek}, G. S.~H., {Im}, M., {Kim}, J., {et~al.} 2024, \apj, 960, 113, \dodoi{10.3847/1538-4357/ad0238}

\bibitem[{{Paek} {et~al.}(2025){Paek}, {Im}, {Jeong}, {Chang}, {Hur}, {Hong}, {Kim}, {Lee}, {Lee}, {Lee}, {Jung}, {Kim}, {Lee}, {Lee}, \& {Kim}}]{2025ApJ...981...38P}
{Paek}, G. S.~H., {Im}, M., {Jeong}, M., {et~al.} 2025, \apj, 981, 38, \dodoi{10.3847/1538-4357/adaf99}

\bibitem[{{Rodr{\'\i}guez} {et~al.}(2019){Rodr{\'\i}guez}, {Meza-Retamal}, {Quirola}, {Olivares}, {Cartier}, {Tucker}, {Soares-Santos}, {Martinez-Vazquez}, {Garcia}, {Herner}, {Annis}, {Palmese}, {Sherman}, {Morgan}, {Bachmann}, {Davis}, \& {Desgw Team}}]{2019GCN.25423....1R}
{Rodr{\'\i}guez}, {\'O}., {Meza-Retamal}, N., {Quirola}, J., {et~al.} 2019, GRB Coordinates Network, 25423, 1

\bibitem[{{Soares-Santos} {et~al.}(2019){Soares-Santos}, {Tucker}, {Allam}, {Annis}, {Garcia}, {Herner}, {Davis}, {Sherman}, {Morgan}, {Vivas}, {Lidman}, {Malik}, \& {Desgw Team}}]{2019GCN.25336....1S}
{Soares-Santos}, M., {Tucker}, D., {Allam}, S., {et~al.} 2019, GRB Coordinates Network, 25336, 1

\bibitem[{Tan \& Le(2019)}]{pmlr-v97-tan19a}
Tan, M., \& Le, Q. 2019, in Proceedings of Machine Learning Research, Vol.~97, Proceedings of the 36th International Conference on Machine Learning, ed. K.~Chaudhuri \& R.~Salakhutdinov (PMLR), 6105--6114.
\newblock \url{https://proceedings.mlr.press/v97/tan19a.html}

\bibitem[{{Troja} {et~al.}(2017){Troja}, {Piro}, {van Eerten}, {Wollaeger}, {Im}, {Fox}, {Butler}, {Cenko}, {Sakamoto}, {Fryer}, {Ricci}, {Lien}, {Ryan}, {Korobkin}, {Lee}, {Burgess}, {Lee}, {Watson}, {Choi}, {Covino}, {D'Avanzo}, {Fontes}, {Gonz{\'a}lez}, {Khandrika}, {Kim}, {Kim}, {Lee}, {Lee}, {Kutyrev}, {Lim}, {S{\'a}nchez-Ram{\'\i}rez}, {Veilleux}, {Wieringa}, \& {Yoon}}]{2017Natur.551...71T}
{Troja}, E., {Piro}, L., {van Eerten}, H., {et~al.} 2017, \nat, 551, 71, \dodoi{10.1038/nature24290}

\bibitem[{{Tucker} {et~al.}(2019){Tucker}, {Allam}, {Wiesner}, {Cartier}, {Garcia}, {Palmese}, {Zenteno}, {Herner}, {Soares-Santos}, {Annis}, {Bachman}, {Makler}, {Sherman}, {Santana}, {Kilpatrick}, {Figueiredo}, {Galar}, {Espinoza}, \& {Desgw Team}}]{2019GCN.25379....1T}
{Tucker}, D., {Allam}, S., {Wiesner}, M., {et~al.} 2019, GRB Coordinates Network, 25379, 1

\bibitem[{{van Dokkum}(2001)}]{2001PASP..113.1420V}
{van Dokkum}, P.~G. 2001, \pasp, 113, 1420, \dodoi{10.1086/323894}

\end{thebibliography}
\bibliographystyle{aasjournal}

\end{document}